\documentclass[aps,prx,twocolumn,superscriptaddress]{revtex4-1}
\usepackage{graphicx,amssymb,amsmath,amsfonts,empheq}
\usepackage[dvipsnames]{xcolor}
\usepackage{color}
\usepackage{braket}
\usepackage{latexsym}
\usepackage{upgreek}
\usepackage{dsfont}
\usepackage{bm}
\usepackage{multirow}
\usepackage{enumitem}
\usepackage[normalem]{ulem}
\usepackage{soul}
\usepackage{url}
\usepackage{hyperref}
\hypersetup{
colorlinks = true,
linkcolor = [rgb]{0,0,0}, 
citecolor = [rgb]{0.13,0.55,0.13},
urlcolor = [rgb]{0.25, 0.41, 0.88}}
\usepackage{bbm}

\usepackage{tikz}
\usepackage{tikz-cd}
\usetikzlibrary{arrows}
\usetikzlibrary{intersections}
\usetikzlibrary{shapes.geometric}
\usetikzlibrary{decorations.pathmorphing, patterns,shapes}
\usetikzlibrary{decorations.markings}

\tikzset{
	partial ellipse/.style args={#1:#2:#3}{
		insert path={+ (#1:#3) arc (#1:#2:#3)}
	}
}

\tikzset{
	mid arrow/.style={postaction={decorate,decoration={
				markings,
				mark=at position .575 with {\arrow[#1]{stealth}}
	}}},
	near arrow/.style={postaction={decorate,decoration={
				markings,
				mark=at position .275 with {\arrow[#1]{stealth}}
	}}},
	far arrow/.style={postaction={decorate,decoration={
				markings,
				mark=at position .800 with {\arrow[#1]{stealth}}
	}}},
}

\pgfdeclarepatternformonly{south west lines}{\pgfqpoint{-0pt}{-0pt}}{\pgfqpoint{3pt}{3pt}}{\pgfqpoint{3pt}{3pt}}{
	\pgfsetlinewidth{0.4pt}
	\pgfpathmoveto{\pgfqpoint{0pt}{0pt}}
	\pgfpathlineto{\pgfqpoint{3pt}{3pt}}
	\pgfpathmoveto{\pgfqpoint{2.8pt}{-.2pt}}
	\pgfpathlineto{\pgfqpoint{3.2pt}{.2pt}}
	\pgfpathmoveto{\pgfqpoint{-.2pt}{2.8pt}}
	\pgfpathlineto{\pgfqpoint{.2pt}{3.2pt}}
	\pgfusepath{stroke}}

\newcommand{\Tr}{\operatorname{Tr}}

\newcommand{\SU}{\operatorname{SU}}

\newcommand{\SL}{\operatorname{SL}}

\newcommand{\calC}{\mathcal{C}}

\newcommand{\calO}{\mathcal{O}}

\newcommand{\figref}[1]{Fig.~\ref{#1}}

\newcommand{\XW}[1]{\textcolor{blue}{#1}}

\definecolor{Acolor}{RGB}{242,120,121}
\definecolor{Bcolor}{RGB}{130,230,130}
\definecolor{Ccolor}{RGB}{153,163,252}

\def\be{\begin{equation}}
\def\ee{\end{equation}}

\begin{document}

\title{Floquet's Refrigerator: Conformal Cooling in Driven Quantum Critical Systems}

\author{Xueda Wen} 
\affiliation{Department of Physics, Harvard University, Cambridge, MA 02138, USA}
\affiliation{Department of Physics, University of Colorado, Boulder, CO 80309, USA}
\author{Ruihua Fan} 
\affiliation{Department of Physics, Harvard University, Cambridge, MA 02138, USA}
\author{Ashvin Vishwanath} 
\affiliation{Department of Physics, Harvard University, Cambridge, MA 02138, USA}

\begin{abstract}

We propose a general method of cooling---periodic driving generated by spatially deformed Hamiltonians---and study it in general one-dimensional quantum critical systems 
described by a conformal field theory.
Our protocol is able to efficiently cool  a finite-temperature Gibbs (mixed) state down to zero temperature at prescribed sub-regions exponentially rapidly in Floquet time cycles.
%
At the same time, entropy and energy are transferred and localized to the complementary regions that shrink with time.
%
We derive these conclusions through an exact analytic solution of the full time evolution of reduced density matrices. We also use numerics in free-fermion lattice models as a benchmark and find remarkable agreement. Such conformal Floquet refrigerators open a promising new route to cooling synthetic quantum systems. 

\end{abstract}

\maketitle

\textit{Introduction:} Quantum phenomena in many-body systems reveal their full glories near zero temperature.
Famous examples in solid state materials including superconductivity and particularly the fractional quantum Hall effects, are not possible except at low enough temperatures~\cite{girvin2019modern}.
Ultracold quantum gases, which can reach the nanokelvin regime, provide a new and  controlled platform to investigate physics in a different realm.
They have already been shown to exhibit novel phenomena, such as Bose-Einstein condensation, Mott-superfluid transition and are turning to the realization of exotic topological phases ~\cite{anderson1995observation,davis1995bose,Greiner2002,BrowaeysSPT,AidelsburgerChiralFloquet,BlochSPT,Semeghini:2021wls,Leonard:2022ndq}.

Yet, currently achieved temperatures in several cold-atom platforms are still too high for the exploration of more exotic quantum phases, such as the high-$T_\text{c}$ superconductivity in Fermi-Hubbard lattice systems.
Finding efficient methods to realize even lower temperatures is a central demand and also one of the major challenges in the contemporary cold-atom experiments.
Besides its practical interest, solutions to the above problem can also provide new theoretical insights into non-ergodic quantum dynamics.

The guiding principle of cooling is to deposit extra energy and entropy in a reservoir that can be later discarded, e.g, bringing a bath with a lower entropy density into thermal contact with the system of interest~\cite{Catani2009,Mazurenko2016,Chiu2018,Kantian2018}.
However, the magnitude of the thermal conductance and the size of the bath place significant limitations on the time and resource efficiency, respectively.
Another method is adiabatic preparation, where the environment serves as the reservoir~\cite{Anders2010}. 
Namely, one initializes a simple state and slowly tunes parameters in the Hamiltonian \emph{uniformly in space} to reach the desired state. 
There are two principal bottlenecks in this approach: the entropy of the initial state; the minimal energy gap of the Hamiltonian as parameters are tuned, which is a fatal issue in the presence of unavoidable gap closing or in the preparation of ground states of critical systems.

\begin{figure}
\centering
\begin{tikzpicture}[x=0.75pt,y=0.75pt,yscale=-0.6,xscale=0.6]

\draw    (56.25,109.5) -- (500.75,109.5) ;
\draw  [color={rgb, 255:red, 0; green, 0; blue, 0 }  ,draw opacity=1 ][fill={rgb, 255:red, 0; green, 0; blue, 0 }  ,fill opacity=1 ] (110,109.25) .. controls (110,105.8) and (112.8,103) .. (116.25,103) .. controls (119.7,103) and (122.5,105.8) .. (122.5,109.25) .. controls (122.5,112.7) and (119.7,115.5) .. (116.25,115.5) .. controls (112.8,115.5) and (110,112.7) .. (110,109.25) -- cycle ;
\draw  [color={rgb, 255:red, 0; green, 0; blue, 0 }  ,draw opacity=1 ][fill={rgb, 255:red, 0; green, 0; blue, 0 }  ,fill opacity=1 ] (267,110.25) .. controls (267,106.8) and (269.8,104) .. (273.25,104) .. controls (276.7,104) and (279.5,106.8) .. (279.5,110.25) .. controls (279.5,113.7) and (276.7,116.5) .. (273.25,116.5) .. controls (269.8,116.5) and (267,113.7) .. (267,110.25) -- cycle ;
\draw  [color={rgb, 255:red, 0; green, 0; blue, 0 }  ,draw opacity=1 ][fill={rgb, 255:red, 0; green, 0; blue, 0 }  ,fill opacity=1 ] (424,110.25) .. controls (424,106.8) and (426.8,104) .. (430.25,104) .. controls (433.7,104) and (436.5,106.8) .. (436.5,110.25) .. controls (436.5,113.7) and (433.7,116.5) .. (430.25,116.5) .. controls (426.8,116.5) and (424,113.7) .. (424,110.25) -- cycle ;
\draw [color={rgb, 255:red, 144; green, 19; blue, 254 }  ,draw opacity=1 ]   (369.5,93) .. controls (389.59,77.56) and (405.84,77.02) .. (427.19,92.53) ;
\draw [shift={(428.5,93.5)}, rotate = 216.87] [fill={rgb, 255:red, 144; green, 19; blue, 254 }  ,fill opacity=1 ][line width=0.08]  [draw opacity=0] (12,-3) -- (0,0) -- (12,3) -- cycle    ;
\draw [color={rgb, 255:red, 144; green, 19; blue, 254 }  ,draw opacity=1 ]   (436.33,91.12) .. controls (455.99,76.75) and (472.16,77) .. (493.5,93) ;
\draw [shift={(434.5,92.5)}, rotate = 322.47] [fill={rgb, 255:red, 144; green, 19; blue, 254 }  ,fill opacity=1 ][line width=0.08]  [draw opacity=0] (12,-3) -- (0,0) -- (12,3) -- cycle    ;
\draw [color={rgb, 255:red, 144; green, 19; blue, 254 }  ,draw opacity=1 ]   (210.83,95) .. controls (230.92,79.56) and (247.17,79.02) .. (268.52,94.53) ;
\draw [shift={(269.83,95.5)}, rotate = 216.87] [fill={rgb, 255:red, 144; green, 19; blue, 254 }  ,fill opacity=1 ][line width=0.08]  [draw opacity=0] (12,-3) -- (0,0) -- (12,3) -- cycle    ;
\draw [color={rgb, 255:red, 144; green, 19; blue, 254 }  ,draw opacity=1 ]   (277.67,93.12) .. controls (297.33,78.75) and (313.49,79) .. (334.83,95) ;
\draw [shift={(275.83,94.5)}, rotate = 322.47] [fill={rgb, 255:red, 144; green, 19; blue, 254 }  ,fill opacity=1 ][line width=0.08]  [draw opacity=0] (12,-3) -- (0,0) -- (12,3) -- cycle    ;
\draw [color={rgb, 255:red, 144; green, 19; blue, 254 }  ,draw opacity=1 ]   (54.83,95) .. controls (74.92,79.56) and (91.17,79.02) .. (112.52,94.53) ;
\draw [shift={(113.83,95.5)}, rotate = 216.87] [fill={rgb, 255:red, 144; green, 19; blue, 254 }  ,fill opacity=1 ][line width=0.08]  [draw opacity=0] (12,-3) -- (0,0) -- (12,3) -- cycle    ;
\draw [color={rgb, 255:red, 144; green, 19; blue, 254 }  ,draw opacity=1 ]   (121.67,93.12) .. controls (141.33,78.75) and (157.49,79) .. (178.83,95) ;
\draw [shift={(119.83,94.5)}, rotate = 322.47] [fill={rgb, 255:red, 144; green, 19; blue, 254 }  ,fill opacity=1 ][line width=0.08]  [draw opacity=0] (12,-3) -- (0,0) -- (12,3) -- cycle    ;


    \node at (262pt,45pt){\small \textcolor[rgb]{0.56,0.07,1} {Entropy flow}}; 
   
    \draw [color={rgb, 255:red, 74; green, 144; blue, 226 }  ,draw opacity=1 ]  [dashed][thick][-stealth](85pt, 88pt)--(75pt,110pt);
    \draw [color={rgb, 255:red, 74; green, 144; blue, 226 }  ,draw opacity=1 ]  [dashed][thick][-stealth](155pt, 88pt)--(170pt,110pt);
   
   \node at (72pt,120pt){\small \textcolor[rgb]{0.82,0.12,0.01}{Heating region}}; 
   
    \node at (202pt,120pt){\small \textcolor[rgb]{0.82,0.12,0.01}{Cooling region}}; 
      
    \draw [xshift=115pt,yshift=76pt][thick](-31pt,0pt)..controls (-30pt,-2pt) and (-29pt,-5pt)..(-28pt,-40pt)..controls (-27pt,-5pt) and (-26pt,-2pt)..(-25pt,0pt);

    \draw [xshift=233pt,yshift=76pt][thick](-31pt,0pt)..controls (-30pt,-2pt) and (-29pt,-5pt)..(-28pt,-40pt)..controls (-27pt,-5pt) and (-26pt,-2pt)..(-25pt,0pt);      
        
    \draw [xshift=351pt,yshift=76pt][thick](-31pt,0pt)..controls (-30pt,-2pt) and (-29pt,-5pt)..(-28pt,-40pt)..controls (-27pt,-5pt) and (-26pt,-2pt)..(-25pt,0pt);    
\end{tikzpicture}
\caption{
Conformal Floquet cooling in a driven quantum critical system. The emergent hot spots (black) absorb entropy from the rest of the system,
resulting in extensive cooling regions. The energy and thermal entropy are accumulated at the hot spots.
}
\label{CoolingCartoon}
\end{figure}
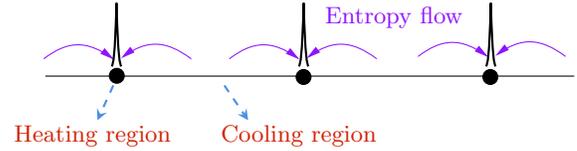

In this work, we engineer a class of Floquet driving protocols that are targeted at efficient cooling of general 1+1D critical systems that can be captured by conformal field theories (CFTs). Note, our results make no distinction between CFTs that are essentially free (e.g. free fermions) or ones that are strongly interacting. Instead, we utilize conformal invariance to arrive at a solution.
The Floquet unitary evolution is generated by \emph{spatially deformed Hamiltonians}, as studied extensively in recent works on Floquet CFTs~\cite{wen2018floquet,Wen_2018,Fan_2020,Lapierre:2019rwj,fan2020General, lapierre2020geometric,han2020classification,2020Lapierre,wen2020periodically,2020Andersen,2021Ageev,2021Das,RandomCFT2021,2022Das_OTOC,2022Bermond,2022Choo}.
There are dynamically generated hot spots serving as the reservoir, which are depicted as the black dots (heating regions) in \figref{CoolingCartoon}.
 They are points in the space and can absorb energy and entropy exponentially fast in time, which cool the rest of the system (cooling region in \figref{CoolingCartoon}) down to zero temperature.
Therefore time and resource efficiency are achieved simultaneously.
We will focus on the cooling of finite temperature Gibbs states. Generalizing our discussion to other initial states is straightforward and we believe the result does not change qualitatively.

The idea of cooling with spatially deformed Hamiltonians was first proposed and termed ``conformal cooling" by \cite{Norman1611}, though in the context of generic 1D quantum systems without any conformal symmetries, and was studied in critical systems only recently~\cite{kuzmin2021probing,goto2021nonequilibrating} (see also \cite{Hyperbolic2019} for similar ideas).
The above-mentioned works only consider quench dynamics for the cooling, while we employ Floquet driving, which is on one hand readily implemented in experiments and on the other hand enriches the flexibility of the protocol. We refer to our protocol as conformal Floquet cooling (CFC) to distinguish it from the previous works. 
We describe the exact calculation of three key quantities, the energy, von Neumann entropy and subregion density matrix, starting with an initial Gibbs state. These new theoretical results cement our understanding of cooling from Floquet conformal drives.

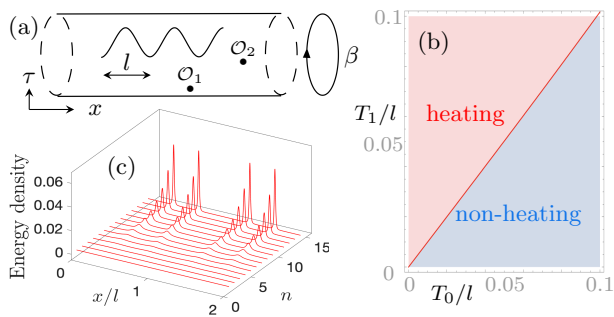
\begin{figure}[t]
\centering

\begin{tikzpicture}[baseline={(current bounding box.center)}][x=0.75pt,y=0.75pt,yscale=-0.5,xscale=0.5]


       \begin{scope}[xshift=-20pt,yshift=-10pt]            
\node[inner sep=0pt] (russell) at (175pt,10pt)
    {\includegraphics[width=.15\textwidth,height=0.2\textwidth]{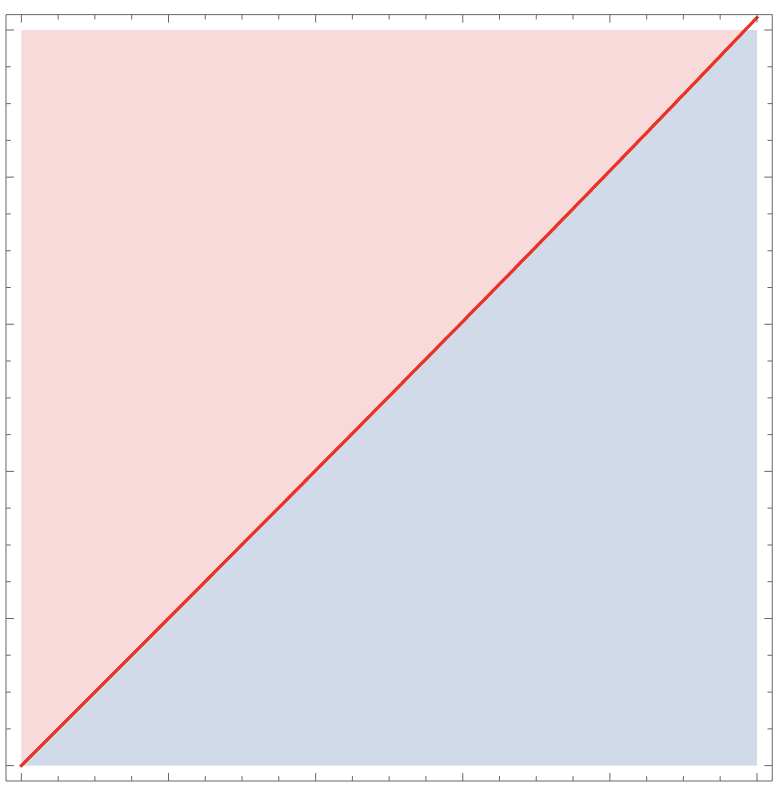}};
    
    \node at (131pt, -40pt){\footnotesize \textcolor[rgb]{0.61,0.61,0.61}{$0$}};
    \node at (129pt, 9pt){\footnotesize \textcolor[rgb]{0.61,0.61,0.61}{$0.05$}};   
        \node at (131pt, 57pt){\footnotesize \textcolor[rgb]{0.61,0.61,0.61}{$0.1$}};    
        
            \node at (126pt, 20pt){\footnotesize $T_1/l$};

               \node at (148pt, 47pt){(b)};               
               \node at (160pt, 20pt){\small\textcolor{red}{heating}};    
               \node at (180pt, -18pt){\small\textcolor[rgb]{0.06,0.45,0.91}{non-heating}};         
    \node at (139pt, -44pt){\footnotesize \textcolor[rgb]{0.61,0.61,0.61}{$0$}};
        \node at (175pt, -44pt){\footnotesize \textcolor[rgb]{0.61,0.61,0.61}{$0.05$}};
                \node at (211pt, -44pt){\footnotesize \textcolor[rgb]{0.61,0.61,0.61}{$0.1$}};

        \node at (155pt, -48pt){\footnotesize $T_0/l$};
    \end{scope}

    \node[inner sep=0pt] (russell) at (30pt,-25pt)
    {\includegraphics[width=.25\textwidth,height=0.17\textwidth]{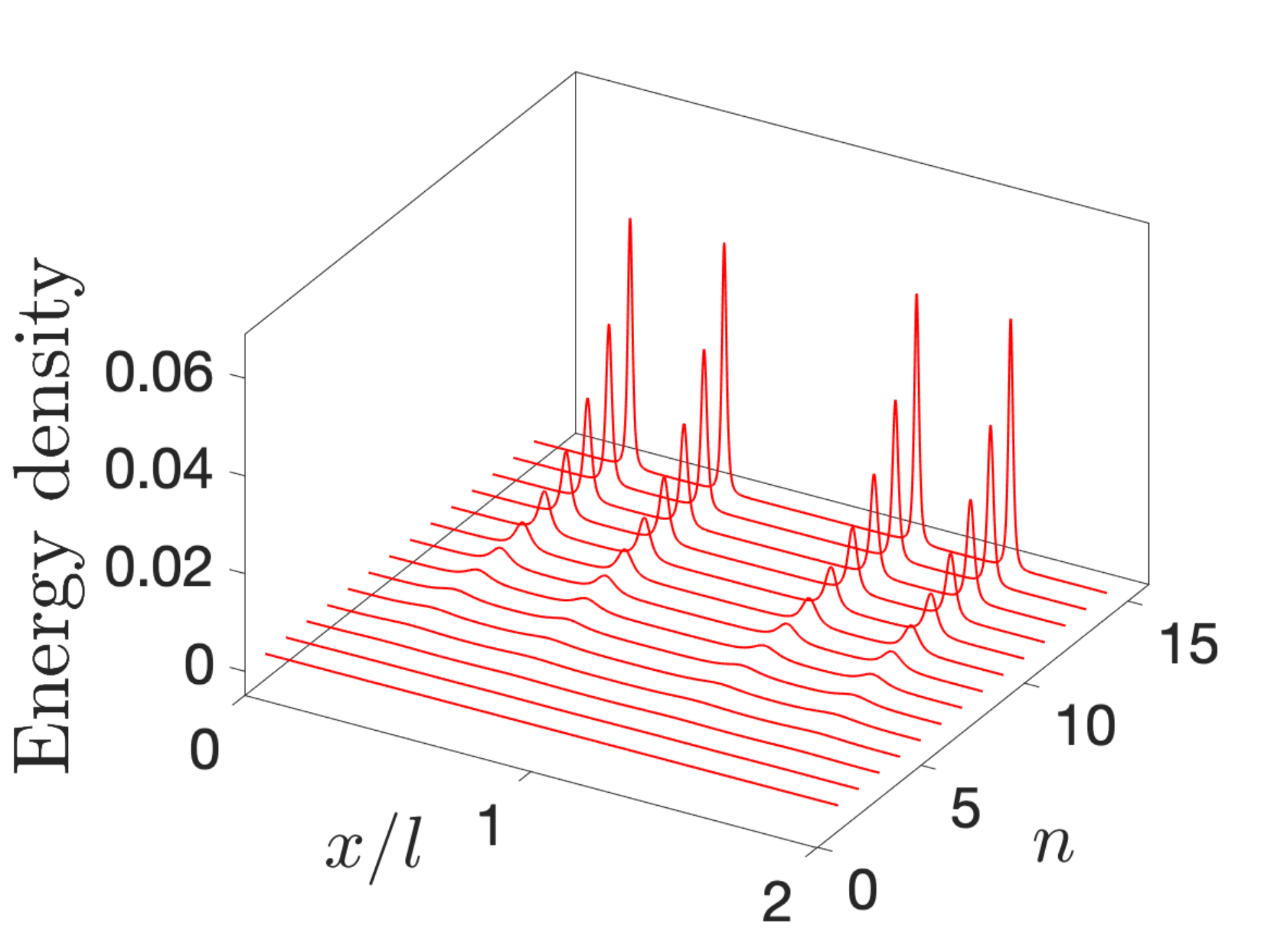}};


 \begin{scope}[xshift=-15pt,yshift=50pt]

\node[inner sep=0pt] (russell) at (50pt,28-45pt)
    {\includegraphics[width=.23\textwidth]{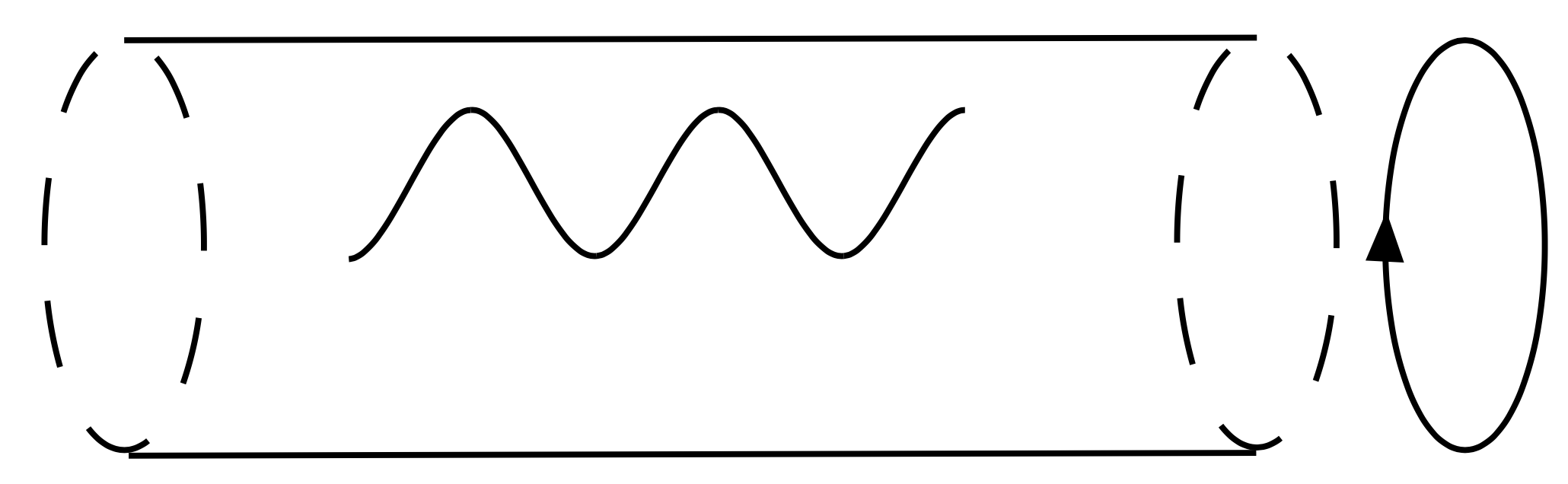}};
                   \node at (112pt, 27-45pt){$\beta$};      
\node at (-12pt,38-45pt){(a)};

   \draw  [-stealth](-10pt, 7-45pt)--(-10pt,15-45pt);
   \draw  [-stealth](-10pt, 7-45pt)--(7pt,7-45pt);
                   \node at (-10pt, 19-45pt){$\tau$};   
                   \node at (15pt, 7-45pt){$x$};   
                   
   \draw  [stealth-stealth](18pt, 20-45pt)--(36pt,20-45pt);
                   \node at (27pt, 25-45pt){\small $l$};   
                   
                                      \node at (71pt, 25-45pt){\tiny $\bullet$};   
                                      
                                       \node at (51pt, 15-45pt){\tiny $\bullet$};   
                                       
                                          \node at (51pt, 20-45pt){\scriptsize $\mathcal{O}_1$};   
                                            \node at (71pt, 30-45pt){\scriptsize $\mathcal{O}_2$};   
                   
            \end{scope}         
        
                        \node at (10pt, -10pt){(c)};
\end{tikzpicture}
\caption{
(a) Path integral representation of the two-point correlation $\langle \mathcal O_1(x_1,\tau_1)\mathcal O_2(x_2,\tau_2)\rangle$
under driving.
(b) Phase diagram for the choice of driving Hamiltonians with $v_0=1$ and $v_1=\cos\frac{2\pi x}{l}$
in \eqref{H_deform}. 
(c) Energy density distribution in the heating phase, where the energy density peaks correspond to the heating regions. We choose $T_0/l=1/50$ and $T_1/l=1/25$ with initial temperature $\beta/l=1/10$.
}
\label{PI_cooling}
\end{figure}

\textit{Setup:}
Consider a general critical system (CFT) initialized in a Gibbs state $\rho_{\text{th}}=e^{-\beta H}/Z$ at a finite temperature $\beta^{-1}$.
The periodic driving is generated by the following spatially deformed Hamiltonians 
\be
\label{H_deform}
H_i=\int \,\, v_i(x)T_{00}(x) dx .
\ee
Here $T_{00}(x)$ is the Hamiltonian density, and $v_i(x)$ characterizes the spatial deformation. 
We denote the uniform Hamiltonian by $H_0 = \int T_{00}(x) dx$.
A minimal setup is a two-step driving as follows:
\be
\begin{tikzpicture}[baseline={(current bounding box.center)}][x=0.75pt,y=0.75pt,yscale=-0.4,xscale=0.4]

\draw [thick](0pt,20pt)--(20pt,20pt);
\draw [thick](20pt,20pt)--(20pt,0pt);
\draw [thick](20pt,0pt)--(40pt,0pt);

\draw [thick](40pt,0pt)--(40pt,20pt);

\draw [thick](40pt,20pt)--(60pt,20pt);
\draw [thick](60pt,20pt)--(60pt,0pt);
\draw [thick](60pt,0pt)--(80pt,0pt);

\draw [thick](80pt,0pt)--(80pt,20pt);

\draw [thick](80pt,20pt)--(100pt,20pt);
\draw [thick](100pt,20pt)--(100pt,0pt);
\draw [thick](100pt,0pt)--(120pt,0pt);

\draw [thick](120pt,20pt)--(120pt,0pt);

\draw [thick](120pt,20pt)--(140pt,20pt);

\node at (150pt, 10pt){$\cdots$};

\node at (8pt,12pt){$H_1$};
\node at (30pt, 7pt){$H_0$};
\end{tikzpicture}
\ee
In each period, the system is driven with Hamiltonian $H_1$ for time $T_1$ and then with $H_0$ for $T_0$.
Our analytical analysis is for systems with an infinite length, while our numerics is done on a finite chain. The two results agree with each other remarkably.
The time evolution of the state can be dictated by the correlation functions, which we study based on the operator evolution~\cite{wen2018floquet,Wen_2018,Fan_2020}.
For the spatially deformed Hamiltonians, the associated unitary evolution is described by a conformal transformation. Let $U(t)=(e^{-iH_0T_0}e^{-iH_1T_1})^n$ be the unitary evolution operator at time $t=n(T_0+T_1)$, $\calO$ a primary operator with conformal dimension $(h,\,\bar h)$. We have
\begin{equation}
\small
\label{OperatorTransf}
U^{\dag}(t)\, \mathcal O(w,\bar{w})\, U(t)=\left(\frac{\partial w_n}{\partial w}\right)^{h}
\left(\frac{\partial \bar{w}_n}{\partial \bar{w}}\right)^{\bar{h}} \mathcal{O}\big(w_n,\bar{w}_n\big),
\end{equation}
where $w=\tau+ix$ is the coordinate of the spacetime cylinder of length $\beta$ in the imaginary time direction (Fig.\ref{PI_cooling}~(a)), and $w_n$ is related to $w$ via
\be
\label{gn_w}
w_n=g_n(w):=\underbrace{g\circ g\circ \cdots g}_{\text{$n$ times}}(w).
\ee
Here, the conformal transformation $g(w)$ is determined by $(H_i, \, T_i)$.
A more thorough characterization is via the reduced density matrix, $\rho_A(t)=\text{Tr}_{\bar A}(\rho(t))$ of some subsystem $A$,
where $\rho(t)=U(t)\, \rho_{\text{th}}\, U^\dag(t)$.
Equivalently, one can use the entanglement Hamiltonian
\be
K_A(t) = - \log \rho_A(t)\,.
\ee
In the following, we will provide both analytical and numerical results of the correlation functions and the entanglement Hamiltonian.

To obtain closed expressions for the time evolution, we adopt the deformation function $v_1 = \cos (2\pi x/l)$, where $l$ is the wavelength of deformation.
For such deformations with a single wavelength, the operator evolution is determined by
\be
\label{w_n}
\small
w_n=\frac{l}{2\pi }\log \left(\Pi_n\cdot e^{\frac{2\pi}{l}w}\right)\,,\, \Pi_n=
\begin{pmatrix}
\alpha_n &\beta_n\\
\beta_n^\ast &\alpha_n^\ast
\end{pmatrix},
\ee
where $\Pi_n$ is an SU(1,1) matrix~\cite{wen2020periodically,RandomCFT2021}.
The phase diagram, as we tune parameters $T_0$ and $T_1$, is shown in \figref{PI_cooling}~(b). 
In the non-heating, heating phases, and at the phase transition, the operator has no fixed point, flows to a fixed point exponentially fast or polynomially fast, respectively. 
We remark that the phase diagram is only determined by the operator evolution in \eqref{OperatorTransf} and independent of the choice of initial state, which is the underlying reason that our cooling method can be applied generally. 
We will show that, paradoxically, cooling occurs in the {\em heating} phase, where there are emergent hot spots that absorb the energy and entropy. The behaviors of the dynamics in the non-heating phase and at the phase transition will also be commented at the very end.

\textit{Heating and cooling regions:}
Let us identify the heating and cooling regions via the time evolution of energy density $E(x,n):=\langle T_{00}(x,n)\rangle$.
It can be separated into a chiral and anti-chiral component, i.e. $2\pi T_{00} = T+ \bar T$.
The result of the chiral part is given by
\be\label{Expectation_finiteT}
\small
\langle T(x,n) \rangle
=-\frac{\pi^2 c}{6\, l^2}+
 \left(\frac{\pi^2 c}{6\beta^2}+\frac{\pi^2 c}{6\, l^2}
\right)\cdot
\left|\alpha_n\, e^{i\frac{2\pi x}{l}}+\beta_n\right|^{-4}
\ee
where $c$ is the central charge, and similarly for the anti-chiral part. As a sanity check, we have $\alpha_n=1$ and $\beta_n=0$ at $n=0$, and we recover the thermal energy density $E(x) = \pi c/6\beta^2$.

\begin{figure}[t]
\centering
\begin{tikzpicture}

    \node[inner sep=0pt] (russell) at (15pt,-85pt)
    {\includegraphics[width=.25\textwidth]{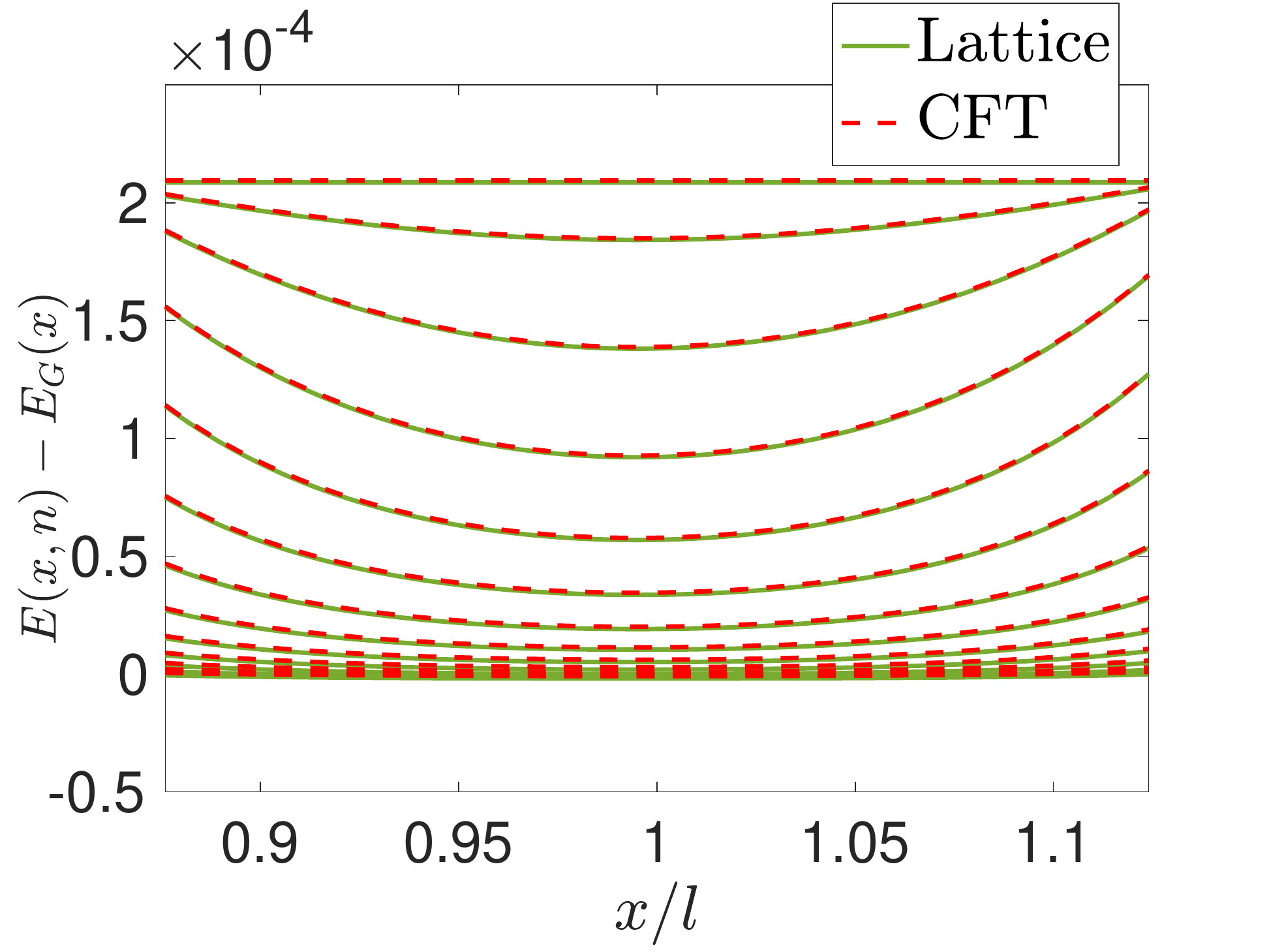}};
        \node[inner sep=0pt] (russell) at (145pt,-85pt)
    {\includegraphics[width=.25\textwidth]{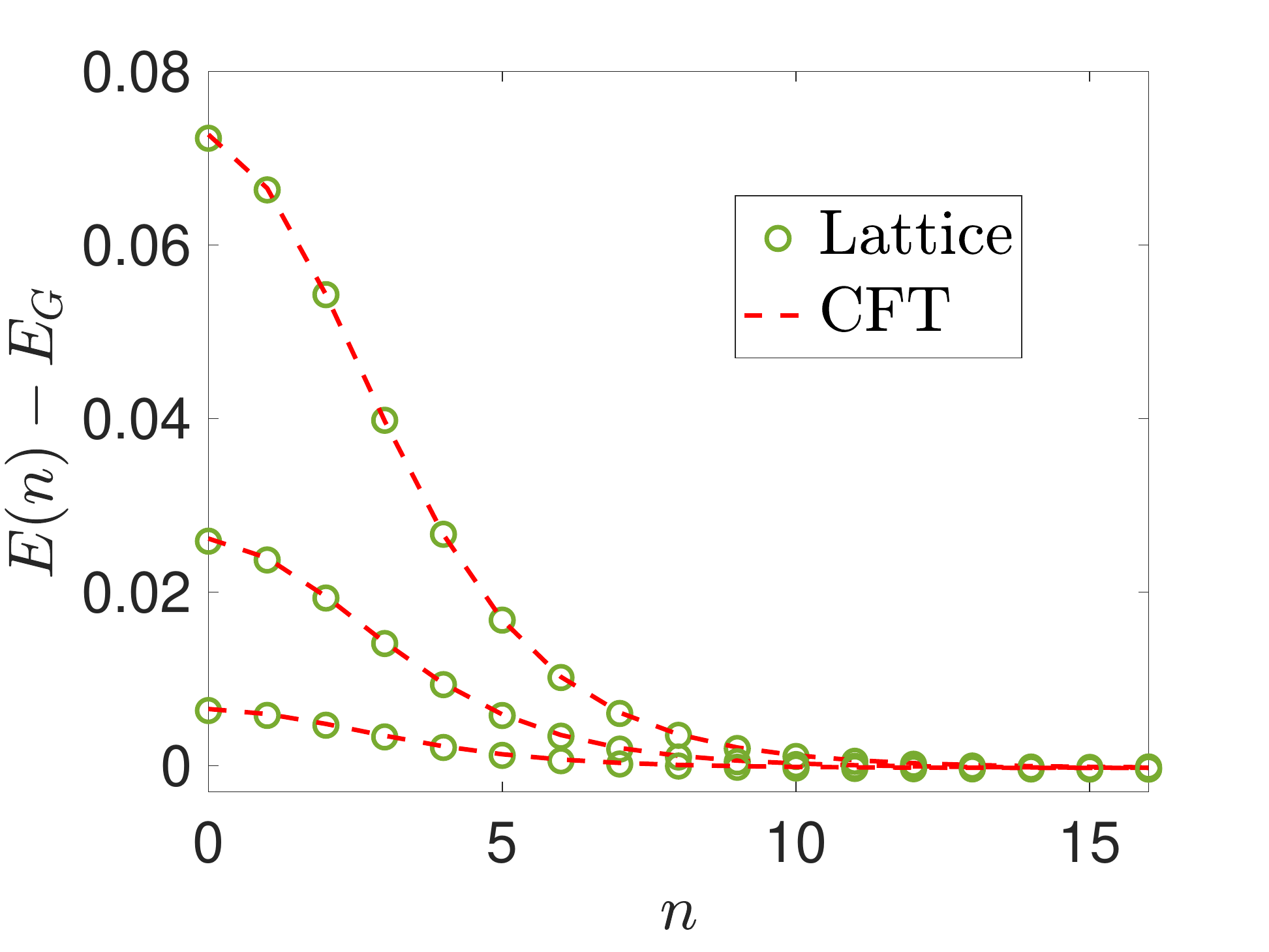}};
    \node[inner sep=0pt] (russell) at (15pt,-180pt)
    {\includegraphics[width=.25\textwidth]{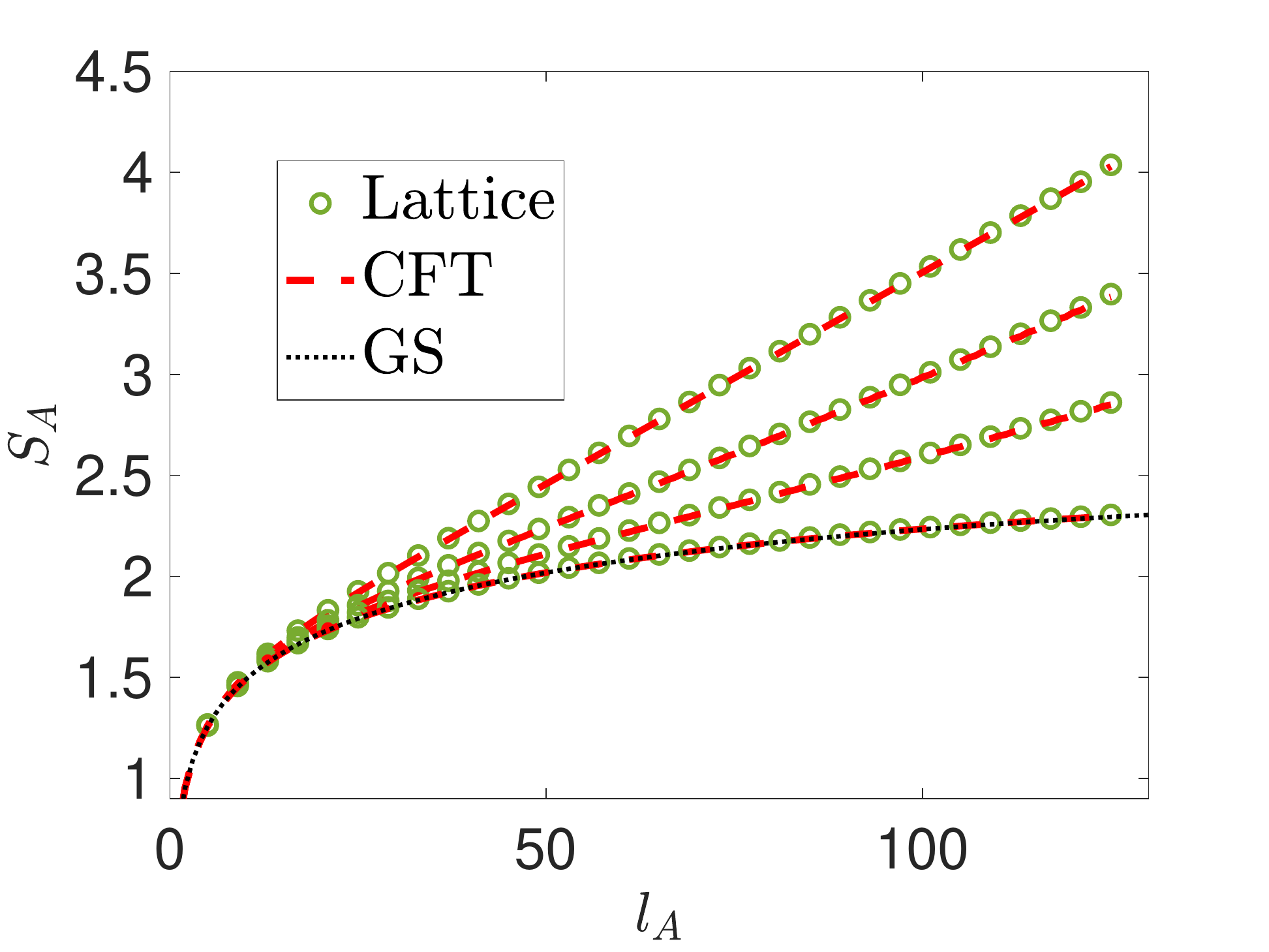}};    
    \node[inner sep=0pt] (russell) at (145pt,-180pt)
    {\includegraphics[width=.25\textwidth]{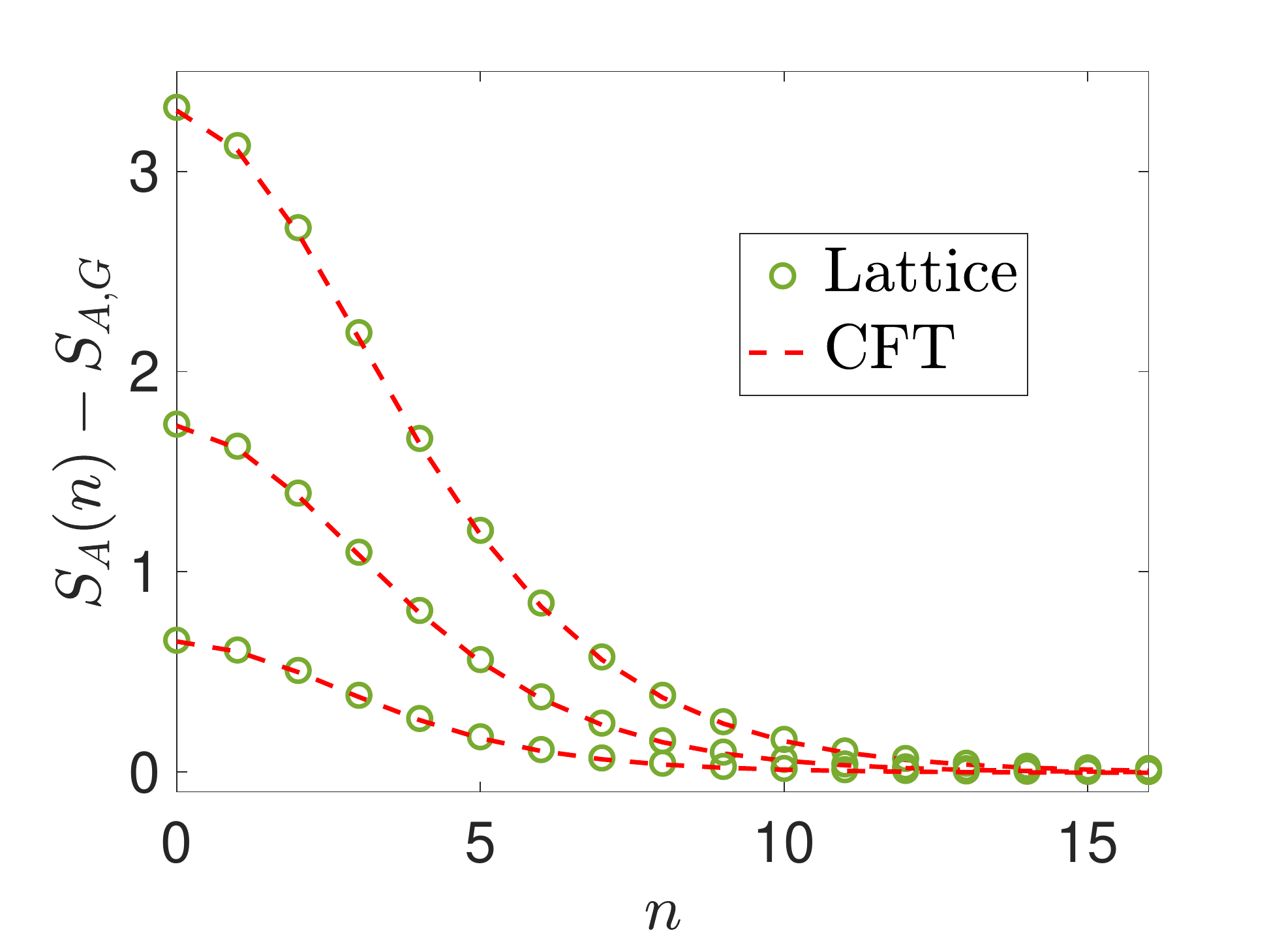}};


             \node at (-18pt, -112pt){(a)};
            \node at (185pt, -102pt){(b)};
            \node at (55pt, -200pt){(c)};          
	    \node at (185pt, -200pt){(d)};

                     
 
 

    \end{tikzpicture}
\caption{
(a) Energy density evolution in the cooling region $A=[\frac{7l}{8}, \frac{9l}{8}]$ after $n$ driving cycles ($n=0,1,\cdots, 12$ from top to bottom). $E_G(x)$ is the ground state energy density.
(b) Evolution of total energy in the cooling region with different initial temperatures ($\beta/l=3/50$, $1/10$, and $1/5$ from top to bottom), obtained by the integral of energy density in (a). 
(c) von Neumann entropy as a function of $l_A=|x_2-x_1|$ centered at $x=l$ in the cooling region after $n$ driving cycles $n$ ($n=0$, $3$, $5$, and $16$ from top to bottom), with $\beta/l=1/10$.
(d) Time evolution of von Neumann entropy for a fixed $l_A/l=1/4$ in (c), with different initial temperatures ($\beta/l=3/50$, $1/10$, and $1/5$ from top to bottom). We take $l=500$ in the lattice model calculation.
}
\label{Fig:EnerghEE}
\end{figure}
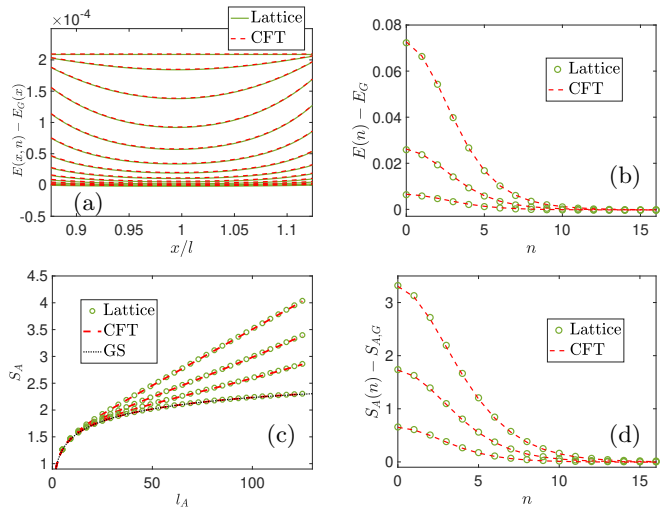

In the heating phase, the magnitudes of both $\alpha_n$ and $\beta_n$ grow exponentially in time $|\alpha_n|\sim|\beta_n|\sim\frac{1}{2}e^{\lambda_L\cdot n}$~\cite{wen2020periodically}. 
Here $\lambda_L$ is the Lyapunov exponent that characterizes the exponential growth of $||\Pi_n||$ in \eqref{w_n}, and will play the role of cooling rate in CFC.
From \eqref{Expectation_finiteT}, one can find energy density peaks located at $x^\ast$ which are determined by
\begin{equation}
\small
    \exp\left(\frac{2\pi i x^\ast}{l}\right)=-\lim_{n\to \infty}\frac{\beta_{n}}{\alpha_{n}}\,.
\end{equation}
The disconnected small neighborhoods of these peaks are the heating regions that will absorb entropy and energy.
Since the width of these peaks shrink exponentially fast in time, so do the area of the heating regions.
The rest of the system is in the cooling region. Specifically, one finds:
\be
\small
    E(x,n)-(-\frac{\pi c}{6\, l^2})\propto \left(\frac{\pi c}{6\beta^2}+\frac{\pi c}{6\, l^2} \right)
    e^{-4\lambda_L  n} \,,\, 
    \lambda_L n\gg 1
\ee
in the late-time regime for $x$ being away from the aforementioned energy peaks.
Namely, in the cooling region, $E(x,n)$ decreases exponentially in time to $-\pi c/6 l^2$, the ground-state energy density of a CFT of length $l$ with \emph{periodic} boundary conditions.
For the total energy within a single wavelength $E(n)=\int_0^l dx \, E(x,n)$, one has
\be
\label{E_average_intro}
E(n) \simeq
\frac{\pi c}{12}
\left(\frac{1}{\beta^2}+\frac{1}{l^2}
\right)\cdot l \cdot
e^{2\lambda_L n} \,,\, \lambda_L n\gg 1.
\ee
Therefore, although the energy in the cooling region decreases in time, the system keeps absorbing energy at and only at those hot spots so that the total energy still grows in time.
See Fig.\ref{PI_cooling} and Fig.\ref{Fig:EnerghEE} for the energy density evolution in different regions.

\textit{von Neumann entropy evolution:}
Let us study the cooling dynamics from the perspective of entanglement.
Based on the operator evolution in \eqref{OperatorTransf}, one can apply the replica trick and find the von-Neumann entropy of a single interval $A=[x_1,x_2]$ to be
\be
\small
\label{SvN_driving_finiteT}
\begin{aligned}
S_A(n)- & S_A(0) \\
=& 
\frac{c}{12}\cdot
\log\left(\frac{
\left|\sinh\frac{\pi }{\beta}(g_{n}(x_1)-g_{n}(x_2))\right|^2 }{
\left|\sinh\frac{\pi }{\beta}(x_{1}-x_{2})\right|^2
\frac{\partial g_{n}(x_1)}{\partial x_1}
\frac{\partial g_{n}(x_2)}{\partial x_2}
}
\right)\\
+&\text{anti-chiral},\\
\end{aligned}
\ee
where $S_A(0)=\frac{c}{3}\cdot \log \left(\frac{\beta}{\pi}\cdot \sinh\frac{\pi(x_2-x_1)}{\beta}\right)$ is the von Neumann entropy of the initial thermal state and $g_n(x)$ is given by \eqref{w_n} with $w=ix$.
In the cooling region, we have
\be
\small
S_{A}(n)= \frac{c}{3}\cdot \log\left(
\frac{l}{\pi}\cdot \sin\frac{\pi(x_2-x_1)}{l}
\right)+ O(e^{-2\lambda_L n}), \,\, \lambda_L n\gg 1.
\ee
The first term corresponds to the ground-state value von Neumann entropy $S_{A,G}$ in a CFT of length $l$ with periodic boundary conditions, consistent with the energy density evolution. 
See Fig.\ref{Fig:EnerghEE} for the von Neumann entropy evolution in the cooling region.
In the heating region, one has
\be
\label{SA_heating}
\small
S_A(n)\simeq\frac{c}{3}\cdot \lambda_L\cdot n+\frac{c}{6}\cdot
\log\left(
\frac{\beta}{\pi}\cdot
\sinh\frac{\pi l}{\beta}
\right)+\text{anti-chiral}
\ee
for $\lambda_L\cdot n\gg 1$.
The first term above arises from the strong entanglement between the energy density peak and its two nearest 
peaks of the same chirality \cite{Fan_2020}. The second term is exactly the entropy (of the chiral component) in a segment of length $l$ in the initial thermal state. Here one can see clearly the initial thermal entropy is shifted and accumulated at the hot spots.

\begin{figure}[t]
\centering
\begin{tikzpicture}
\node[inner sep=0pt] (russell) at (15pt,10pt)
    {\includegraphics[width=.25\textwidth]{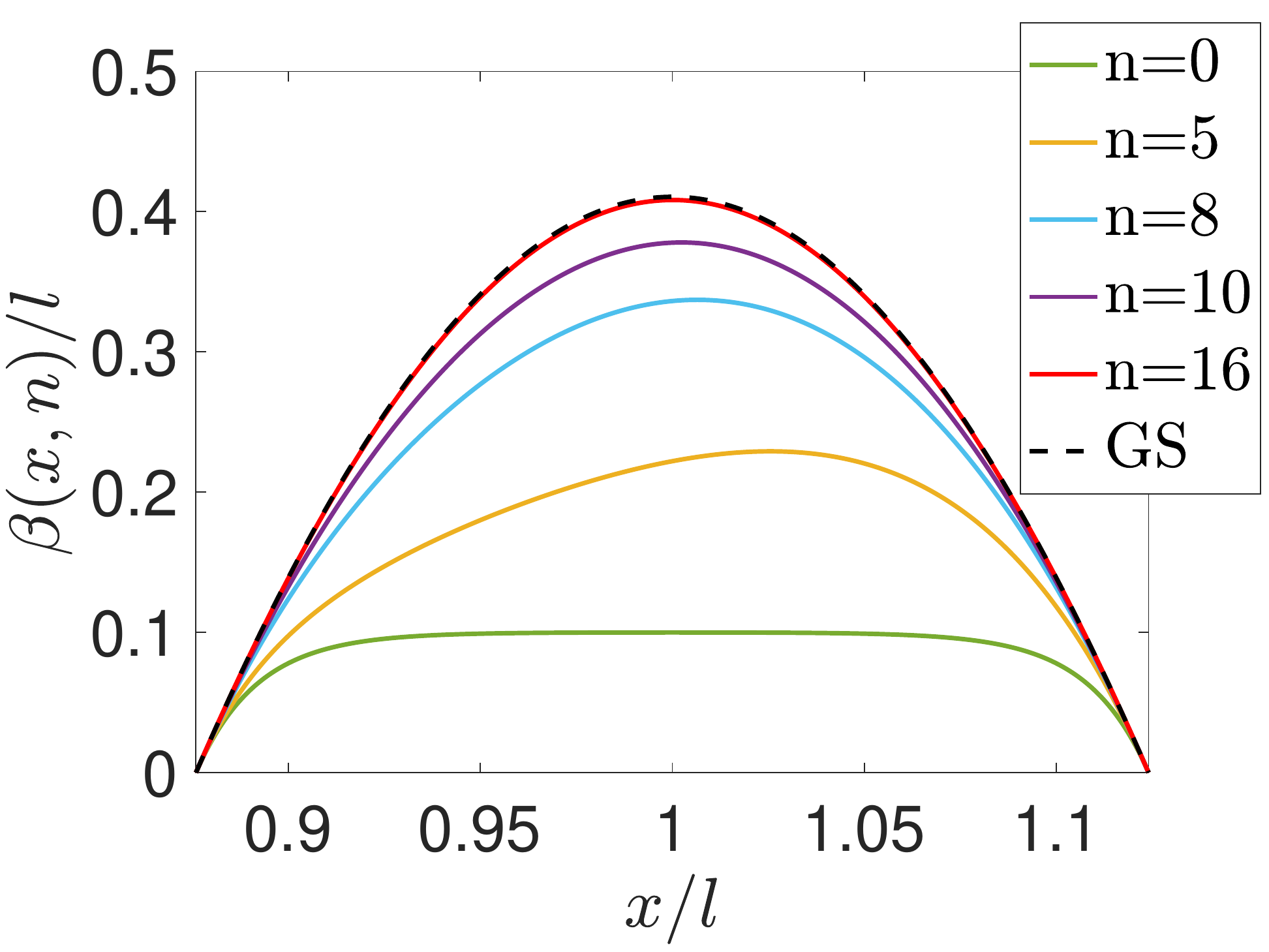}};
    \node[inner sep=0pt] (russell) at (145pt,13pt)
    {\includegraphics[width=.25\textwidth]{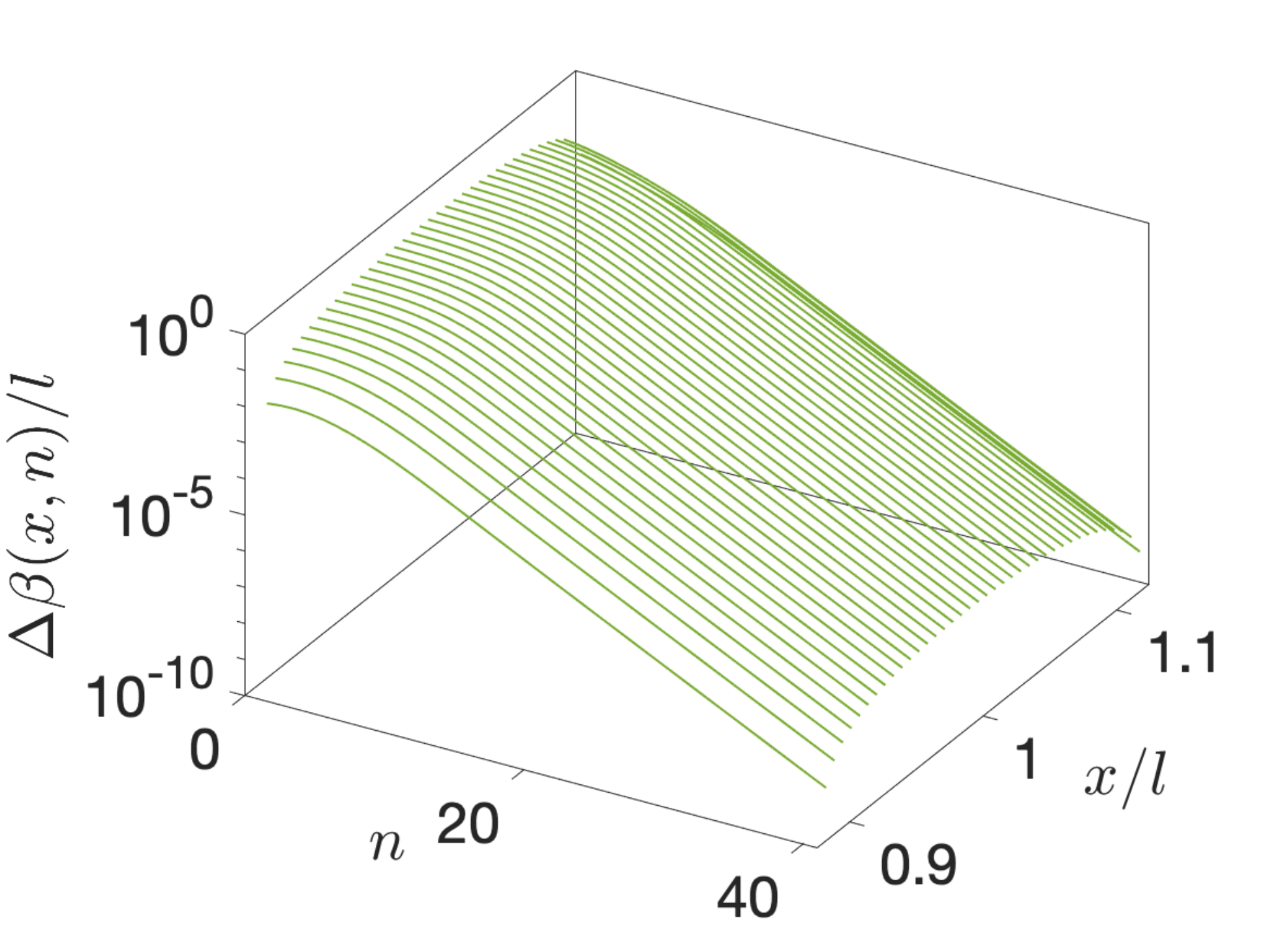}};
    \node[inner sep=0pt] (russell) at (15pt,-85pt)
    {\includegraphics[width=.25\textwidth]{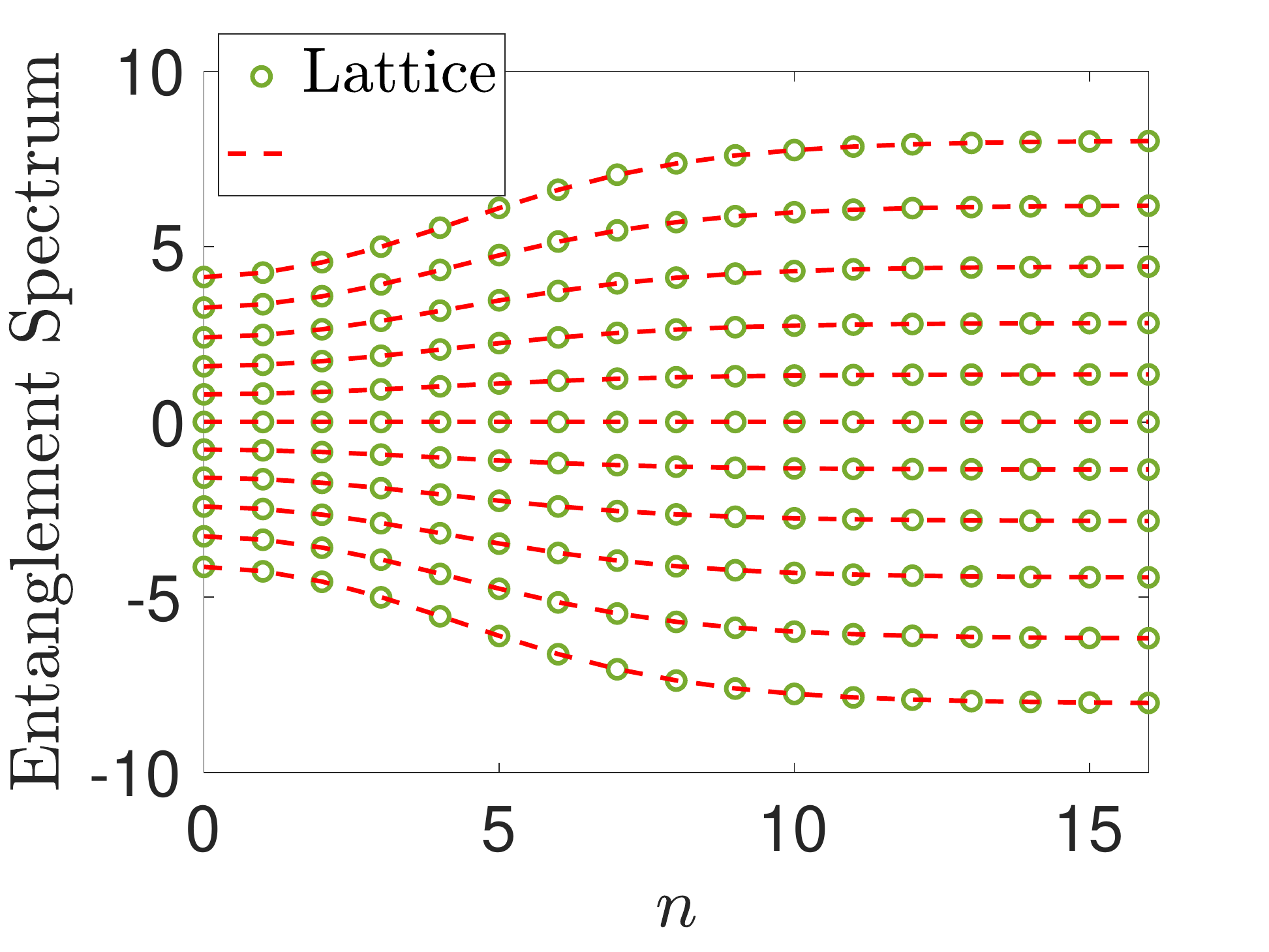}};
        \node[inner sep=0pt] (russell) at (145pt,-85pt)
    {\includegraphics[width=.25\textwidth]{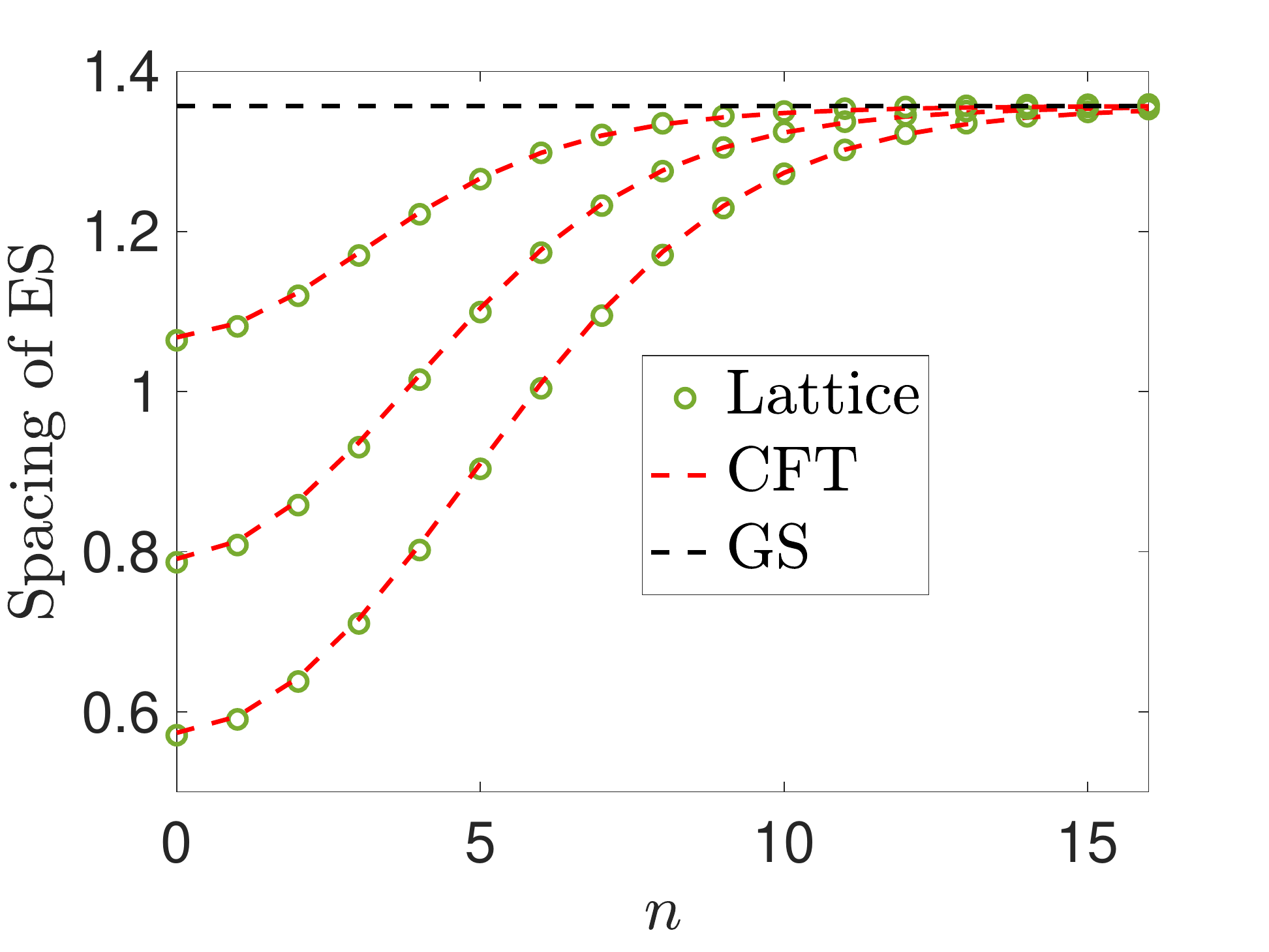}};
    
                \node at (-10pt, -53pt){\tiny Eq.\ref{KA_n}};
    
            \node at (-18pt, 42pt){(a)};
            \node at (182pt, 32pt){(b)};    
             \node at (-18pt, -107pt){(c)};
                          \node at (185pt, -107pt){(d)};
    \end{tikzpicture}
\caption{
(a) 
Time evolution of the effective inverse temperature $\beta(x,n)$ of the (chiral component of) entanglement Hamiltonian in the cooling region.
The dashed line corresponds to the ground state (GS) value in \eqref{GS_KA}.
(b) The difference $\Delta\beta(x,n)=\beta_{\text{GS}}(x)-\beta(x,n)$ as a function of $n$ rapidly approaches zero.
(c) Entanglement spectrum evolution in the cooling region in (a).
(d) Time evolution of the spacing of entanglement spectrum (energy difference of the lowest two levels) in the cooling region,
with different initial temperatures (from top to bottom: $\beta/l=1/5$, $1/10$, and $3/50$). Note the close agreement between the CFT and lattice simulations.
}
\label{Fig:EH}
\end{figure}
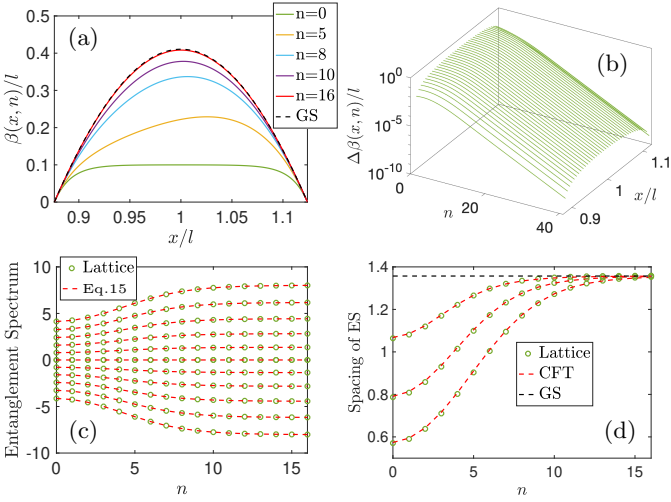

\textit{Entanglement Hamiltonian evolution:}
To fully characterize the cooling dynamics, we study the time evolution of $\rho_A$, or equivalently, the entanglement Hamiltonian $K_A$, by generalizing the approach in  \cite{2016Cardy_Tonni}.
A related study for the pure excited states was recently considered in \cite{Das_2018,2020Kabat}. 
The main idea is to map the path-integral representation of the reduced density matrix $\rho_A$ to a cylinder via a conformal transformation $\zeta=f(w)$~\cite{2016Cardy_Tonni}. 
Then the entanglement Hamiltonian can be obtained through a pull-back of the translation generator around this $\zeta$-cylinder. Here, we have 
\be
\label{EH_ConformalMap}
f(w)=\log\left(
\frac{e^{2\pi i g_n(w)/\beta}-e^{2\pi i g_n(w_1)/\beta} }{
e^{2\pi i g_n(w_2)/\beta}-e^{2\pi i g_n(w)/\beta}
}
\right),
\ee
where $g_n(w)$ is given in \eqref{gn_w}, and $w_{1,2}$ denote the locations of the two boundary points of subsystem $A$. The $\zeta$-cylinder is of length $2\pi$ in the $\text{Im}(\zeta)$ direction which is periodic, and of length $W$ in $\text{Re}(\zeta)$ direction. Then the entanglement Hamiltonian is of the form $K_A=2\pi \int_A \frac{T(x)}{f'(x)}dx+2\pi \int_A \frac{\overline T(x)}{\bar f'(x)}dx$.
It is found that after $n$ cycles of driving, 
\be
\label{KA_n}
K_A(n)=\int_{x_1}^{x_2}\beta(x,n)\, T(x)+\text{anti-chiral},
\ee
where the inverse ``local temperature" $\beta(x,n)$ has the form
\be
\small
\label{EH_finiteT_floquet}
\begin{split}
\beta(x,n)=2\beta\cdot \frac{\sinh\frac{\pi(g_n(x_2)-g_n(x))}{\beta}\sinh\frac{\pi(g_n(x)-g_n(x_1))}{\beta}}{g_n'(x)\cdot \sinh\frac{\pi (g_n(x_2)-g_n(x_1))}{\beta}}. 
\end{split}
\ee
Before the driving, i.e., $n=0$, $K_A$ reduces to the entanglement Hamiltonian in a Gibbs state at finite temperature $\beta^{-1}$. 
In the cooling region and at late time $\lambda_L n\gg 1$, one can find~\cite{SM}
\be
\small
\label{GS_KA}
\beta(x,n)=
2l \cdot  \frac{\sin\frac{\pi(x_2-x)}{l} \sin\frac{\pi(x-x_1)}{l}}{\sin\frac{\pi(x_2-x_1)}{l}} + O(e^{-2\lambda_L n})\,,
\ee
where the first term, denoted as $\beta_{\text{GS}}(x)$, is exactly the inverse local temperature for $K_A$ in the ground state of a CFT of length $l$ with periodical boundary conditions. 
This is, again, consistent with the previous analysis.
A sample plot of $\beta(x,n)$ can be found in Fig.\ref{Fig:EH} (a), and its exponential convergence to $\beta_{\text{GS}}(x)$ is in Fig.\ref{Fig:EH} (b).
Note that $\beta(x,n)$ is in general asymmetric because it only characterizes the chiral component of $K_A$.

Next, we consider the entanglement spectrum. It is determined by the length $W$ of the $\zeta$-cylinder as
$2\pi^2(-\frac{c}{24}+\Delta_j)/W$ up to a global constant, where $\Delta_j$ are dimensions of the boundary operators consistent with the boundary conditions imposed at the two entangling points~\cite{2016Cardy_Tonni}. After $n$ cycles of driving, one has
\be
\small
\label{W_length}
\begin{split}
W
=&\frac{1}{2}
\log\left[
\frac{
\left(\frac{\beta}{\pi}\sinh\frac{\pi (g_{n}(x_2)-g_{n}(x_1))}{\beta}\right)^2
}{
\frac{\partial g_{n}(x_1)}{\partial x_1} \cdot \frac{\partial g_{n}(x_2)}{\partial x_2}
 \cdot \epsilon^2
}\right]+O(\epsilon)
+\text{anti-chiral},
\end{split}
\ee
where $\epsilon$ is a non-universal UV cutoff. 
Similar to the behavior of $K_A$,
for $\lambda_L\cdot n\gg 1$, one can show that the entanglement spectrum also
approaches to the ground state value exponentially fast.
A comparison with the lattice simulations can be found in Fig.\ref{Fig:EH} (d).
Another way to obtain the entanglement spectrum is to evaluate the energy spectrum of $K_A$ in \eqref{EH_finiteT_floquet} directly.
In the case where the subsystem is chosen to be symmetric with respect to the deformation of driving Hamiltonians, 
one can find $K_A=\frac{1}{2\pi}\int_{x_1}^{x_2}\beta(x,n)T_{00}(x)$. One can put  $K_A$ on a lattice and obtain the entanglement 
spectrum by diagonalizing $K_A$ directly, as shown in Fig.\ref{Fig:EH} (c). 
In short, in the cooling region, the entanglement Hamiltonian/spectrum approaches the ground state value exponentially fast in time.
As a remark, there are also interesting features in the entanglement Hamiltonian/spectrum evolution in the heating region, which reflect the hot spot structure in Fig.\ref{CoolingCartoon} and Fig.\ref{PI_cooling}.
For instance, the local temperature $\beta^{-1}(x,n)$ at the hot spots grow exponentially in time as $e^{2\lambda_L n}$~\cite{SM}.

\textit{Non-heating phase and the phase transition:}
In the non-heating phase, the energy density and von Neumann entropy of any interval oscillate in time, so does the entanglement Hamiltonian.
At the transition, the cooling effect is qualitative the same as but quantitatively different from the heating phase. One can show that the energy density still exhibits peaks with polynomial growth in time, and the von Neumann entropy of the heating region grows logarithmically. Therefore, the cooling is less effective than that in the heating phase.

\textit{Lattice calculation:}
To verify the CFT calculation, we simulate the cooling process on a free-fermion lattice model.
The undeformed Hamiltonian is $H_0=-\frac{1}{2}\sum_{j=1}^L  c_j^\dag c_{j+1}+h.c$ defined on a lattice of length $L=1000$ with periodic boundary conditions.
At half filling, the low energy physics of this lattice model is described by a $c=1$ Dirac fermion CFT.
By preparing the system in a Gibbs state of finite temperature $\beta^{-1}$, 
we drive the system with $H_i=-\frac{1}{2}\sum_{j=1}^L v_i(j) c_j^\dag c_{j+1}+h.c $,
where $v_i(j)$ is the discrete version of $v_i(x)$ in \eqref{H_deform}.
As an illustration, we choose $v_0(x)=1$ and $v_1(x)=\cos(\frac{2\pi x}{l})$, with $l=L/2$.
The comparison of lattice and CFT calculations can be found in Fig.\ref{PI_cooling}, \ref{Fig:EnerghEE}, and \ref{Fig:EH}, where the agreement is remarkable. At later times, the energy and entropy growth will saturate due to the finite cutoff of the lattice, deviating from CFT predictions.

\textit{Discussion:} 
Now we consider general deformations $v_i(x)$ in the driving Hamiltonians \eqref{H_deform} and argue that the cooling effect is a generic feature. 
The essential ingredient in CFC is the emergent hot spots that can absorb energy and entropy.
For general choices of $v(x)$, it is always possible to have a heating phase with those emergent hot spots~\cite{lapierre2020geometric,fan2020General}.
To see the cooling effect, note that in the cooling region, $g_n(x)$ and $g_n(x_{1(2)})$ in \eqref{EH_finiteT_floquet} will flow to the \textit{same} fixed point $x_{\bullet}$ exponentially fast 
as $g_n(x_i)-g_n(x_\bullet)=u(x_i)e^{-2 \lambda_L n}$ for $\lambda_L\cdot n\gg 1$~\cite{fan2020General}. 
Then one can find $\beta(x,n)$ in \eqref{EH_finiteT_floquet} becomes independent of both the temperature and the driving cycles $n$,
and in general corresponds to the entanglement Hamiltonian in the ground state of an \emph{inhomogeneous} CFT~\cite{2018Tonni}.
Detailed features are left for future study.
Another interesting generalization is to apply our CFC method to highly excited (pure) states, e.g., a product state. Technically, studying CFC with such initial states amounts to replacing the cylinder in Fig.\ref{PI_cooling} by a strip with appropriate boundary conditions~\cite{Calabrese_2005}.
There is an interesting relation between our CFC method and the moving mirror setup in the cavity electrodynamics, where parametric driving also decreases energy density to the ground state value of a system with smaller sizes~\cite{2019Martin}. It will be interesting to further investigate the cooling effect in the moving mirror setup more explicitly, and the relation between the two protocols.
For our CFC method, the time and resource efficiency is determined by the hotspots, which are easily tuned by changing the driving frequencies. This flexibility and the cooling efficiency can bring prospects of experimentation realization of the ground states of gapless 1D systems~\cite{Schleier2019}.

\textit{Acknowledgments} 
We thank Francisco Machado, Ivar Martin, and Shinsei Ryu for helpful discussions. We thank Yingfei Gu for collaboration on related projects and helpful discussion.
XW, RF and AV are supported by a Simons Investigator award (AV) and by the Simons Collaboration on Ultra-Quantum Matter, which is a grant from the Simons Foundation (651440, AV). 
XW is also supported by the Simons Collaboration on Ultra-Quantum Matter, which is a grant from the Simons Foundation (651440, MH, XW). AV and RF are supported by NSF-DMR 2220703.

\bibliography{CoolingRef.bib}

\onecolumngrid
\newpage
\begin{center}
\textbf{\large Supplemental Materials:\\ \vspace{5pt} }
\end{center}
\tableofcontents

\newpage

\section{Operator evolution in driven conformal field theories}

Here we give a brief introduction to the operator evolution as used in the main text. 
More details can be found in, e.g., \cite{wen2020periodically,han2020classification,fan2020General}.
We consider the deformed Hamiltonian
\begin{equation}
\label{Ht}
H_i=\int_0^L v_i(x) \, T_{00}(x) dx \,,
\end{equation}
where $L$ is the total length of the system. By taking $L\to \infty$, $H_i$ is reduced to the same Hamiltonian in \eqref{H_deform}.
When there is a single wavelength in the deformation, $v_i(x)$ can be chosen of the general form  $v_i(x)=a_i+b_i\cos(\frac{2\pi x}{l}+c_i)$,
which can be rewritten as
\be\label{Envelope_F2}
v_i(x)=a_i^0+a_i^+\cos \frac{2\pi x}{l}+a_i^-\sin\frac{2\pi x}{l}. 
\quad a_i^0,a_i^+, a_i^-\in\mathbb R,
\ee
By choosing periodical boundary conditions, we have $L=q\cdot l$, where $q\in \mathbb Z$.
More generally, one can also deform the momentum density $T_{01}(x)$ in space in a similar way \cite{wen2020periodically}.
One can find the deformed Hamiltonian $H_i$ is a linear combination of $\{L_0,L_{\pm q}\}$ and $\{\bar{L}_0,\bar{L}_{\pm q}\}$, which are the generators of the $\SL^{(q)}(2,\mathbb R)$ subgroup.
To be concrete, we have $H=H_{\text{chiral}}+H_{\text{anti-chrial}}$,
with
\be
\label{Hdeform_Virasoro}
H_{i,\text{chiral}}=\frac{2\pi}{L}\left(a^0 L_0+a^+ L_{q,+}
+a^- L_{q,-}
\right)-\frac{\pi c}{12 L}, 
\ee
where we have defined
$
L_{q,+}:=\frac{1}{2}(L_q+L_{-q})$ and $L_{q,-}:=\frac{1}{2i}(L_q-L_{-q}).
$
One can further define the quadratic Casimir element: 
$
c^{(2)}:= -(a^0)^2 + (a^+)^2 + (a^-)^2
$
\cite{han2020classification,ishibashi2015infinite,ishibashi2016dipolar},
based on which one can classify the deformed Hamiltonians in
Eq.\eqref{Hdeform_Virasoro} into three types:
\be\label{HamiltonianType}
\left\{
\begin{split}
&c^{(2)}<0:\quad \text{Elliptic Hamiltonian},\\
&c^{(2)}=0:\quad \text{Parabolic Hamiltonian},\\
&c^{(2)}>0:\quad \text{Hyperbolic Hamiltonian}.\\
\end{split}
\right.
\ee
Next, we consider the operator evolution 
$e^{H \tau}\mathcal{O}(w,\bar{w}) e^{-H \tau}\to \mathcal O(w',\bar w')$ with the Hamiltonian in 
Eq.\eqref{Hdeform_Virasoro} for an imaginary time interval $\tau$.
The operator evolution is determined by a M\"obius transformation:
\be\label{Mobius_Transf}
z'=\frac{a z+b}{c z+d}, \quad
\text{where} 
\underbrace{
\begin{pmatrix}
a &b\\
c &d
\end{pmatrix}
}_{\text{denoted as } M} \in \SL(2,\mathbb C) \,.
\ee
In the $w$-cylinder in the main text, we have
\be
\label{OP_w_cylinder}
w'=\frac{l}{2\pi}\ln \frac{
a\, e^{\frac{2\pi w}{l}}+b
}{c\, e^{\frac{2\pi w}{l}}+d}.
\ee
More concretely, after an (imaginary) time duration $\tau$,
depending on the types of driving Hamiltonians, we have
\be
\begin{split}
&\text{Elliptic}: 
M(\tau)=\begin{pmatrix}
-\cosh{\left( \frac{\calC\pi \tau}{l} \right)} - \frac{a^0}{\calC} \sinh{\left( \frac{ \calC\pi  \tau}{l} \right)} &- \frac{a^+ + ia^-}{\calC} \sinh{\left( \frac{\calC \pi \tau}{l} \right)}\\
\frac{a^+ - ia^-}{\calC} \sin{\left( \frac{\calC \pi \tau}{l} \right)} &-\cosh{\left( \frac{\calC\pi T}{l} \right)} + \frac{a^0}{\calC} \sinh{\left( \frac{ \calC\pi  \tau}{l} \right)}
\end{pmatrix},\\
&
\text{Parabolic}: 
M(\tau)=\begin{pmatrix}
-1-   \frac{a^0 \pi \tau}{l} &-\frac{(a^+ + ia^-)\pi \tau }{l}\\
\frac{(a^+ - ia^-)\pi \tau }{l} &-1+   \frac{a^0 \pi \tau}{l}
\end{pmatrix},\\
&
\text{Hyperbolic}: 
M(\tau)=\begin{pmatrix}
-\cos{\left( \frac{\calC\pi  \tau}{l} \right)} - \frac{a^0}{\calC} \sin{\left( \frac{\calC\pi  \tau}{l} \right)} & - \frac{a^+ + ia^-}{\calC} \sin{\left( \frac{\calC\pi  \tau}{l} \right)}\\
  \frac{a^+ - ia^-}{\calC} \sin{\left( \frac{\calC\pi  \tau}{l} \right)} &-\cos{\left( \frac{\calC\pi  \tau}{l} \right)} + \frac{a^0}{\calC} \sin{\left( \frac{\calC\pi  \tau}{l} \right)}
\end{pmatrix},
\end{split}
\ee
where $\calC=(|-(a^0)^2+(a^+)^2+(a^-)^2|)^{1/2}$ 
and $l=L/q$.
Next, by performing analytical continuation $\tau=it$, the matrices $M$ above become $\SU(1,1)$ matrices
$M(\tau\to it)=\begin{pmatrix}
\alpha &\beta\\
\beta^* &\alpha^*
\end{pmatrix}$ as
follows:\cite{han2020classification}
\be\label{MobiusThreeTypeAppendix}
\begin{split}
&\text{Elliptic:}\quad
\alpha= -\cos{\left( \frac{\calC\pi t}{l} \right)} - i \frac{a^0}{\calC} \sin{\left( \frac{ \calC\pi  t}{l} \right)},\quad
\beta= -i \frac{a^+ + ia^-}{\calC} \sin{\left( \frac{\calC \pi t}{l} \right)},\\
&\text{Parabolic:}\quad
\alpha=-1- i  \frac{a^0 \pi t}{l},\quad
\beta=-i\frac{(a^+ + ia^-)\pi t }{l},\\
&\text{Hyperbolic:}    \quad
\alpha = -\cosh{\left( \frac{\calC\pi  t}{l} \right)} - i\frac{a^0}{\calC} \sinh{\left( \frac{\calC\pi  t}{l} \right)},\quad
\beta= -i \frac{a^+ + ia^-}{\calC} \sinh{\left( \frac{\calC\pi  t}{l} \right)}.
\end{split}
\ee
Note $\text{SU}(1,1)\cong \text{SL}(2,\mathbb R)$, both are subgroups of $\text{SL}(2,\mathbb C)$. This isomorphism is expected since we start from a $\SL(2,\mathbb R)$ action on the states. 
One can check explicitly that for elliptic, parabolic,
and hyperbolic Hamiltonians, the corresponding 
$\SU(1,1)$ matrices in Eq.\eqref{MobiusThreeTypeAppendix} 
have the properties $|\text{Tr}(M)|<2$, 
$|\text{Tr}(M)|=2$, and $|\text{Tr}(M)|>2$ respectively,
as expected.  

In a periodically driven CFT, the phase diagram is determined only by the operator evolution within one period. 
Suppose there are $p$ discrete steps of driving within each period, then the operator evolution after one period is 
determined by 
$
M_1M_2\cdots M_p,
$
which is also a $\SU(1,1)$ matrix.
Then the phase diagram of a Floquet CFT is determined by
\be
\label{PhaseDiagram_criteria}
\left\{
\begin{split}
&\text{Tr}(M_1M_2\cdots M_p)<2,\quad \text{non-heating phase},\\
&\text{Tr}(M_1M_2\cdots M_p)=2,\quad \text{phase transition},\\
&\text{Tr}(M_1M_2\cdots M_p)>2,\quad \text{heating phase}.\\
\end{split}
\right.
\ee
Let us denote the matrix elements of $M_1M_2\cdots M_p$ and $\Pi_n:=(M_1M_2\cdots M_p)^n$ as follows
\begin{equation}\label{M1Mp}
M_1M_2\cdots M_p = 
\begin{pmatrix}
\alpha_p &\beta_p\\
\beta_p^* &\alpha_p^*
\end{pmatrix}\,, \quad 
\Pi_n=(M_1M_2\cdots M_p)^n
=\begin{pmatrix}
\alpha_{np} &\beta_{np}\\
\beta_{np}^* &\alpha_{np}^*
\end{pmatrix} : = \begin{pmatrix}
\alpha_p &\beta_p\\
\beta_p^* &\alpha_p^*
\end{pmatrix}^n \,.
\end{equation}

It is convenient to use the fixed points of the M\"obius transformation to characterize the transformation when we repeat it multiple times.
For this purpose, we rewrite the M\"obius transformation into the following form
 \be\label{FixedPoint01}
\frac{z'-\gamma_1}{z'-\gamma_2}=\eta \cdot \frac{z-\gamma_1}{z-\gamma_2}\,,
\ee
where  $\gamma_{1,2}$ are the fixed points, and $\eta$ is the multiplier. For $M_1M_2\cdots M_p$ parametrized in \eqref{M1Mp}, we have 
\be\label{FixedPoint_Eta}
\gamma_{1,2}=\frac{1}{2\beta_p^*}\Big[(\alpha_p-\alpha_p^*\mp \sqrt{(\alpha_p+\alpha_p^*)^2-4})\Big],
\ee
\be\label{Multiplier_eta}
\eta
= \frac{\text{Tr}(M_1\cdots M_p)+\sqrt{ [\text{Tr}(M_1\cdots M_p)]^2-4} }{\text{Tr}(M_1\cdots M_p)-\sqrt{ [\text{Tr}(M_1\cdots M_p)]^2-4}}, \qquad \text{where }\text{Tr}(M_1\cdots M_p)=\alpha_p+\alpha_p^*.
\ee
For parabolic class when $|\Tr(M_1\cdots M_p)|=2$, we have $\gamma_1=\gamma_2=\gamma$, the transformation \eqref{FixedPoint01} becomes trivial and we need to consider
\begin{equation}
\label{ParabolicFixPoint}
\frac{1}{z'-\gamma}=\frac{1}{z-\gamma}+ \beta_p^*\,, \quad {\text{where}} \quad \gamma=\frac{\alpha_p-\alpha_p^*}{2\beta_p^*} \,.
\end{equation}
By repeating $n$ times, we simply need to modify $\eta \rightarrow \eta^n$ for $|\Tr (M_1\cdots M_p)| \neq 2 $ case and $\beta_p^* \rightarrow n \beta_p^*$ for $|\Tr (M_1\cdots M_p)| = 2 $.  Therefore, we have a simple expression for the matrix elements of $\Pi_n$ defined in \eqref{M1Mp}:
\be\label{AlphaBeta}
\alpha_{np}=\frac{\eta^{-\frac{n}{2}}\gamma_1-\eta^{\frac{n}{2}}\gamma_2}{\gamma_1-\gamma_2},\quad 
\beta_{np}=\frac{(\eta^{\frac{n}{2}}-\eta^{-\frac{n}{2}})\cdot \gamma_1\gamma_2}{\gamma_1-\gamma_2} \quad \text{when} \quad |\text{Tr}(M_1\cdots M_p)|\neq2
\ee
\be\label{AlphaBeta2}
\alpha_{np}=1+n\gamma\beta_p^*, \quad \beta_{np}=-n\gamma^2\beta_p^*
\quad \text{when} \quad |\text{Tr}(M_1\cdots M_p)|=2 \,.
\ee 
These matrix elements, i.e., $\alpha_{np}$ ($\beta_{np}$) and $\alpha_{np}^\ast$ ($\beta_{np}^\ast$) 
correspond to $\alpha_n$ ($\beta_n$) and $\alpha_n^\ast$ ($\beta_n^\ast$) in \eqref{w_n} in the main text.

\bigskip

 One can define the Lyapunov exponent $\lambda_L=\lim_{n\to \infty}\frac{1}{n} || \Pi_n||$ to characterize the growth of the norm of $\Pi_n$.
 $\lambda_L$ also characterize the operator evolution during the driving. Now $\lambda_L$ only depends on $\eta$ as follows
\be\label{Lyapunov_Eta}
\lambda_L
=\frac{1}{2}\log\left( \text{max}\left\{|\eta|, \, |\eta|^{-1}\right\}\right)
=\log
\Big|
\frac{|\text{Tr}(M_1\cdots M_p)|+ \sqrt{|\text{Tr}(M_1\cdots M_p)|^2-4}}{2}
\Big| \,. 
\ee

\subsection{A concrete example}

In the minimal setup, we consider only $p=2$ driving steps within one period:

\begin{eqnarray}\label{FloquetSetup}
\small
\begin{tikzpicture}[baseline={(current bounding box.center)}]
\node at (-18pt, 7pt){$H(t)$:};
\draw [thick](0pt,20pt)--(20pt,20pt);
\draw [thick](20pt,20pt)--(20pt,0pt);
\draw [thick](20pt,0pt)--(40pt,0pt);

\draw [thick](40pt,0pt)--(40pt,20pt);

\draw [thick](40pt,20pt)--(60pt,20pt);
\draw [thick](60pt,20pt)--(60pt,0pt);
\draw [thick](60pt,0pt)--(80pt,0pt);

\draw [thick](80pt,0pt)--(80pt,20pt);

\draw [thick](80pt,20pt)--(100pt,20pt);
\draw [thick](100pt,20pt)--(100pt,0pt);
\draw [thick](100pt,0pt)--(120pt,0pt);

\draw [thick](120pt,20pt)--(120pt,0pt);

\draw [thick](120pt,20pt)--(140pt,20pt);
\draw [thick](140pt,20pt)--(140pt,0pt);
\draw [thick](140pt,0pt)--(160pt,0pt);

\node at (10+0pt,27pt){$H_1$};
\node at (30pt, 7pt){$H_0$};

\draw [>=stealth,->] (120pt, -10pt)--(145pt,-10pt);
\node at (150pt, -10pt){$t$};
\node at (85pt, -20pt){Periodic driving};


\end{tikzpicture}
\end{eqnarray}

We drive with $H_1$ for a time duration $T_1$ first, and then with $H_0$ for a time duration $T_0$ second, and then repeat this driving process. For the choices of $H_1$ and $H_0$, we consider SL$_2$ deformed Hamiltonians on an infinite line:
\be\label{H0_H1}
\begin{split}
H_0=&\int_{-\infty}^\infty   T_{00}(x) dx,\quad
H_1=
\int_{-\infty}^\infty  \cos\left(\frac{2\pi  x}{l}\right) T_{00}(x) dx.
\end{split}
\ee
That is, we choose $a_j^0=1$, $a_j^+=a_j^-=0$ in $H_0$, and $a_j^0=a_j^-=0$ and
$a_j^+=1$ in $H_1$ in the deformation in \eqref{Envelope_F2}.
The corresponding M\"obius transformations $M_0$ and $M_1$ have the following form
(after analytical continuation to real time):
\begin{equation}\label{MobiusM}
M_0= \begin{pmatrix}
e^{i\frac{\pi T_0}{l}} &0 \\
0 & e^{-i\frac{\pi T_0}{l}} 
\end{pmatrix},
\quad
M_1= \begin{pmatrix}
\cosh\left(\frac{\pi T_1}{l}\right) &i\sinh\left(\frac{\pi T_1}{l}\right) \\
-i\sinh\left(\frac{\pi T_1}{l}\right) & \cosh\left(\frac{\pi T_1}{l}\right)
\end{pmatrix}.
\end{equation}
Then the M\"obius transformation after one cycle of driving is
\be
M_1M_0=
\begin{pmatrix}
\cosh\left(\frac{\pi T_1}{l}\right)e^{i\frac{\pi T_0}{l}} &i\sinh\left(\frac{\pi T_1}{l}\right)e^{-i\frac{\pi T_0}{l}} \\
-i\sinh\left(\frac{\pi T_1}{l}\right)e^{i\frac{\pi T_0}{l}} & \cosh\left(\frac{\pi T_1}{l}\right)e^{-i\frac{\pi T_0}{l}}
\end{pmatrix}.
\ee
The M\"obius transformation after $n$ cycles of driving is $M^n$.
The phase diagram of this periodically driven CFT can be obtained based on the criteria in \eqref{PhaseDiagram_criteria}, 
with part of the phase diagram shown in Fig.\ref{PI_cooling}. 
Depending on the driving parameters $(T_0/l, T_1/l)$, we have two different phases (heating and non-heating phases) with the phase transitions inbetween.

To illustrate the feature of time evolution in different phases, 
we choose the driving times $T_0/l=1/50$ and $T_1/l=1/25$ in the heating phase, and $T_0/l=1/10$ and $T_1/l=1/50$ in the non-heating phase.

\section{Time evolution of two point correlation function}

Here we give details on the time evolution of two-point functions of arbitrary primary operators of conformal dimensions $(h,\bar h)$. This can be applied to the entanglement entropy evolution by considering two-point function of twist operators.
The behavior of operator evolution in the heating phase is of particular interest.
Let us start from the operator evolution in \eqref{w_n} in the main text, which can be rewritten as follows:
\be
z_n=\frac{\alpha_n z+\beta_n}{\beta_n^\ast z+\alpha^\ast_n}, \quad z_n=e^{\frac{2\pi w_n}{l}}, \quad z=e^{\frac{2\pi w}{l}}, 
\ee
where $w_n=0+ix_n$ and $w=0+ix$. In the heating phase, the above equation can be rewritten as
 \be\label{FixedPoint}
\frac{z_n-\gamma_1}{z_n-\gamma_2}=\eta^n \cdot \frac{z-\gamma_1}{z-\gamma_2}\,,
\ee
with
\be\label{AlphaBeta}
\alpha_{n}=\frac{\eta^{-\frac{n}{2}}\gamma_1-\eta^{\frac{n}{2}}\gamma_2}{\gamma_1-\gamma_2},\quad 
\beta_{n}=\frac{(\eta^{\frac{n}{2}}-\eta^{-\frac{n}{2}})\cdot \gamma_1\gamma_2}{\gamma_1-\gamma_2}. 
\ee
Note that we have either $0<\eta<1$ or $\eta>1$ in the heating phase. Without loss of generality, we can assume $0<\eta<1$.
Then $\gamma_1$ is the stable fixed point, and $\gamma_2$ is the unstable fixed point.
Note also that $\eta$ is related to the Lyapunov exponent $\lambda_L$ as [see \eqref{Lyapunov_Eta}]:
\be
\eta=e^{-2\lambda_L}.
\ee
For later use, the term $\left|\alpha_n\, e^{i\frac{2\pi x}{l}}+\beta_n\right|$ can be approximated by 
\be
\label{formula_abs01}
\begin{split}
\left|\alpha_n\, e^{i\frac{2\pi x}{l}}+\beta_n\right|=\left|\alpha_n\, z+\beta_n\right|=&\left|\frac{\eta^{-\frac{n}{2}}\gamma_1(z-\gamma_2)+\eta^{\frac{n}{2}}\gamma_2(\gamma_1-z)}{\gamma_1-\gamma_2}\right|
\simeq
\eta^{-\frac{n}{2}}
\left|\frac{\gamma_1(z-\gamma_2)}{\gamma_1-\gamma_2}\right|, \quad 0<\eta<1,\quad n\gg 1.\\
\end{split}
\ee
Also, we will need to evaluate $z_n$ or $x_n$. One has
\be
\label{zn_OP}
z_n=\gamma_1+\eta^n\cdot \frac{(z-\gamma_1)(\gamma_1-\gamma_2)}{(z-\gamma_2)-\eta^n(z-\gamma_1)},\quad 0<\eta<1,
\ee 
or equivalently
\be
\label{xn_OP}
x_n=\frac{l}{2\pi i}\log\left(\gamma_1+\eta^n\cdot \frac{(e^{\frac{2\pi i x}{l}}-\gamma_1)(\gamma_1-\gamma_2)}{(e^{\frac{2\pi i x}{l}}-\gamma_2)-\eta^n(e^{\frac{2\pi i x}{l}}-\gamma_1)}
\right),\quad 0<\eta<1,
\ee 
As $n\to \infty$ or more precisely $\eta^n\ll 1$, we have
\be
\label{zn_heating}
z_n\simeq \gamma_1+\eta^n\cdot \frac{(z-\gamma_1)(\gamma_1-\gamma_2)}{(z-\gamma_2)},\quad 0<\eta<1.
\ee
In the $w$ coordinate, we have
\be
\label{x1x2}
x_{n,2}-x_{n,1}\simeq \pm\,  \eta^n \cdot \frac{l}{2\pi}\cdot  \left|  \frac{\gamma_1-\gamma_2}{\gamma_1}\cdot \left(
\frac{z_2-\gamma_1}{z_2-\gamma_2}-\frac{z_1-\gamma_1}{z_1-\gamma_2}
\right)\right| \quad \text{mod }l, \quad 0<\eta<1.
\ee
Here the sign $\pm$ depends on the relative locations of $x_{n,2}$ and $x_{n,1}$ mod $l$.

Note that in the specific case that $x_{n,1}$ corresponds to the stable fixed point $x_\bullet$ toward which $x_{n,2}$ flows, we have
\be
\label{x_fixedpoint_approach}
|x_{n,2}-x_\bullet|\simeq
\eta^n \cdot \frac{l}{2\pi}\cdot  \left|  \frac{\gamma_1-\gamma_2}{\gamma_1}\cdot 
\frac{z_2-\gamma_1}{z_2-\gamma_2}
\right|, \quad \eta=e^{-2\lambda_L}.
\ee
In other words, $x_n$ is exponentially close to the fixed point.
In addition, based on \eqref{zn_heating}, we have
\be
\label{diff_xn}
\frac{\partial x_n}{\partial x}\simeq \eta^n\cdot \left|
\frac{(\gamma_1-\gamma_2)^2\cdot z}{(z-\gamma_2)^2\cdot \gamma_1}
\right|,\quad 0<\eta<1.
\ee
Note that the above formula only works for $z\neq \gamma_2$. For $z=\gamma_2$, we have
\be
\label{Diff_unstable}
\frac{\partial x_n}{\partial x}=\frac{1}{\eta^n}=e^{2\lambda_L n}.
\ee
This will be used in studying the ``local temperature" $\beta(x,n)$ in the entanglement Hamiltonian.

\subsection{Zero temperature}

Let us consider the time evolution of two-point correlation function where the initial state is the ground state of a uniform CFT of length $L$ with periodic boundary conditions.
Denoting the wavefunction after $n$ steps of driving as $|\Psi_n\rangle$, we have
\be
\label{2point_correlation_general}
\begin{split}
&\langle \Psi_n|\calO(w_1,\bar{w}_1)
\calO(w_2,\bar{w}_2)|\Psi_n\rangle
=\prod_{i=1,2}
\left(\frac{\partial w_{n,i}}{\partial w_i}\right)^h 
\prod_{i=1,2}
\left(\frac{\partial \bar{w}_{n,i}}{\partial \bar{w}_{i}}\right)^{\bar{h}}
\langle 
\calO(w_{n,1},\bar{w}_{n,1})
\calO(w_{n,2},\bar{w}_{n,2})
\rangle_w.
\end{split}
\ee
Here $w_{1,2}=0+ix_{1,2}$.  Note that if there is no driving, then we have
\be
\label{2point_groundstate}
\langle 
\calO(w_{1},\bar{w}_{1})
\calO(w_{2},\bar{w}_{2})
\rangle_w=
\left(\frac{2\pi}{L}\right)^{2h+2\bar h}
\left(4\sin^2\frac{\pi }{L}(x_1-x_2)\right)^{-(h+\bar h)}.
\ee
Under the time-dependent driving, one has
\be
\begin{split}
&\langle 
\calO(w_{n,1},\bar{w}_{n,1})
\calO(w_{n,2},\bar{w}_{n,2})
\rangle_w
=
\prod_{i=1,2}
\left(\frac{\partial z_{n,i}}{\partial w_{n,i}}\right)^h 
\prod_{i=1,2}
\left(\frac{\partial \bar{z}_{n,i}}{\partial \bar{w}_{n,i}}\right)^{\bar{h}}
\langle 
\calO(z_{n,1},\bar{z}_{n,1})
\calO(z_{n,2},\bar{z}_{n,2})
\rangle_z\\
=&
\left(\frac{2\pi}{L}\right)^{2h}
\left(\frac{2\pi}{L}\right)^{2\bar h}
\prod_{i=1,2}\left(e^{\frac{2\pi}{L}w_{n,i}}\right)^h
\prod_{i=1,2}\left(e^{\frac{2\pi}{L}\bar w_{n,i}}\right)^{\bar h}
\left(e^{\frac{2\pi}{L}w_{n,1}}-e^{\frac{2\pi}{L}w_{n,2}}\right)^{-2h}
\left(e^{\frac{2\pi}{L}\bar w_{n,1}}-e^{\frac{2\pi}{L}\bar w_{n,2}}\right)^{-2\bar h}.
\end{split}
\ee
Also, we have 
\be
\label{w'}
\frac{\partial w_n}{\partial w}=\frac{1}{\left|\alpha_n\, e^{i\frac{2\pi x}{l}}+\beta_n\right|^2},\quad \frac{\partial \bar w_n}{\partial \bar w}=\frac{1}{\left|\alpha'_n\, e^{-i\frac{2\pi x}{l}}+\beta'_n\right|^2}.
\ee
From now on, to simplify the discussion, let us only consider the chiral component. The anti-chiral component can be discussed in a simialr way. Eq.\eqref{2point_correlation_general} has the following expression:
\be
\label{2point_holomorphic_ground}
\begin{split}
&\langle \Psi_n|\calO(w_1,\bar{w}_1)
\calO(w_2,\bar{w}_2)|\Psi_n\rangle\\
=&
\frac{1}{\left|\alpha_n\, e^{i\frac{2\pi x_1}{l}}+\beta_n\right|^{2h}}
\cdot
\frac{1}{\left|\alpha_n\, e^{i\frac{2\pi x_2}{l}}+\beta_n\right|^{2h}}
\cdot
\left(\frac{\frac{\pi}{L}}{\sin\frac{\pi }{L}|x_{n,1}-x_{n,2}|}\right)^{2h}\times
\text{anti-holomorphic},
\end{split}
\ee
where $x_{n,1}$ and $x_{n,2}$ are defined through \eqref{xn_OP}.

\subsection{Finite temperature}

At finite temperature, the operator is inserted in the $w$-cylinder as sketched in Fig.\ref{PI_cooling} in the main text.
That is, $w$ is periodic in the imaginary time direction with length $\beta$.
One can map this $w$-cylinder to a complex $z$-plane through the following conformal map:
\be\label{ConformalMap_finiteT_b}
z=e^{\frac{2\pi i}{\beta}w},\quad \bar{z}=e^{-\frac{2\pi i}{\beta}\bar{w}}\, ,
\ee
where $w=\tau+ix$ and $\bar w=\tau-ix$.
Without any driving, the equal-time 2-point function at $\tau=0$ is:
\be
\label{2point_thermal}
\begin{split}
\langle \mathcal{O}(w_1,\bar{w}_1)\mathcal{O}(w_2,\bar{w}_2)\rangle_{w\text{-cylinder}}
=&
\left(\frac{\partial z_1}{\partial w_1}\right)^{h}
\left(\frac{\partial \bar{z}_1}{\partial \bar{w}_1}\right)^{\bar{h}}
\left(\frac{\partial z_2}{\partial w_2}\right)^{h}
\left(\frac{\partial \bar{z}_2}{\partial \bar{w}_2}\right)^{\bar{h}}
\cdot
\langle \mathcal{O}(z_1,\bar{z}_1)\mathcal{O}(z_2,\bar{z}_2)\rangle_{z-\text{plane}}\\
=&\left(\frac{2\pi }{\beta}\right)^{2h}
\left(\frac{2\pi }{\beta}\right)^{2\bar h}
\left[4\sinh^2\left(\frac{\pi}{\beta}(x_1-x_2)\right)\right]^{-(h+\bar h)}
\end{split}
\ee
Note to compare this expression with \eqref{2point_groundstate} where the initial state is the ground state.

Now let us introduce the periodic driving. After $n$ steps of driving, the two-point function becomes
\be
\label{2point_correlation_finiteT}
\begin{split}
\langle U_n^\dag \calO(w_1,\bar{w}_1)
\calO(w_2,\bar{w}_2) U_n \rangle
=&\prod_{i=1,2}
\left(\frac{\partial w_{n,i}}{\partial w_i}\right)^h 
\prod_{i=1,2}
\left(\frac{\partial \bar{w}_{n,i}}{\partial \bar{w}_{i}}\right)^{\bar{h}}
\langle 
\calO(w_{n,1},\bar{w}_{n,1})
\calO(w_{n,2},\bar{w}_{n,2})
\rangle_{\text{cylinder}}\\
=&
\frac{1}{\left|\alpha_n\, e^{i\frac{2\pi x_1}{l}}+\beta_n\right|^{2h}}
\cdot
\frac{1}{\left|\alpha_n\, e^{i\frac{2\pi x_2}{l}}+\beta_n\right|^{2h}}
\cdot
\left(\frac{\frac{\pi}{\beta}}{\sinh\frac{\pi }{\beta}|x_{n,1}-x_{n,2}|}\right)^{2h}\times
\text{anti-chiral}.
\end{split}
\ee
It is helpful to compare the above result with \eqref{2point_holomorphic_ground} where the initial state is the ground state.
Comparing to \eqref{2point_holomorphic_ground}, we have replaced $\frac{\pi}{L}\to \frac{\pi}{\beta}$ and $\sin(\cdots)\to \sinh(\cdots)$.
Note that the first two terms in \eqref{2point_correlation_finiteT} are purely contributed by the driving, and the third term is contributed by both the driving and the initial thermal state.

In the heating phase, the concrete result in \eqref{2point_correlation_finiteT} depends on whether there are energy density peaks in the region $[x_1,\,x_2]$, as follows. 

\begin{enumerate}

\item \textit{There are energy density peaks in $[x_1,x_2]$}

In this case, $x_{n,1}$ and $x_{n,2}$ will flow to different stable fixed points. Suppose there are $m$ (chiral) peaks within $[x_1,x_2]$, then one has $x_2-x_1\simeq ml$ (where $m\in\mathbb Z$) in the long time driving limit. 
Then one has
\be
\label{2point_correlation_finiteT_withPeaks}
\begin{split}
\langle U_n^\dag \calO(w_1,\bar{w}_1)
\calO(w_2,\bar{w}_2) U_n \rangle
\simeq&\,
\frac{1}{\left|\alpha_n\, e^{i\frac{2\pi x_1}{l}}+\beta_n\right|^{2h}}
\cdot
\frac{1}{\left|\alpha_n\, e^{i\frac{2\pi x_2}{l}}+\beta_n\right|^{2h}}
\cdot
\left(\frac{\frac{\pi}{\beta}}{\sinh\frac{\pi m l}{\beta}}\right)^{2h}\times
\text{anti-chiral}\\
\propto&\,
e^{-\lambda_L\cdot 4h\cdot n}\cdot
\left(\frac{\frac{\pi}{\beta}}{\sinh\frac{\pi m l}{\beta}}\right)^{2h}\times
\text{anti-chiral},
\end{split}
\ee
where we have considered $ |\alpha_n| \sim |\beta_n| \sim \frac{1}{2} e^{\lambda_L n}$ in the heating phase.
In this case, the two point correlation function decays exponentially fast in time.

In particular, for the heating region, which is a small region that contains only one energy density peak, one has 
\be
\langle U_n^\dag \calO(w_1,\bar{w}_1)\calO(w_2,\bar{w}_2) U_n \rangle
\propto 
e^{-\lambda_L\cdot 4h\cdot n}\cdot
\left(\frac{\frac{\pi}{\beta}}{\sinh\frac{\pi  l}{\beta}}\right)^{2h}\times
\text{anti-chiral}.
\ee

\item \textit{There is no energy density peak in $[x_1,x_2]$}

This case corresponds to the cooling region as discussed in the main text. $x_{n,1}$ and $x_{n,2}$ will flow to the \textit{same} stable fixed point.  In the long time driving limit $\lambda_L n\gg1$, by using \eqref{formula_abs01} and \eqref{x1x2}, one can obtain
\be
\label{2point_driving_longtime}
\begin{split}
\langle U^\dag \calO(w_1,\bar{w}_1)\calO(w_2,\bar{w}_2) U \rangle
\simeq& \frac{1}{\left| \frac{l}{2\pi} (e^{\frac{i2\pi x_2}{l}}-e^{\frac{i2\pi x_1}{l} }) \right|^{2h}}\times
\text{anti-chiral}
=
 \frac{1}{\left( \frac{l}{\pi}\sin\frac{\pi(x_2-x_1)}{l}\right)^{2h}}\times
\text{anti-chiral},
\end{split}
\ee
which is the result of two-point correlation function in the ground state of a CFT of length $l$ 
with periodic boundary conditions.

\end{enumerate}

\section{Energy evolution}

In this section, we give more details and results on the energy evolution in the heating phase where one can observe conformal cooling.

\subsection{No driving}

For later use, let us first review the case without driving first.
For a thermal state at finite temperature $\beta^{-1}$, the path-integral representation of one point function is illustrated as follows:
\begin{eqnarray}\label{cylinder}
\small
\begin{tikzpicture}[baseline={(current bounding box.center)}]

\draw (0pt,0pt) ellipse (8pt and 20pt);
\draw (100pt,0pt) ellipse (8pt and 20pt);
\draw (0pt,20pt)--(100pt,20pt);
\draw (0pt,-20pt)--(100pt,-20pt);

\draw [>=stealth,->] (115pt, 0pt)--(115pt,15pt);
\draw [>=stealth,->] (115pt, 0pt)--(130pt,0pt);
\node at (135pt,0pt){$x$};
\node at (120pt,15pt){$\tau$};

\node at (30pt,0pt){$\bullet$};
\node at (30pt,-8pt){$\mathcal O$};


\end{tikzpicture}
\end{eqnarray}
Here we consider an infinite system with $L\to \infty$.  On this $w$-cylinder which is periodic in $\tau$ direction, we have $w=\tau+ix$, $\tau\in [0,\beta]$, and $x\in (-\infty,\infty)$.
Now we consider the conformal mapping
\be\label{ConformalMap_finiteT}
z=e^{\frac{2\pi i}{\beta}w},\quad \bar{z}=e^{-\frac{2\pi i}{\beta}\bar{w}}\, ,
\ee
which maps the $w$-cylinder to a complex $z$-plane.
Then the stress tensor transforms as
\be\label{T_transform_02}
\begin{split}
T_{\text{cyl}}(w)=&\left(\frac{dw}{dz}\right)^{-2}\left[
T_{\text{plane}}(z)-\frac{c}{12}\{w,z\}
\right]\\
\end{split}
\ee
based on which we have
\be
\begin{split}
\langle T_{\text{cyl}}(w)\rangle=&-\frac{c}{12}\{w,z\}\cdot\left(\frac{dw}{dz}\right)^{-2}
=-\frac{c}{12}\cdot \frac{1}{2z^2}\cdot \left(
\frac{\beta}{2\pi i}\cdot\frac{1}{z}
\right)^{-2}
=\frac{\pi^2 c}{6\beta^2},
\end{split}
\ee
where we have considered $\langle T_{\text{plane}}(z)\rangle=0$. 
Then the energy density is
\be
\frac{1}{2\pi}\Big(
\langle T_{\text{cyl}}(w)\rangle+\langle \bar T_{\text{cyl}}(\bar w)\rangle
\Big)=\frac{\pi c}{6 \beta^2}.
\ee

\subsection{With driving}

One can study the time evolution of stress tensors 
based on the operator evolution. Note that $T(x)$ ($\bar T(x)$) is not 
a primary field, the operator evolution becomes
\be\label{OP_evolution_finiteT}
U_n^{\dag}\,T(w)\, U_n=\left(\frac{\partial w_n}{\partial w}\right)^{2}
T\big(w_n\big)+\frac{c}{12}\text{Sch}(w_n,w),
\ee
where the last term represents the Schwarzian derivative.
Note that the operator evolution on $w$-cylinder is described by
\be
w_n=\frac{l}{2\pi}\ln \frac{
\alpha_n e^{\frac{2\pi w}{l}}+\beta_n
}{\beta_n^\ast e^{\frac{2\pi w}{l}}+\alpha_n^\ast}.
\ee
Then one can obtain
\be\label{Expectation_finiteT0}
\begin{split}
\langle U_n^{\dag}\,T(w)\, U_n\rangle=\left(\frac{\partial w_n}{\partial w}\right)^{2}
\cdot \frac{\pi^2 c}{6\beta^2}+\frac{c}{12}\text{Sch}(w_n,w),
\end{split}
\ee
which can be further written as
\be\label{Expectation_finiteT_SM}
\begin{split}
\langle U_n^{\dag}\,T(w)\, U_n\rangle
=&-\frac{\pi^2 c}{6\, l^2}+
 \left(\frac{\pi^2 c}{6\beta^2}+\frac{\pi^2 c}{6\, l^2}
\right)\cdot
\frac{1}{\left|\alpha_n\, e^{i\frac{2\pi x}{l}}+\beta_n\right|^4}.
\end{split}
\ee
In the non-heating phase, since $\alpha_n$ and $\beta_n$ will simply oscillate in time, so does the energy density. In the heating phase, we have
 \begin{equation}
      |\alpha_n| \sim |\beta_n| \sim \frac{1}{2} e^{\lambda_L n} \qquad \text{for} \quad \lambda_L n\gg 1.
 \end{equation}
 Then in the cooling region, one has
 \be
 \langle U_n^{\dag}\,T(w)\, U_n\rangle  -\left( -\frac{\pi^2 c}{6\, l^2}\right)\propto
 \left(\frac{\pi^2 c}{6\beta^2}+\frac{\pi^2 c}{6\, l^2}
\right)\cdot
e^{-4\lambda_L\cdot n}, \quad \lambda_n\cdot n\gg 1.
 \ee
In other words, the energy-momentum density will approach the steady value $-\frac{\pi^2 c}{6\,l^2}$ exponentially fast in time.
Based on \eqref{Expectation_finiteT}, one can further obtain the total energy within a single unit cell $[0,l]$:
\be\label{Expectation_finiteT_integral}
\begin{split}
\int_0^l\langle U_n^{\dag}\,T(w)\, U_n\rangle\,dx
=&-\frac{\pi^2 c}{6\, l}+
 \left(\frac{\pi^2 c}{6\beta^2}+\frac{\pi^2 c}{6\, l^2}
\right)\cdot
l\cdot(|\alpha_n|^2+|\beta_n|^2).
\end{split}
\ee
Then the total energy in a unit cell $[0, l]$ is
\be
\begin{split}
E_{\text{unit cell}}:=&\int_0^l \langle U_n^\dag T_{00}(x) U_n \rangle \, dx=\frac{1}{2\pi}\int_0^l \langle U_n^\dag \left(T(w)+\overline T(\bar w)\right) U_n \rangle \, dx\\
=&
-\frac{\pi c}{6\,l}+ \left(\frac{\pi c}{6\beta^2}+\frac{\pi c}{6\, l^2}
\right)\cdot
l \cdot \left(|\alpha_n|^2+|\beta_n|^2\right).
\end{split}
\ee
For $\lambda_L n\gg 1$, then we have
\be
\label{E_average}
E_{\text{unit cell}}(n)-E_{\text{unit cell}}(n=0)\propto
\frac{\pi c}{12}
\left(\frac{1}{\beta^2}+\frac{1}{l^2}
\right)\cdot l\cdot
e^{2\lambda_L\cdot n}.
\ee
The results depend on the two length scales $\beta$ and $l$, which are the \textit{only} two length scales in this problem.
One can find the energy growth is amplified by the finite temperature in the initial state.

\begin{figure}[h]
\begin{center} 
\includegraphics[width=2.2in]{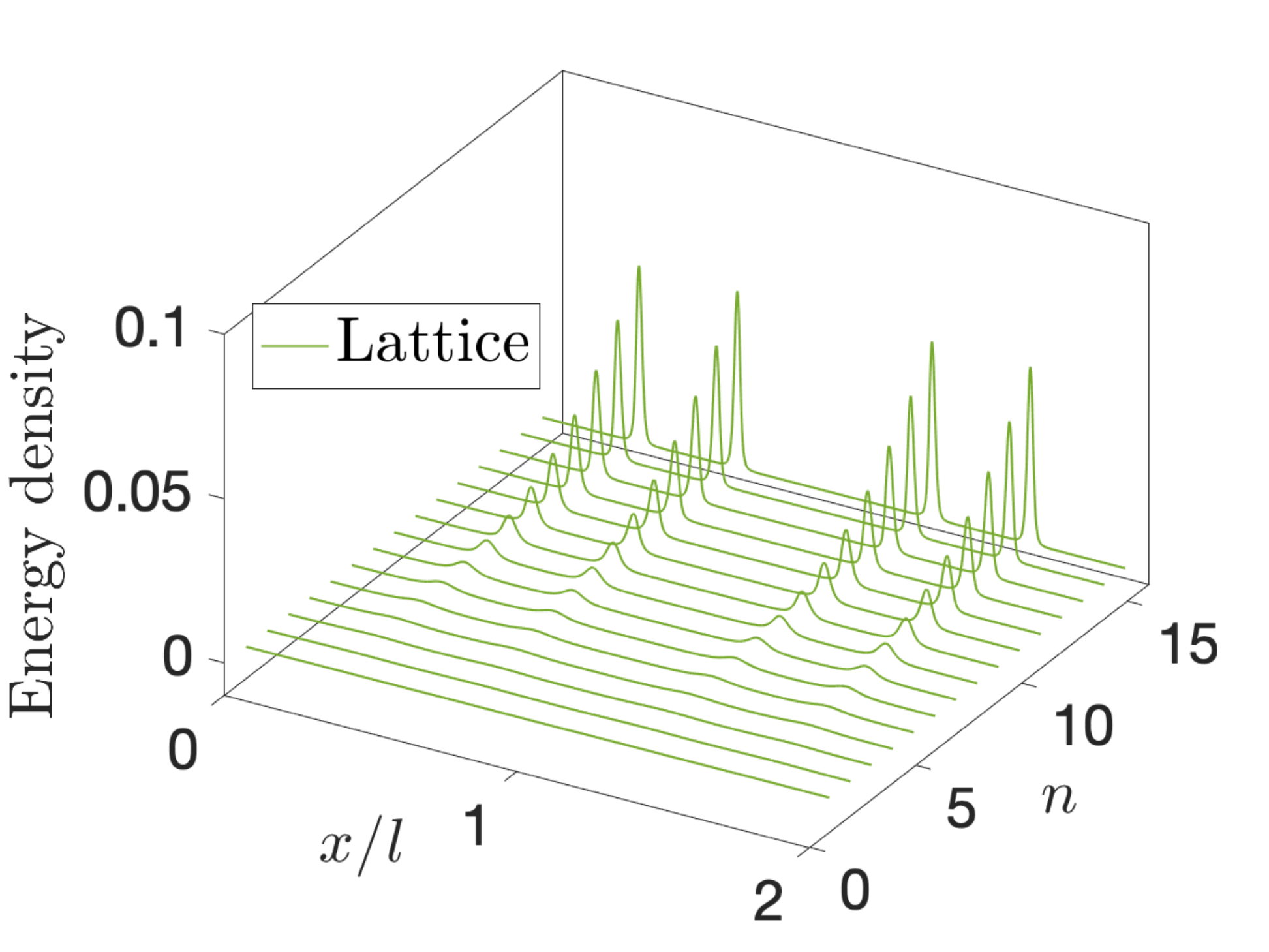}
\includegraphics[width=2.2in]{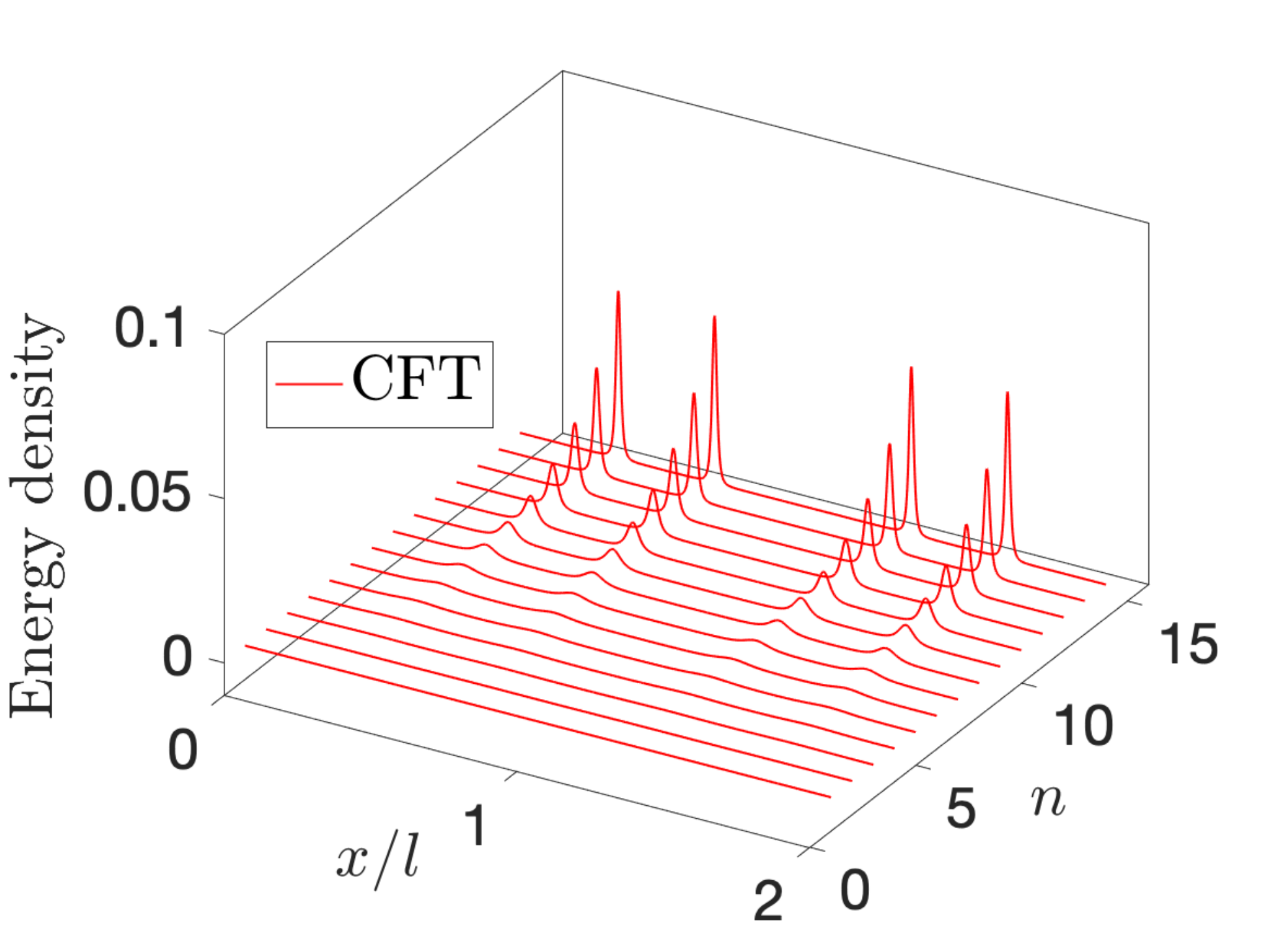}
\end{center}
\caption{
Energy density evolution in two unit cells $[0,2l]$ in Floquet CFT
starting from a Gibbs state with $\beta/l=1/10$.
We choose $L=2l =1000$  in the lattice model. 
The CFT result is plotted according to \eqref{Expectation_finiteT}.
The driving parameters are the same as Fig.\ref{PI_cooling}.
}
\label{EnergyDensity_Floquet_TwoCell}
\end{figure}

\begin{figure}[h]
\begin{center} 
\includegraphics[width=2.2in]{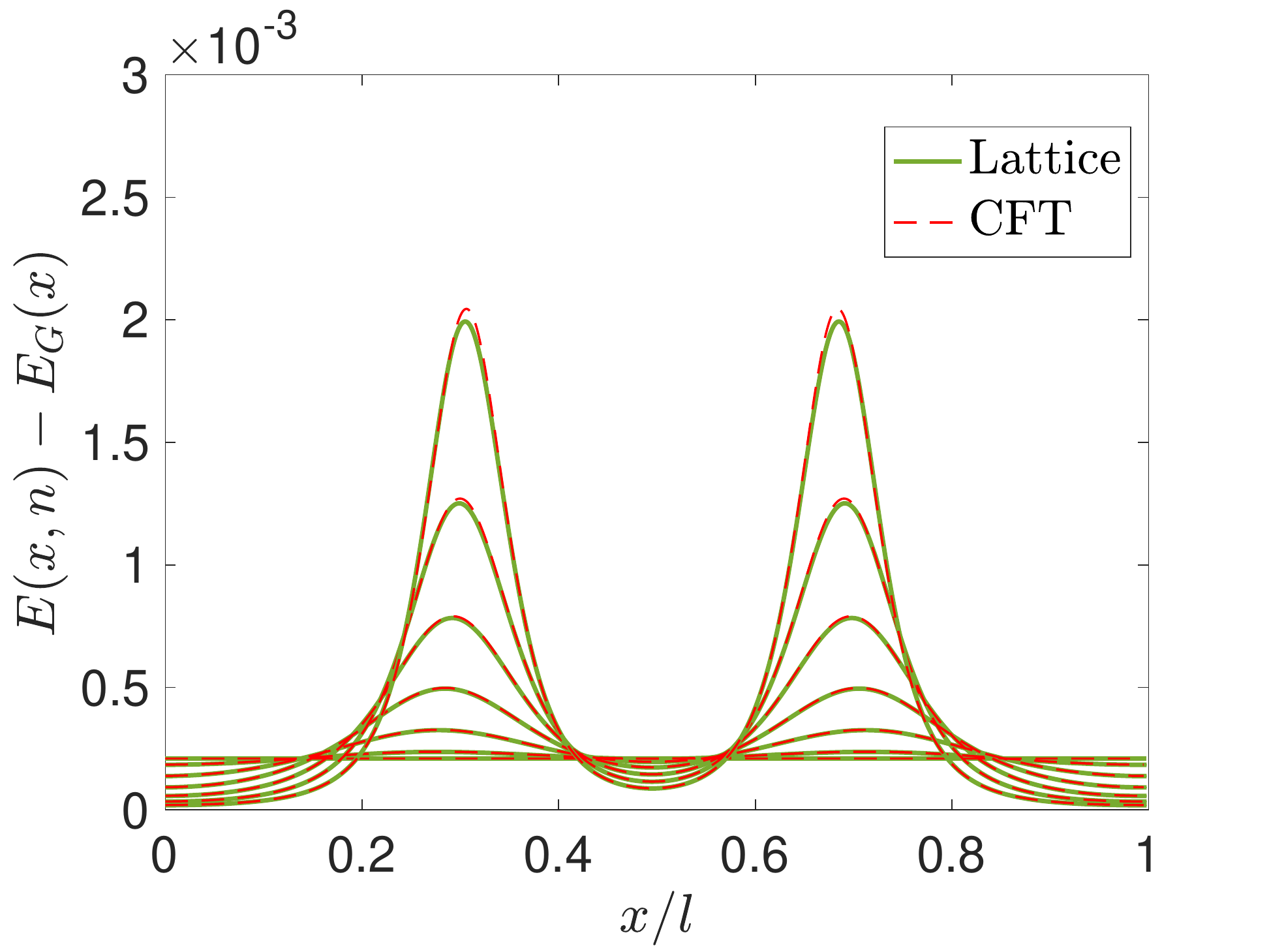}
\includegraphics[width=2.2in]{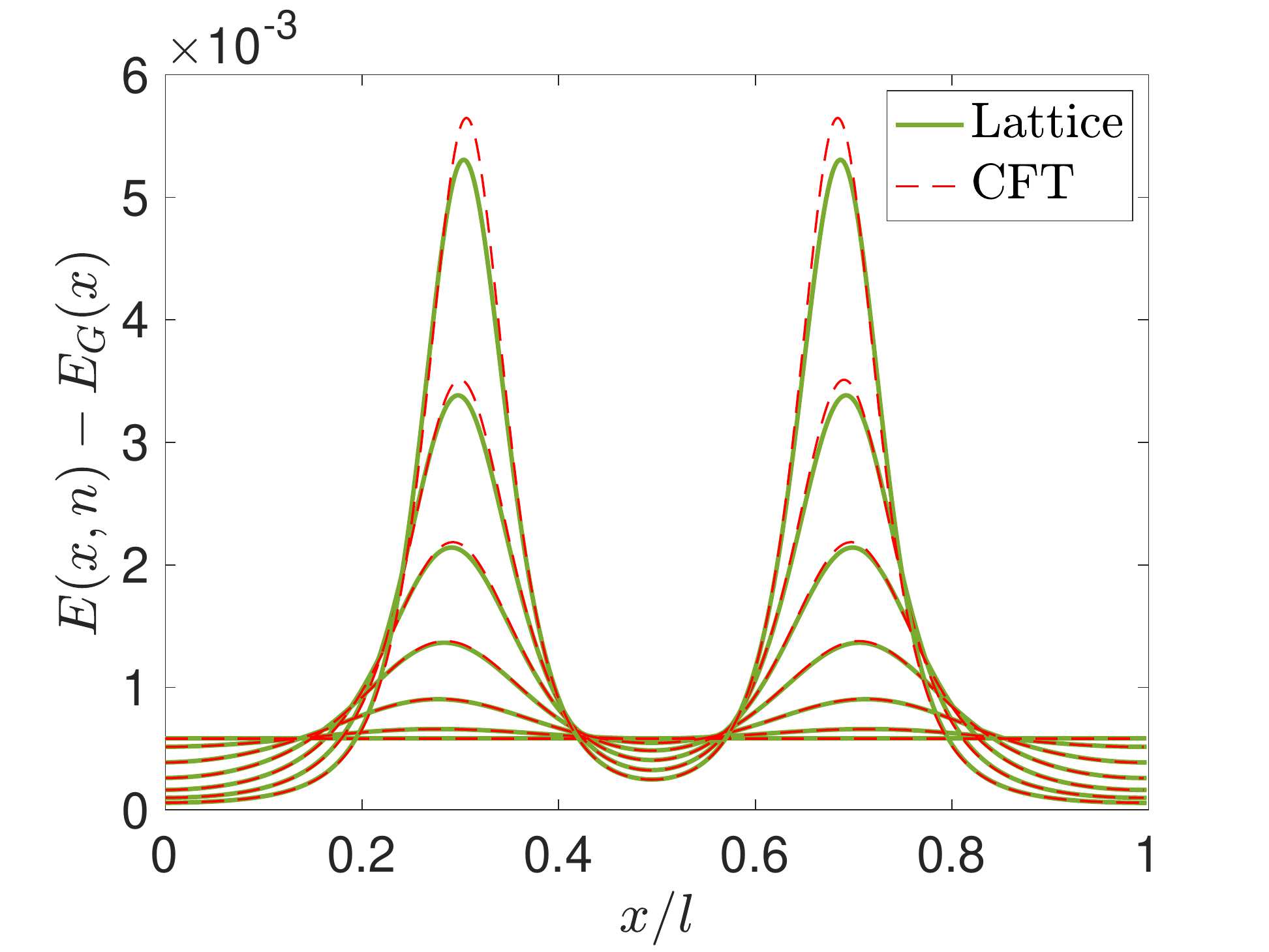}
\end{center}
\caption{
Energy density evolution in a unit cell $[0,l]$ in Floquet CFT
starting from different initial temperatures, with $\beta/l=1/10$ (left) and $\beta/l=3/50$ (right).
Here we take the driving cycles $n=0, 1, \cdots, 6$.
We choose $L=2l =1000$  in the lattice model. 
Here $E_G(x)$ is the ground state energy density of the undriven system.
The CFT result is plotted according to \eqref{Expectation_finiteT}.
}
\label{EnergyDensity_Floquet_UnitCell}
\end{figure}

\begin{figure}[h]
\begin{center} 
\includegraphics[width=2.2in]{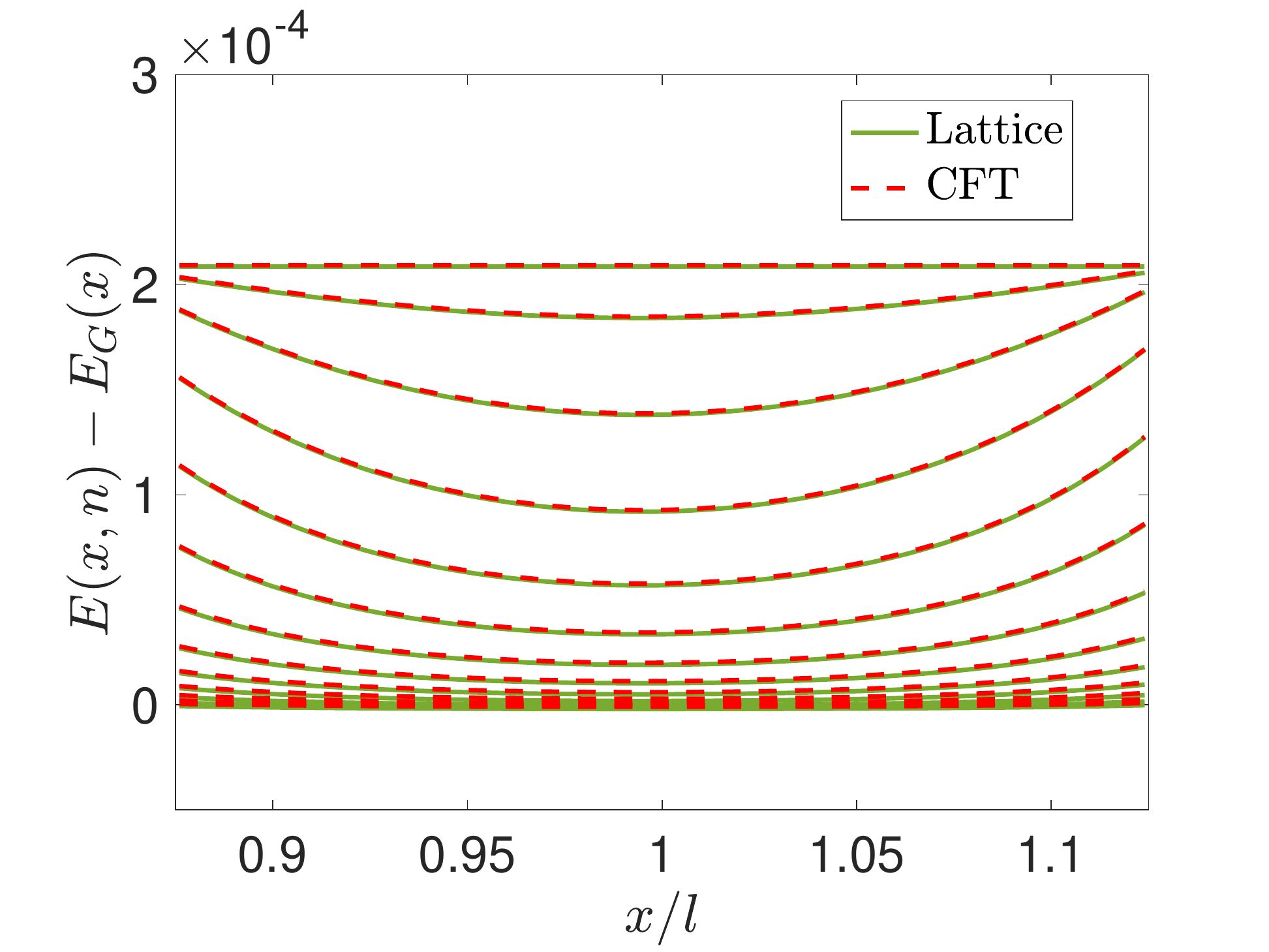}
\includegraphics[width=2.2in]{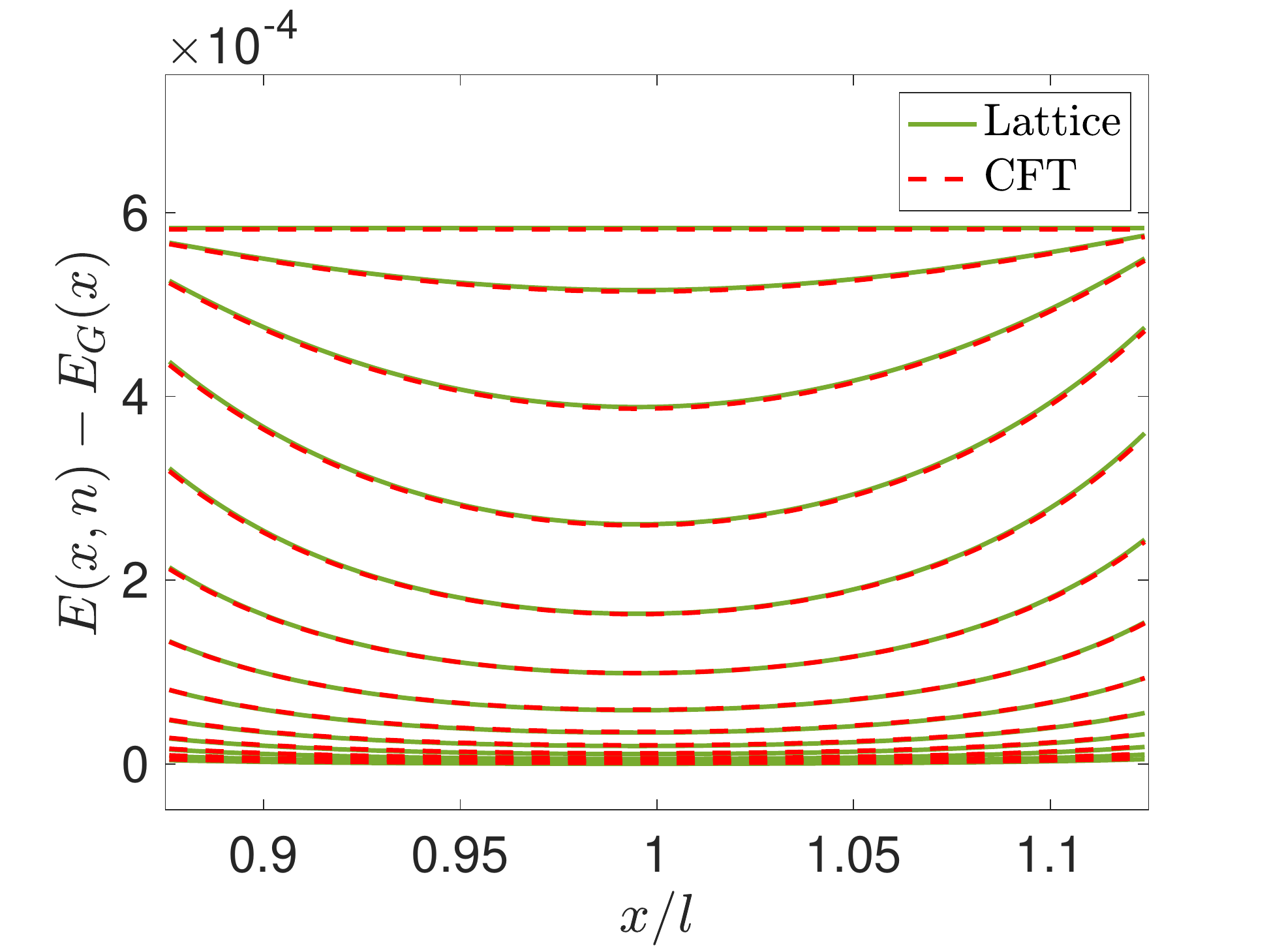}
\end{center}
\caption{
Energy density evolution in (part of) cooling region in a Floquet CFT
starting from different initial temperatures, with $\beta/l=1/10$ (left) and $\beta/l=3/50$ (right).
Here we take the driving cycles $n=0, 1, \cdots, 12$.
We choose $L=2l =1000$  in the lattice model. 
Here $E_G(x)$ is the ground state energy density of the undriven system.
The CFT result is plotted according to \eqref{Expectation_finiteT}.
}
\label{EnergyDensity_Floquet_cooling}
\end{figure}

A sample plot of the energy density evolution within two wavelengths, one single wavelength, and within the cooling region 
can be found in Fig.\ref{EnergyDensity_Floquet_TwoCell},
\ref{EnergyDensity_Floquet_UnitCell} and Fig.\ref{EnergyDensity_Floquet_cooling} respectively.
The comparison of lattice and CFT results in the cooling region is very well.
Near the energy density peaks, one can see deviations between the lattice and CFT results as $n$ grows, because
the higher energy modes play a role here.
This deviation is as expected because the CFT approximation works well only in the low energy limit.

\section{Entanglement entropy evolution}

The entanglement entropy can be studied by calculating the two-point correlation function of twist operators\cite{CC2004}.
Starting from a Gibbs state at finite temperature $\beta^{-1}$, let us first consider the case without driving. Based on \eqref{2point_thermal}, we have 
\be
\begin{split}
&\langle \mathcal{T}(w_1,\bar{w}_1) \overline{\mathcal{T}}(w_2,\bar{w}_2)\rangle_{w}
=\left(\frac{2\pi}{\beta}\right)^{4h} \cdot
\left[2\sinh\left(\frac{\pi}{\beta}(x_2-x_1)\right)\right]^{-4 h}.
\end{split}
\ee
where $h=\bar{h}=\frac{c}{24}(m-\frac{1}{m})$ are the conformal dimensions of the twist operator.
Then one can obtain the Renyi entropy:
\be
\label{Sn_thermal_initial}
\begin{split}
S_A^{(m)}=&\frac{1}{1-m}\log \langle \mathcal{T}(w_1,\bar{w}_1) \overline{\mathcal{T}}(w_2,\bar{w}_2)\rangle_{w}
=\frac{c}{6}\cdot\frac{m+1}{m}\cdot \log \left(
\frac{\beta}{\pi}\cdot \sinh\left[\frac{\pi}{\beta}(x_2-x_1)\right] 
\right),
\end{split}
\ee
and the von-Neumann entropy 
\be
\label{SA_thermal_initial}
\begin{split}
S_A
=\frac{c}{3}\cdot\log \left(
\frac{\beta}{\pi}\cdot \sinh\left[\frac{\pi}{\beta}(x_2-x_1)\right] 
\right).
\end{split}
\ee
Then one has
\be
S_A\simeq \frac{\pi c}{3\beta}\cdot (x_2-x_1), \quad \text{if}\,\, x_2-x_1\gg \beta.
\ee
This corresponds to the high-temperature limit, where we have a volume-law of entanglement entropy.
On the other hand, if $\beta\gg x_2-x_1$, which correspond to the low temperature limit, we have
\be
\begin{split}
S_A\simeq  \frac{c}{3}\log (x_2-x_1), \quad \text{if}\,\, x_2-x_1\ll \beta,
\end{split}
\ee
which corresponds to the result of zero-temperature case.

Now let us consider the effect of driving. 
The two-point correlation function of twist operators $\langle U_n^\dag \mathcal T(w_1,\bar{w}_1)
\mathcal T(w_2,\bar{w}_2) U_n \rangle$ has the same expression as \eqref{2point_correlation_finiteT} with $h=\bar{h}=\frac{c}{24}(m-\frac{1}{m})$.
Then one has
\be
\begin{split}
S_A^{(m)}=&\frac{c}{12}\cdot\frac{m+1}{m}\cdot 
\log\left(
\frac{\beta}{\pi}\cdot \left|\alpha_n\, e^{i\frac{2\pi x_1}{l}}+\beta_n\right|\cdot \left|\alpha_n\, e^{i\frac{2\pi x_2}{l}}+\beta_n\right|\cdot
\left|\sinh\frac{\pi }{\beta}(x_{n,1}-x_{n,2})\right|
\right)
+\text{anti-chiral}.
\end{split}
\ee
The von-Neumann entropy is obtained by taking $m=1$ in the above formula:
\be
\label{SvN_driving_finiteT}
\begin{split}
S_A
=&
\frac{c}{6}\cdot
\log\left(
\left|\alpha_n\, e^{i\frac{2\pi x_1}{l}}+\beta_n\right|\cdot \left|\alpha_n\, e^{i\frac{2\pi x_2}{l}}+\beta_n\right|
\right)+\frac{c}{6}\cdot
\log\left(
\frac{\beta}{\pi}\cdot
\left|\sinh\frac{\pi }{\beta}(x_{n,1}-x_{n,2})\right|
\right)+\text{anti-chiral}.\\
\end{split}
\ee
The first term is purely contributed by the driving and is independent of the initial state. 
The second term is contributed by both the initial state and the driving.
Note that the above formula can be rewritten in a general way as
\be
\label{SvN_driving_finiteT_general}
\begin{split}
S_A
=&
\frac{c}{12}\cdot
\log\left[\frac{\left(
\frac{\beta}{\pi}\cdot
\left|\sinh\frac{\pi }{\beta}(x_{n,1}-x_{n,2})\right|
\right)^2 }{
\frac{\partial x_{n,1}}{\partial x_1}\cdot 
\frac{\partial x_{n,2}}{\partial x_2}
}
\right]+\text{anti-chiral},
\end{split}
\ee
which works for general deformations $v(x)$ in the deformed driving Hamiltonians.

\subsection{Cooing region in the heating phase}

For the cooling region, there is no energy-density peak in $A=[x_1,x_2]$. In this case, $x_{n,1}$ and $x_{n,2}$ will flow to the same stable fixed point, i.e, $|x_{n,1}-x_{n,2}|\to 0$ in the limit $\lambda_L\cdot n\gg 1$.  Then \eqref{SvN_driving_finiteT} becomes 
\be
\label{SvN_driving_finiteT_appendix2}
\begin{split}
S_A
\simeq&
\frac{c}{6}\cdot
\log\left(
\left|\alpha_n\, e^{i\frac{2\pi x_1}{l}}+\beta_n\right|\cdot \left|\alpha_n\, e^{i\frac{2\pi x_2}{l}}+\beta_n\right|
\right)+\frac{c}{6}\cdot
\log\left(
\left|
x_{n,1}-x_{n,2}\right|
\right)+\text{anti-chiral}.\\
\end{split}
\ee
Note that the temperature $\beta$ has disappeared, which reflects the cooling effect.
Then from \eqref{xn_OP}, we have
\be
\begin{split}
S_A(n)\simeq & \frac{c}{6}\cdot\log\left(
\eta^{-n} \left|\frac{\gamma_1(e^{i\frac{2\pi x_1}{l}}-\gamma_2)}{\gamma_1-\gamma_2}\right|\cdot
\left|\frac{\gamma_1(e^{i\frac{2\pi x_2}{l}}-\gamma_2)}{\gamma_1-\gamma_2}\right|
\right)\\
&+\frac{c}{6}\cdot\log\left(
\eta^n \cdot \frac{l}{2\pi}\cdot  \left|  \frac{\gamma_1-\gamma_2}{\gamma_1}\cdot \left(
\frac{e^{i\frac{2\pi x_2}{l}}-\gamma_1}{e^{i\frac{2\pi x_2}{l}}-\gamma_2}-\frac{e^{i\frac{2\pi x_1}{l}}-\gamma_1}{e^{i\frac{2\pi x_1}{l}}-\gamma_2}
\right)\right|
\right)+\text{anti-chiral},
\end{split}
\ee
which can be further simplified as 
\be
S_A(n)\simeq \frac{c}{6}\cdot \log\left|
\frac{l}{2\pi} \cdot \gamma_1\cdot \left(e^{i\frac{2\pi x_2}{l}}-e^{i\frac{2\pi x_1}{l}}\right)
\right|+\text{anti-chiral}=\frac{c}{3}\cdot \log\left(
\frac{l}{\pi}\cdot \sin\frac{\pi(x_2-x_1)}{l}
\right).
\ee
That is, the entanglement entropy becomes the same as that in the ground state of a CFT of length $l$ with periodic boundary conditions.
In fact, by doing Taylor expansion in \eqref{SvN_driving_finiteT_appendix2}, one can further check that 
\be
S_A(n)\simeq \frac{c}{3}\cdot \log\left(
\frac{l}{\pi}\cdot \sin\frac{\pi(x_2-x_1)}{l}
\right)+O(e^{-2\lambda_L n}).
\ee
That is, the entanglement entropy approaches the ground state value exponentially fast in time.

\begin{figure}[h]

\begin{center} 
\includegraphics[width=2.2in]{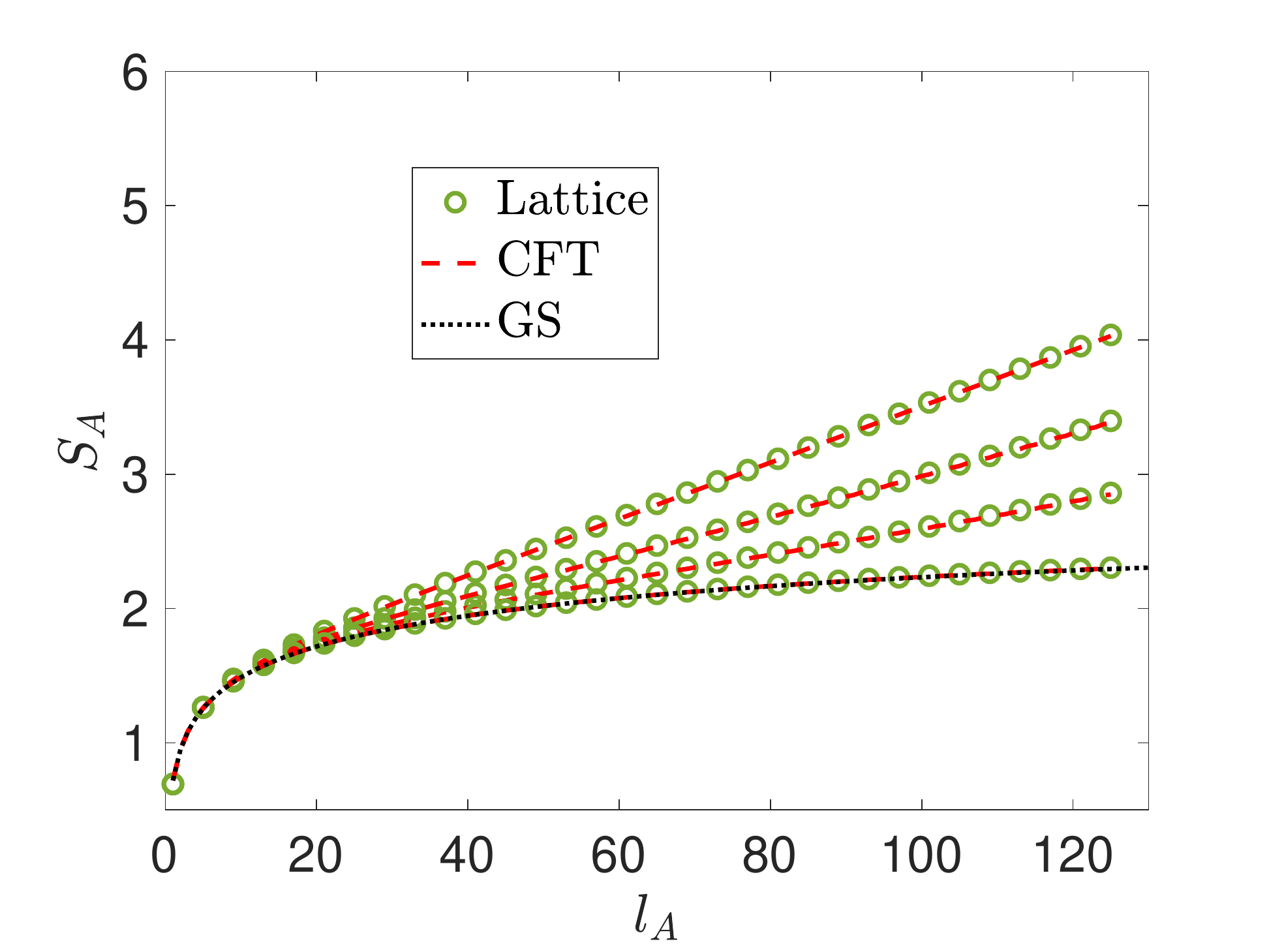}
\includegraphics[width=2.2in]{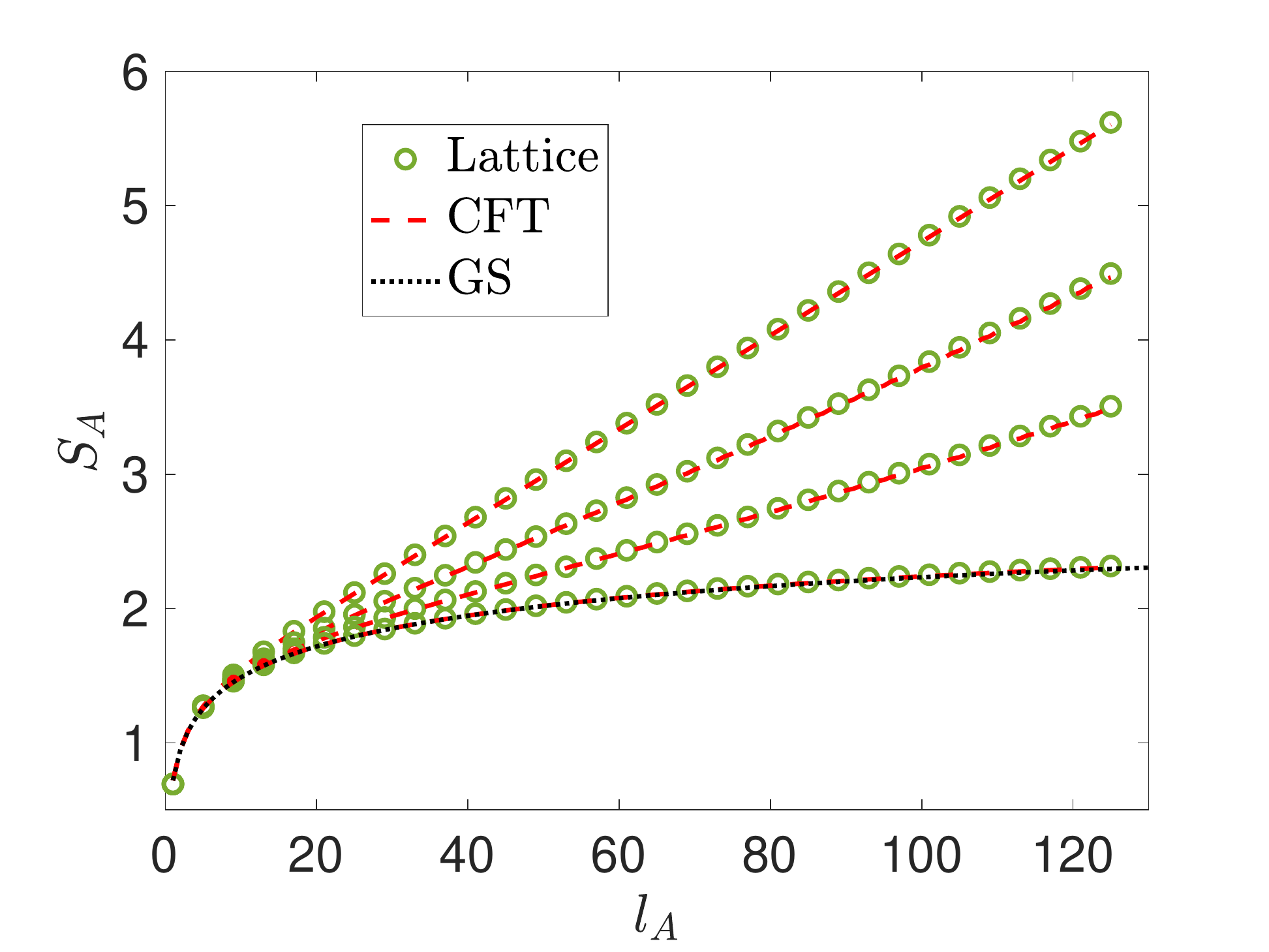}
\includegraphics[width=2.2in]{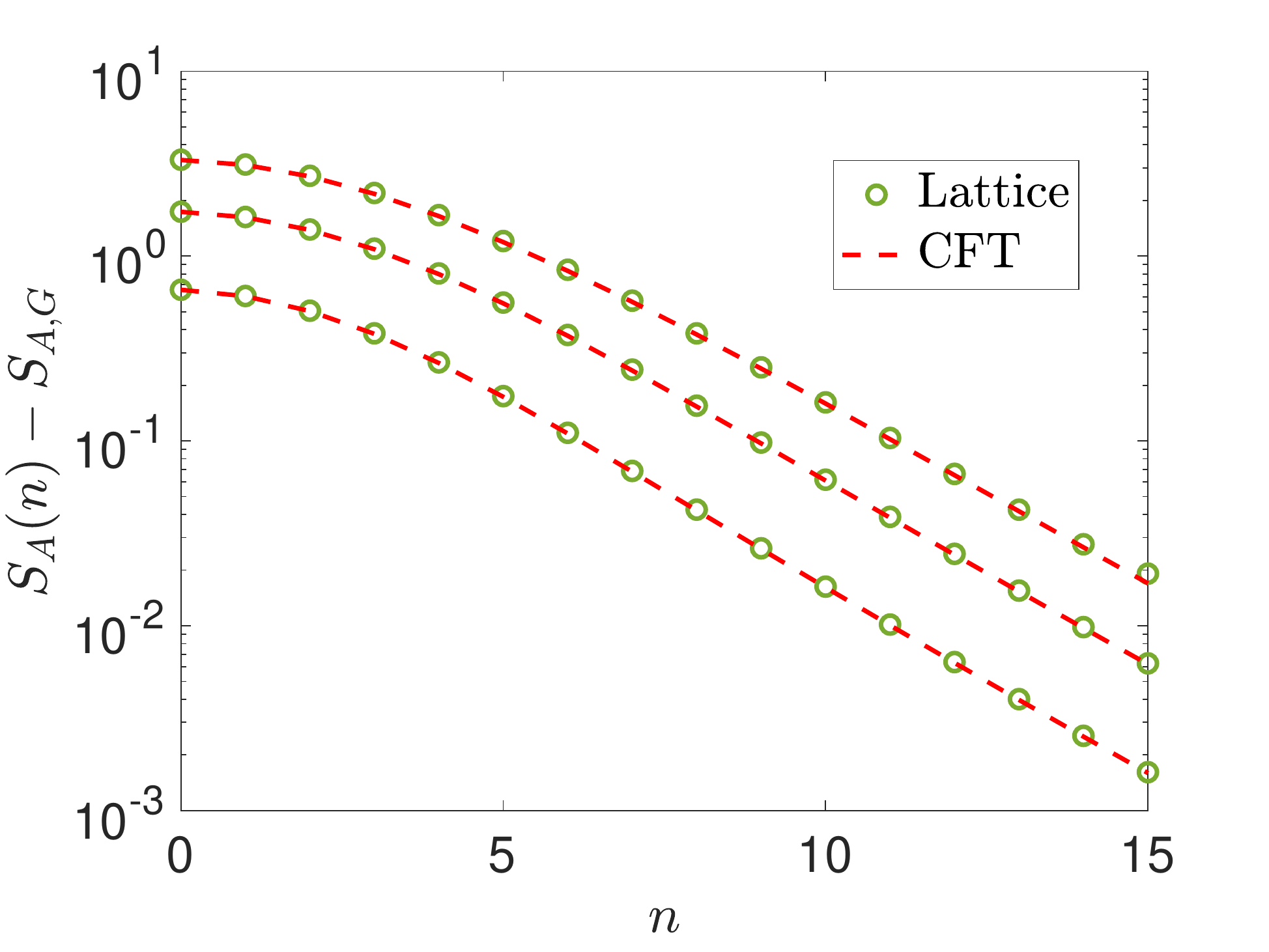}
\end{center}
\caption{
Entanglement entropy as a function of $l_A=|x_2-x_1|$ centered at $x=l$ in the cooling region after $n$ driving cycles $n$ ($n=0$, $3$, $5$, and $16$ from top to bottom), with  $\beta/l=1/10$ (left) and $\beta/l=3/50$ (middle).
Right: A replot of Fig.\ref{Fig:EnerghEE} (d) in the main text with a log-linear scale.
The temperatures are $\beta/l=3/50$, $1/10$, and $1/5$ from top to bottom.
}
\label{EE_cooling_Floquet}
\end{figure}

A sample plot of entanglement entropy evolution as well as its exponential approaching to the ground state value  in the cooling region can be found in Fig.\ref{EE_cooling_Floquet}.

\subsection{Heating region in the heating phase}

Now let us consider how the initial (thermal) entropy is accumulated near the energy density peak.
 We consider the heating region where there is a single (chiral) energy-density peak in the subsystem $A=[x_1,x_2]$. In this case, the operators initially at $x_1$ and $x_2$ will flow to different stable fixed points. One has
\be
\label{x1x2_b}
|x_{n,2}-x_{n,1}|\simeq l-\eta^n \cdot \frac{l}{2\pi}\cdot  \left|  \frac{\gamma_1-\gamma_2}{\gamma_1}\cdot \left(
\frac{z_2-\gamma_1}{z_2-\gamma_2}-\frac{z_1-\gamma_1}{z_1-\gamma_2}
\right)\right|, \quad 0<\eta<1.
\ee
Then based on \eqref{SvN_driving_finiteT}, in the long time driving limit $\lambda_L\cdot n\gg 1$, one has
\be
\label{SA_complete}
\begin{split}
S_A(n)=&\frac{c}{3}\cdot \lambda_L\cdot n+\frac{c}{6}\cdot
\log\left(
\frac{\beta}{\pi}\cdot
\sinh\frac{\pi l}{\beta}
\right)+\frac{c}{6}\cdot\log\left(
\left|\frac{\gamma_1(e^{i\frac{2\pi x_1}{l}}-\gamma_2)}{\gamma_1-\gamma_2}\right|\cdot
\left|\frac{\gamma_1(e^{i\frac{2\pi x_2}{l}}-\gamma_2)}{\gamma_1-\gamma_2}\right|
\right)
+\text{anti-chiral}.\\
\end{split}
\ee
Here $\lambda_L=\frac{1}{2}\log\frac{1}{\eta}$. Note that the last term is independent of the initial state -- it only depends on the 
choice of subsystem as well as the driving.
See \eqref{FixedPoint_Eta} for the expression of $\gamma_{1}$ and $\gamma_2$, which is only determined by the driving within a single driving cycle. Note that \eqref{SA_complete} holds for an arbitrary small interval $A$ as long as it contains a single (chiral) energy density peak. In particular, the second term $\frac{c}{6}\cdot \log\left(\frac{\beta}{\pi}\cdot \sinh\frac{\pi l}{\beta}\right)$ corresponds to the 
initial entropy of the chiral components in an interval of length $l$ [See \eqref{Sn_thermal_initial}]. This means the initial (thermal) entropy in a unit cell of length $l$ is accumulated near the energy density peaks. A sample plot of entanglement entropy evolution of the subsystem containing energy density peaks are shown in Fig.\ref{EE_nonHeating_Floquet}.

\subsection{Non-heating phase}

\begin{figure}[h]
\begin{center} 
\includegraphics[width=2.2in]{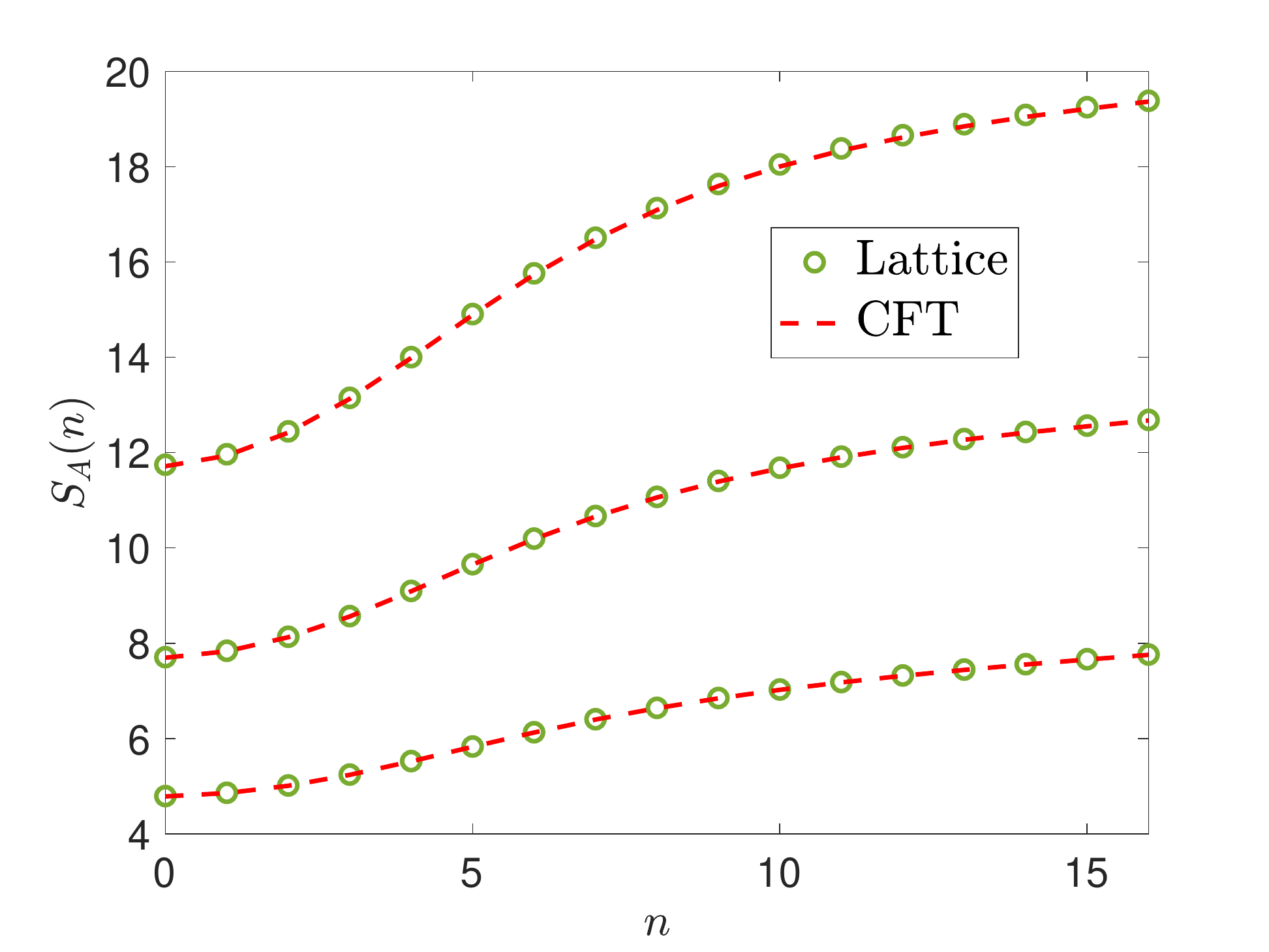}
\includegraphics[width=2.2in]{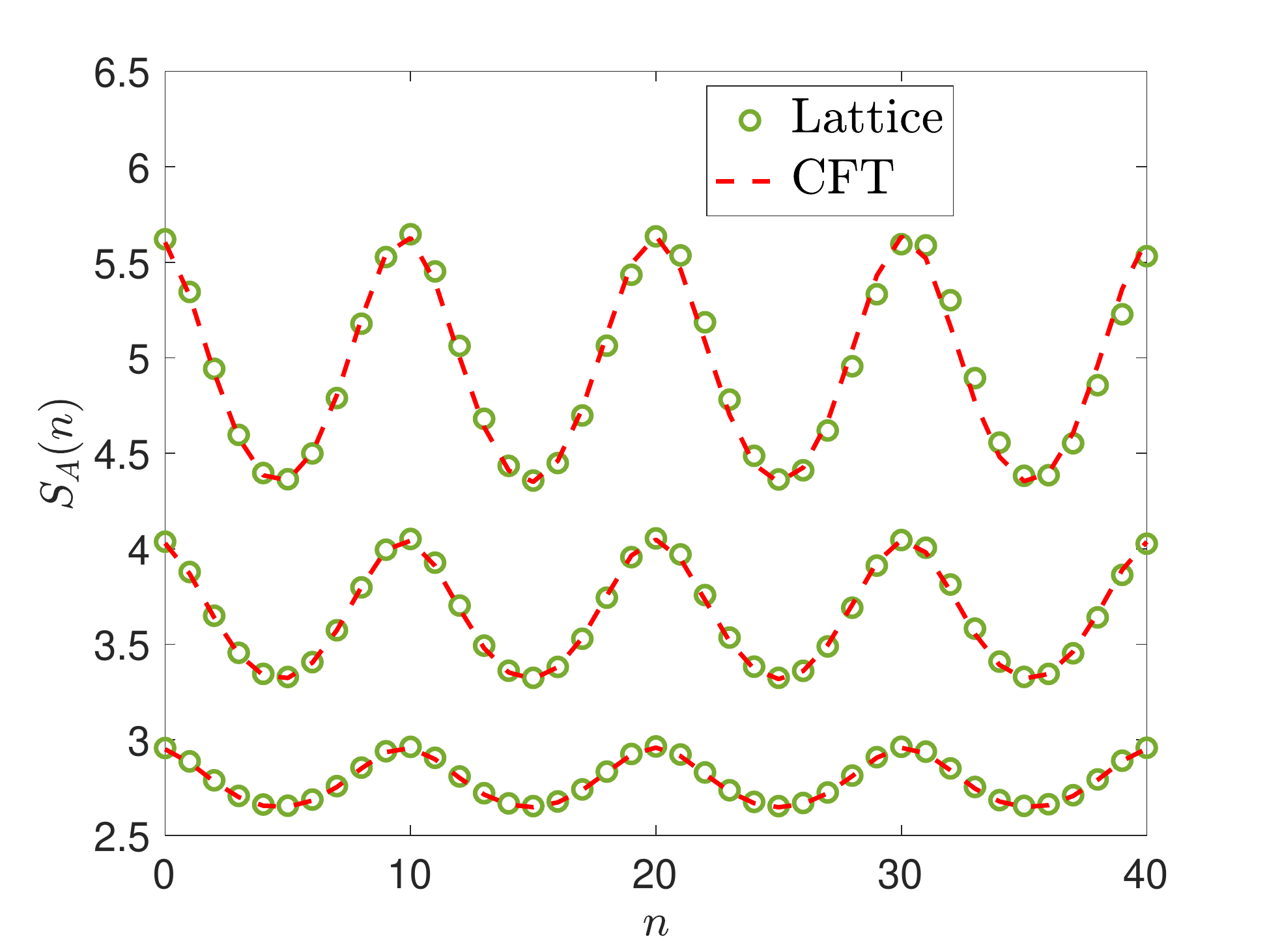}
\end{center}
\caption{
Left:
Entanglement entropy evolution for the subsystem $A=[l/5, 4l/5]$ (in which there are one chiral and one anti-chiral energy density peaks)
that contains the heating regions in the heating phase. From top to bottom, we have $\beta/l=3/50$, $1/10$, and $1/5$, respectively.
Right:
Entanglement entropy evolution for the subsystem $A=[\frac{7l}{8}, \frac{9l}{8}]$ in the non-heating phase.
From top to bottom, we have $\beta/l=3/50$, $1/10$, and $1/5$, respectively.
In the lattice model, we choose $L=2l=1000$.
The driving parameters are $T_0/l=1/10$ and $T_1/l=1/50$.
}
\label{EE_nonHeating_Floquet}
\end{figure}

In the non-heating phase, $\alpha_n$, $\beta_n$ and $(x_{n,2}-x_{n,1})$ in Eq.\eqref{SvN_driving_finiteT} simply oscillate in time. 
Therefore, the entanglement entropy $S_A$ will also oscillate in time.
Interestingly, it is noted that the finite temperature will enhance the amplitude of oscillations, because of the second term in \eqref{SvN_driving_finiteT}. A sample plot of entanglement entropy evolution can be found in Fig.\ref{EE_nonHeating_Floquet}.

\section{Entanglement Hamiltonian/spectrum evolution}

\subsection{General results}

To study the entanglement Hamiltonian evolution in our setup of driven CFTs, we need to generalize the method in Ref.\cite{2016Cardy_Tonni}. In recent works\cite{Das_2018,2020Kabat}, for different motivations, people studied the entanglement Hamiltonian of excited states that are related to the ground state through conformal transformations. This is related to driven CFTs starting from the ground state. Here we generalize to the case with a thermal initial state.

\begin{figure}[h]
\begin{center} 
\includegraphics[width=2.2in]{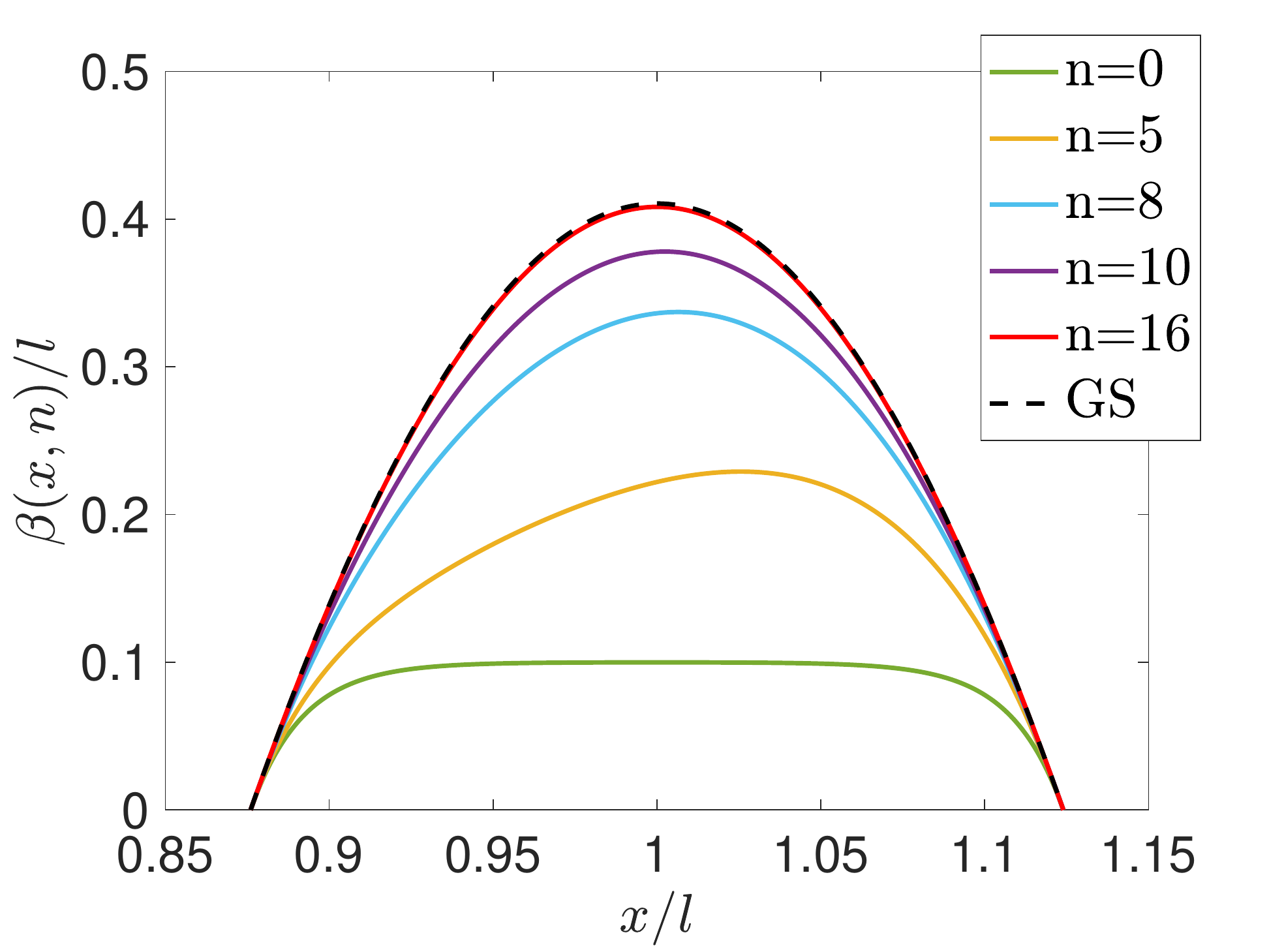}
\includegraphics[width=2.2in]{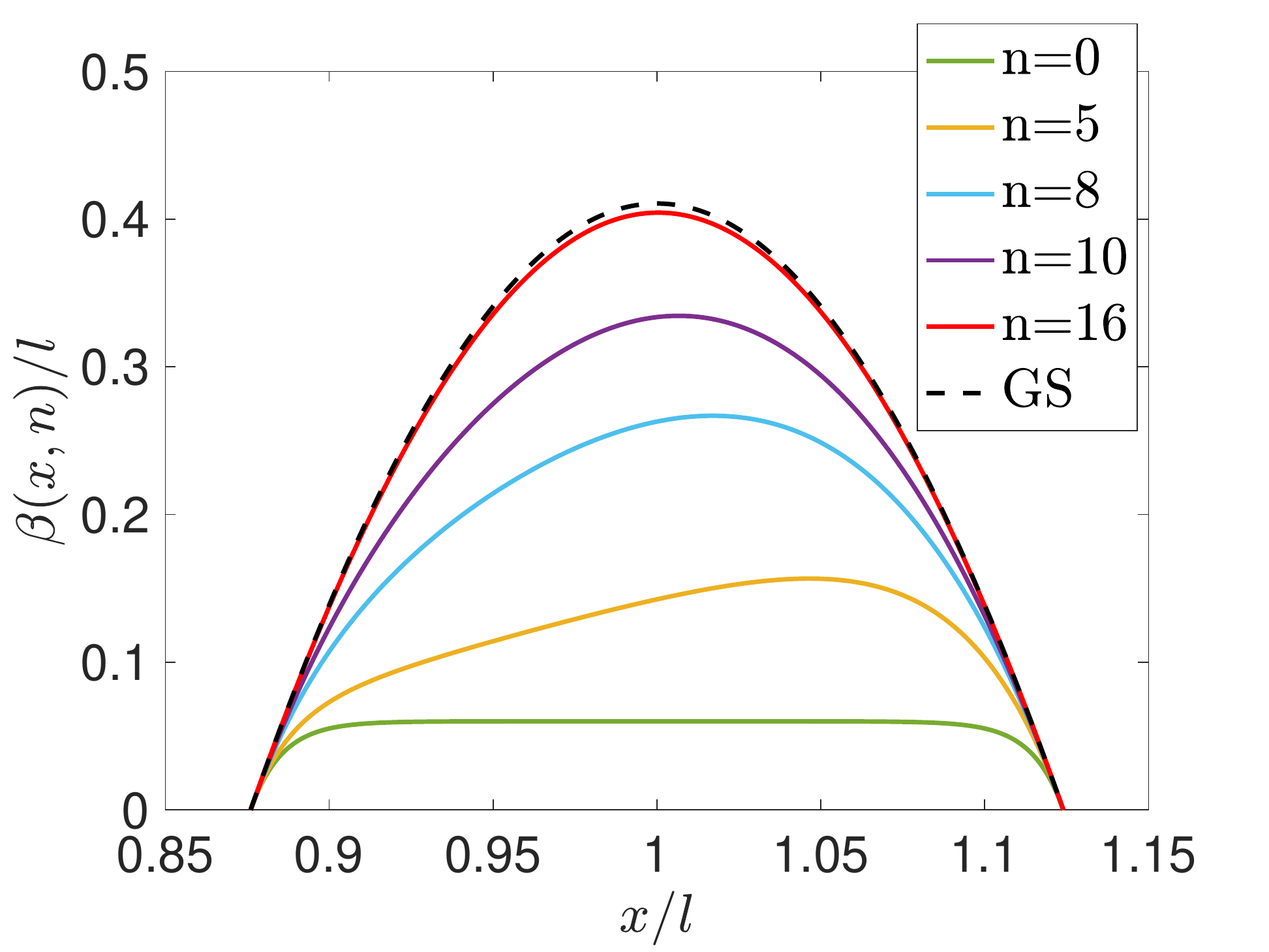}
\end{center}
\caption{
Time evolution of the effective inverse temperature $\beta(x,n)$ of the (chiral component of) entanglement Hamiltonian in the cooling region, after different cycles of driving starting from 
a thermal initial state at $\beta/l=1/10$ (left) and $\beta/l=3/50$ (right).
The dashed lines correspond to the ground state value.
The driving parameters are $T_0/l=1/50$ and $T_1/l=1/25$.
}
\label{EH_CoolingRegion}
\end{figure}

Let us give a very brief review of the approach used in Ref.\cite{2016Cardy_Tonni} first.
Consider a subsystem $A$ in a larger system, it is known that the entanglement entropy is UV divergent. To make the UV regularization, one can remove two small disks of radius $\epsilon$ around the two entangling points between $A$ and $\bar A$ in the euclidean space-time. 
Then one imposes conformal boundary conditions along the two holes.
The analysis of Ref.\cite{2016Cardy_Tonni}  is valid only for the cases in which the spacetime after removing the disks is topologically equivalent to a cylinder. Then one can find a conformal mapping $\zeta=f(w)$ that sends the $w$-spacetime to a $\zeta$-cylinder. This $\zeta$-cylinder is periodic and of length $2\pi$ in Im$(\zeta)$ direction, and of length $W$ in the Re$(\zeta)$ direction.
Then the entanglement Hamiltonian can be obtained through a pull-back of the translation generator around this $\zeta$-cylinder.
Then the entanglement Hamiltonian can be obtained as 
\be
\label{KA_general}
K_A=2\pi \int_A \frac{T(x)}{f'(x)}dx+2\pi \int_A \frac{\overline T(x)}{\bar f'(x)}dx.
\ee
In our driven CFTs, the conformal map $\zeta=f(w)$ is given in \eqref{EH_ConformalMap}, which we rewrite here:
\be
\label{fw_SM}
f(w)=\log\left(
\frac{e^{2\pi i g_n(w)/\beta}-e^{2\pi i g_n(w_1)/\beta} }{
e^{2\pi i g_n(w_2)/\beta}-e^{2\pi i g_n(w)/\beta}
}
\right),
\ee
and similarly for $\bar f(\bar w)$. Here $w_n=g_n(w)$ describes the operator evolution.  
Then based on \eqref{KA_general}, one can obtain
\be
\label{KA_n_SM}
K_A(n)=\int_{x_1}^{x_2}\beta(x,n)\, T(x)+\text{anti-chiral},
\ee
where the ``local temperature" $\beta(x,n)$ has the form
\be
\label{EH_finiteT_floquet_SM}
\begin{split}
\beta(x,n)=2\beta\cdot \frac{\sinh\frac{\pi(g_n(x_2)-g_n(x))}{\beta}\sinh\frac{\pi(g_n(x)-g_n(x_1))}{\beta}}{g_n'(x)\cdot \sinh\frac{\pi (g_n(x_2)-g_n(x_1))}{\beta}}. 
\end{split}
\ee
Note that before the driving, i.e., at $n=0$, one has
\be
\label{EH_finiteT_floquet_SM_ground}
\begin{split}
\beta(x,n=0)=2\beta\cdot \frac{\sinh\frac{\pi(x_2-x )}{\beta}\sinh\frac{\pi(x-x_1)}{\beta}}{\sinh\frac{\pi (x_2-x_1)}{\beta}},
\end{split}
\ee
which corresponds to the entanglement Hamiltonian in a Gibbs state at finite temperature $\beta^{-1}$.
Note that when $|x_i-x|\gg \beta$ and $|x_2-x_1|\gg \beta$, one has $\beta(x,n=0)\simeq \beta$ (See Fig.\ref{EH_CoolingRegion}).

Now let us consider the general formula of entanglement spectrum evolution. As studied in \cite{2016Cardy_Tonni},
the entanglement spectrum is determined by the length $W$ of the $\zeta$-cylinder as
\be
\label{W_spectrum}
2\pi^2(-\frac{c}{24}+\Delta_j)\cdot\frac{1}{W}
\ee
up to a global constant, where $\Delta_j$ are dimensions of the boundary operators consistent with the boundary conditions imposed at the two entangling points.

\begin{figure}[h]
\begin{center} 
\includegraphics[width=2.2in]{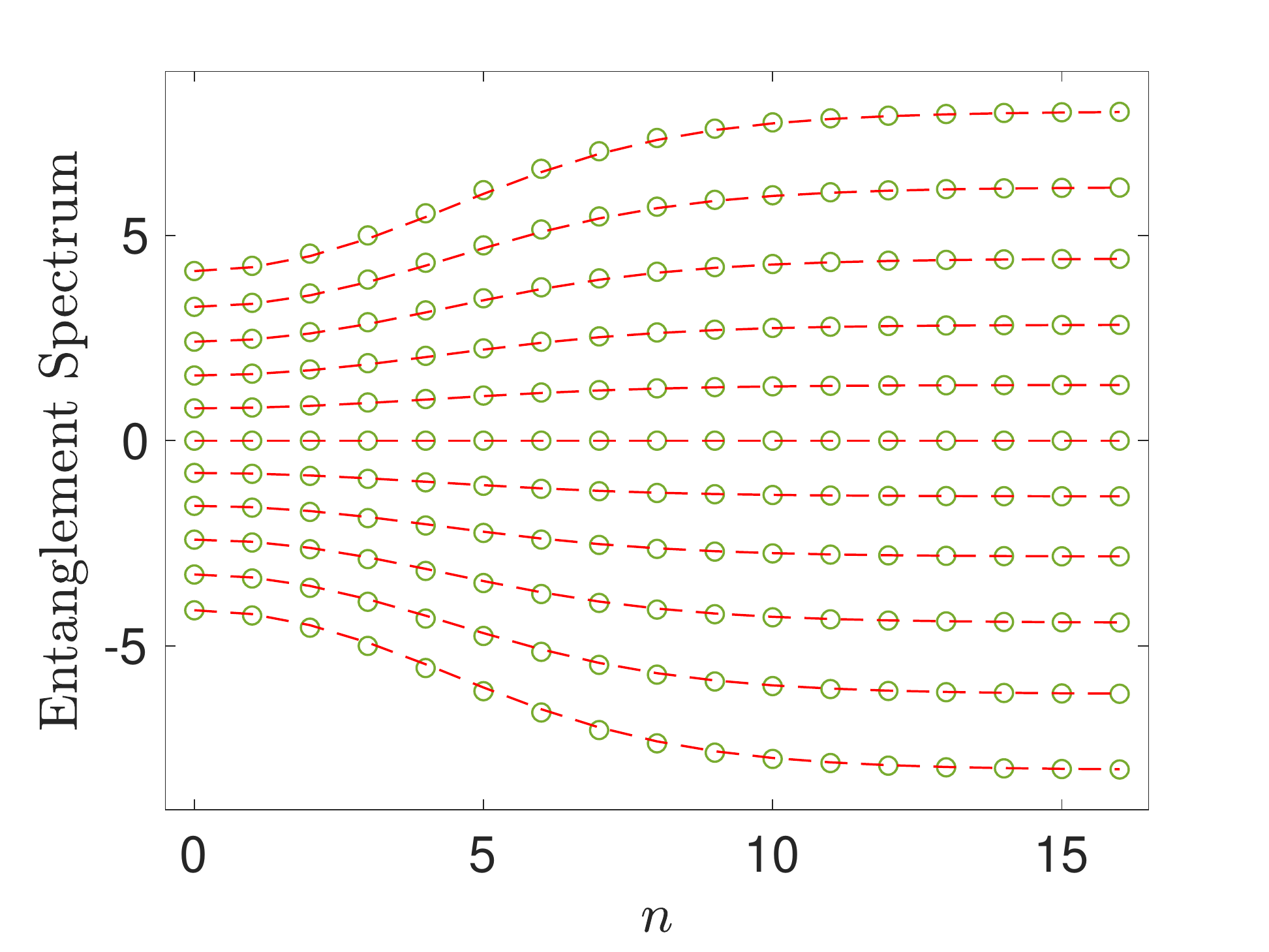}
\includegraphics[width=2.2in]{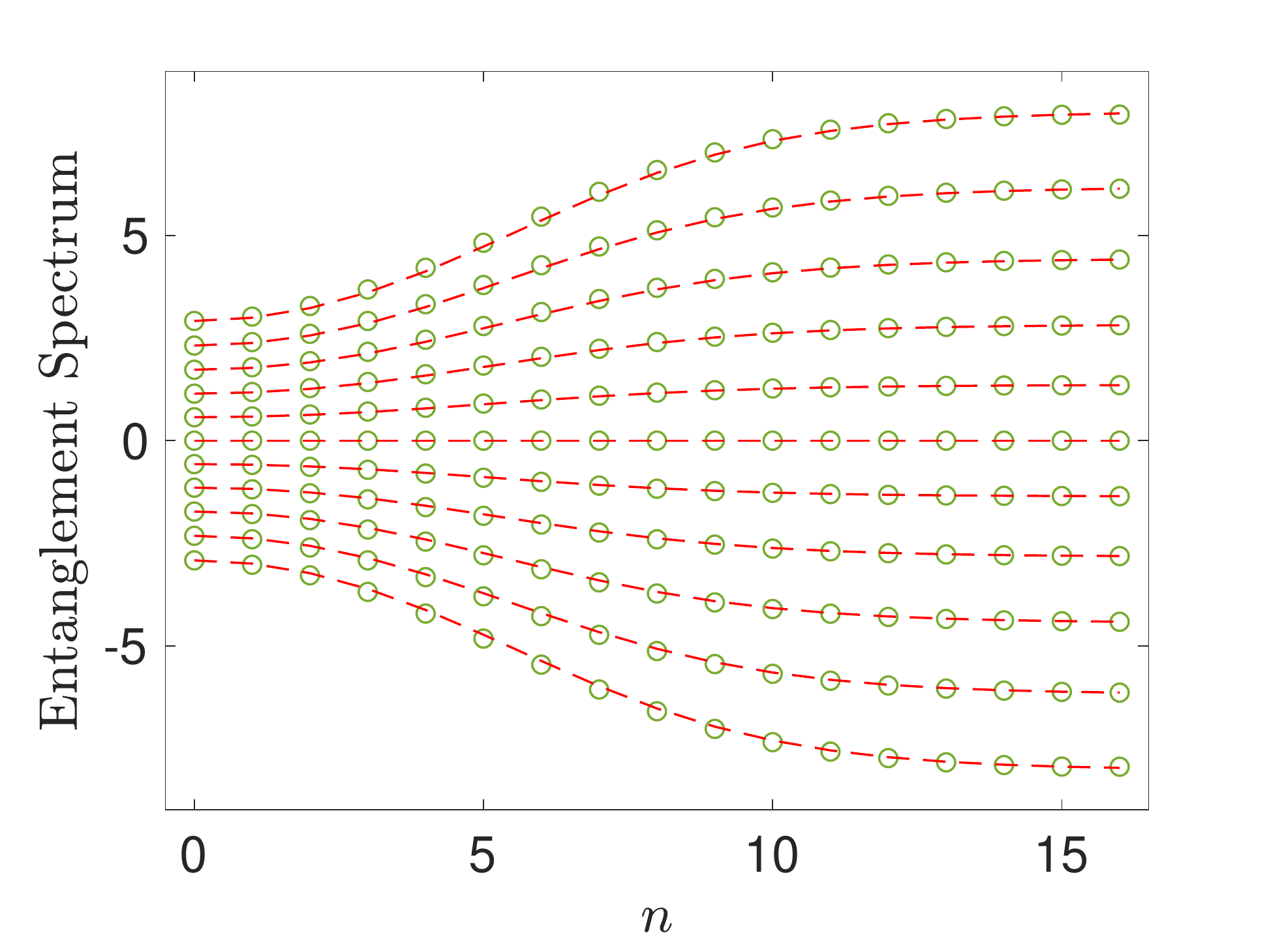}
\end{center}
\caption{
Time evolution of entanglement spectrum in the cooling region during the cooling process in a Floquet CFT. Here we choose $\beta/l=1/10$ (left) and $\beta/l=3/50$ (right).
The subsystem is of length $l_A=l/4=125$ centered at $x=l$.
The green lines are lattice results based on the calculation of correlation matrix.
The red dashed lines are obtained from diagonalizing $K_A$ in \eqref{KA_n} and \eqref{EH_finiteT_floquet} (which is obtained from CFT calculations) on the lattice directly. Parameters are the same as Fig.\ref{Fig:EH} (c).
}
\label{ES_CFT_coolingFloquet}
\end{figure}

The length $W$ of the $\zeta$-cylinder can be evaluated as follows. Based on \eqref{fw_SM} as well as the anti-holomorphic part:
\be
\bar f(w)=\log\left(
\frac{e^{2\pi i \bar g_n(\bar w)/\beta}-e^{2\pi i \bar g_n(\bar w_1)/\beta} }{
e^{2\pi i g_n(\bar w_2)/\beta}-e^{2\pi i g_n(\bar w)/\beta}
}
\right),
\ee
where $w=0+ix$ and $\bar{w}=0-ix$. 
Now we have 
\be
\begin{split}
\mathcal W=&f(i(x_2-\epsilon))-f(i(x_1+\epsilon))
=
\log \left[\frac{(e^{\frac{i 2\pi}{\beta}g_n(w_2)}-e^{\frac{i 2\pi}{\beta}g_n(w_1)})^2 }{\frac{\partial (e^{\frac{i 2\pi}{\beta}g_n(w_1)})}{\partial w_1} \cdot 
\frac{\partial (e^{\frac{i 2\pi}{\beta}g_n(w_2) })}{\partial w_2} \cdot \epsilon^2
}\right]+O(\epsilon).
\end{split}
\ee
Similarly, one can obtain
\be
\begin{split}
\overline{\mathcal W}=&\overline{f(i(x_2-\epsilon))}-\overline{f(i(x_1+\epsilon))}
=
\log \left[\frac{(e^{-\frac{i 2\pi}{\beta}\bar g_n(\bar w_2)}-e^{-\frac{i 2\pi}{\beta}\bar g_n(\bar w_1)})^2  }{\frac{\partial (e^{-\frac{i 2\pi}{\beta}\bar g_n(\bar w_1)})}{\partial \bar w_1} \cdot
\frac{\partial (e^{-\frac{i 2\pi}{\beta}\bar g_n(w_2) })}{\partial \bar w_2} \cdot \epsilon^2
}\right]+O(\epsilon).
\end{split}
\ee
Note that $\mathcal W$ and $\overline{\mathcal W}$ are in general complex in the imaginary time evolution.
The length of the cylinder is $W=\frac{1}{2}(\mathcal W+\bar{\mathcal W})$, which is real.
After the analytical continuation to the real time, one can obtain 
\be
\label{W_length_SM}
\begin{split}
W
=&\frac{1}{2}
\log\left[
\frac{
\left(\frac{\beta}{\pi}\sinh\frac{\pi (g_{n}(x_2)-g_{n}(x_1))}{\beta}\right)^2
}{
\frac{\partial g_{n}(x_1)}{\partial x_1} \cdot \frac{\partial g_{n}(x_2)}{\partial x_2}
 \cdot \epsilon^2
}\right]+O(\epsilon)
+\text{anti-chiral}.
\end{split}
\ee
Then one can observe the entanglement spectrum based on \eqref{W_spectrum}.
Furthermore, one can show that $W$ and the entanglement entropy $S_A$ are related by\cite{2016Cardy_Tonni}
\be
S_A(n)\simeq \frac{c}{6}W.
\ee
Next, we will discuss the features of entanglement Hamiltonian/spectrum evolution in different cases.

\subsection{Cooling region in the heating phase}
\label{SM:Cooling_EH}

In the cooling region, $g_n(x_1)$, $g_n(x_2)$, and $g_n(x)$ in \eqref{EH_finiteT_floquet_SM} will flow to the \textit{same} fixed point during the driving.  The distances among $g_n(x_1)$, $g_n(x_2)$, and $g_n(x)$ decrease in time as $e^{-2\lambda_L n}$, and are much smaller than $\beta$ for $\lambda_Ln\gg 1$. See Eq.\eqref{x_fixedpoint_approach}. Then $\beta(x,n)$ in \eqref{EH_finiteT_floquet} becomes
\be
\beta(x,n)= \pi\cdot \frac{(g_n(x_2)-g_n(x))(g_n(x)-g_n(x_1))}{g_n'(x)\cdot  (g_n(x_2)-g_n(x_1)) } +O(e^{-2\lambda_L n}).
\ee
Based on \eqref{x1x2} and \eqref{diff_xn}, $\beta(x,n)$ can be rewritten as
\be
\label{EH_finiteT_coolingFloquet_SM}
\beta(x,n)= l\cdot \frac{\left|\frac{z_2-\gamma_1}{z_2-\gamma_2}-\frac{z-\gamma_1}{z-\gamma_2}\right|\cdot
\left|\frac{z-\gamma_1}{z-\gamma_2}-\frac{z_1-\gamma_1}{z_1-\gamma_2}\right|}
{\left|\frac{\gamma_1-\gamma_2}{(z-\gamma_2)^2}\right|\cdot  \left|\frac{z_2-\gamma_1}{z_2-\gamma_2}-\frac{z_1-\gamma_1}{z_1-\gamma_2}\right|}+O(e^{-2\lambda_L n}),
\ee
which can be further simplified as
\be\label{EH_ground_finiteL}
\beta(x,n)=
2l \cdot\frac{
\sin\frac{\pi(x_2-x)}{l}\cdot \sin\frac{\pi(x-x_1)}{l}
}{
\sin\frac{\pi(x_2-x_1)}{l}
}+O(e^{-2\lambda_L n}).
\ee
This corresponds to nothing but the entanglement Hamiltonian in the ground state of a CFT of length $l$ with \textit{periodic} boundary conditions.

A sample plot of the time evolution of $\beta(x,n)$ starting from different initial temperatures are shown in Fig.\ref{EH_CoolingRegion}.
The exponential approaching to the ground state value can be found in Fig.\ref{Fig:EH} (b).

As a remark, based on the above discussion, it is straightforward to consider an initial state as the ground state of an infinite CFT.
Then the entanglement Hamiltonian evolves from 
\be
\label{EH_ground_infinite}
K_A(n=0)= \int_{x_1}^{x_2} dx \frac{(x_2-x)(x-x_1)}{x_2-x_1} T(x)+\text{anti-chiral},
\ee
to \eqref{EH_ground_finiteL} in the limit $\lambda_L n\gg 1$.

\begin{figure}[h]
\begin{center} 
\includegraphics[width=2.2in]{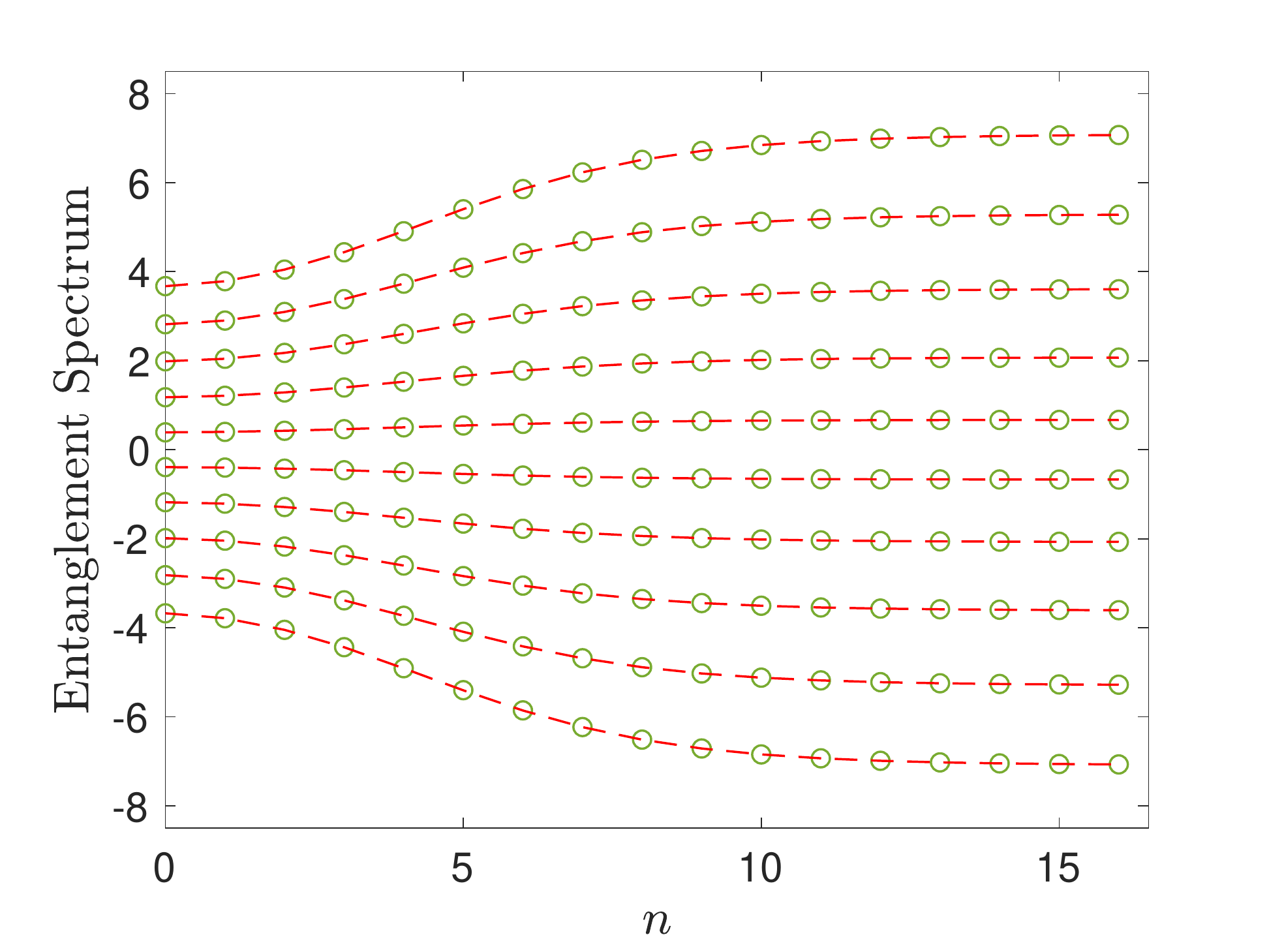}
\includegraphics[width=2.2in]{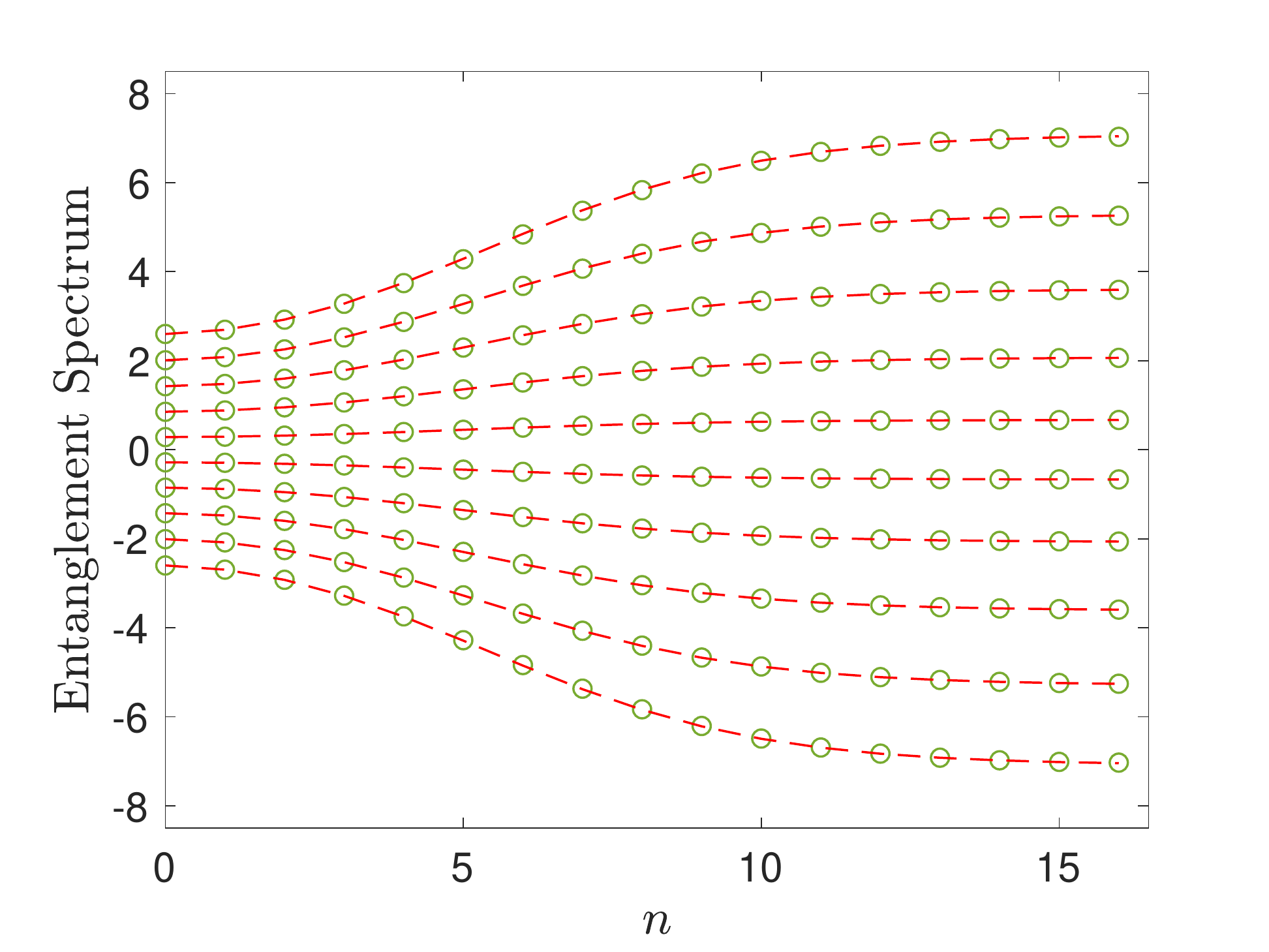}
\end{center}
\caption{
Entanglement spectrum evolution in the cooling region during the cooling process in a Floquet CFT. 
The parameters are the same as Fig.\ref{ES_CFT_coolingFloquet}, except that now we choose $l_A=l/4=125\to l_A=l/4+1=126$, i.e., there are even number of sites in subsystem $A$ now. In this case, there is no zero mode in the single-particle entanglement Hamiltonian.
}
\label{ES_CFT_coolingFloquet_even}
\end{figure}

For the entanglement spectrum, we consider two ways to obtain its time evolution.
One way is to use \eqref{W_spectrum} and \eqref{W_length_SM}. See the plot in Fig.\ref{W_length} (d).
The other way is to use \eqref{KA_n_SM} and \eqref{EH_finiteT_floquet_SM} by diagonalizing $K_A$ directly.
Then the entanglement spectrum corresponds to the energy spectrum of $K_A$.
One can define $K_A$ on a lattice (e.g., lattice free fermion). In general, this is not easy to do because one needs to know the lattice version of $T(x)$ and $\overline T(x)$ respectively. However, if the subsystem is chosen symmetric respect to the deformation in the driving Hamiltonian, then $K_A$ can be written as $K_A=\int_{x_1}^{x_2}\beta(x,n)\, T_{00}(x)$ where $T_{00}(x)$ is the Hamiltonian density.
For example, the red dashed lines in Fig.\ref{ES_CFT_coolingFloquet} are obtained by defining $K_A$ on a lattice free fermion with $T_{00}(x)=-\frac{1}{2}c_x^\dag c_{x+1}+h.c$. 

Careful readers may notice there is a zero mode in the entanglement spectrum in Fig.\ref{ES_CFT_coolingFloquet}.
This can be easily understood based on the form of $K_A$. In the free fermion lattice model, there are only nearest neighbor hopping in $K_A$. One can show explicitly that the single-particle eigenenergy of $K_A$ appear in pairs $(E_i,-E_i)$. In particular, there is a single zero mode if there are odd number of sites in subsystem $A$. If one chooses even number of sites in $A$, then there is no zero mode in the entanglement spectrum, as shown in Fig.\ref{ES_CFT_coolingFloquet_even}.

\subsection{Heating region in the heating phase: local temperature at the energy peak}

In the heating region, different from the behavior in the cooling region, $g_n(x_2)$ and $g_n(x_1)$ in \eqref{EH_finiteT_floquet_SM} will flow to different fixed points $x_{\bullet}$ and $x_{\bullet}+l$ respectively. In this case, the finite temperature $\beta^{-1}$ cannot be canceled out from $\beta(x,n)$ even in the long time limit $\lambda_L n\gg 1$.

\begin{figure}[t]
\centering
\begin{tikzpicture}
    \node[inner sep=0pt] (russell) at (15pt,-85pt)
    {\includegraphics[width=.35\textwidth]{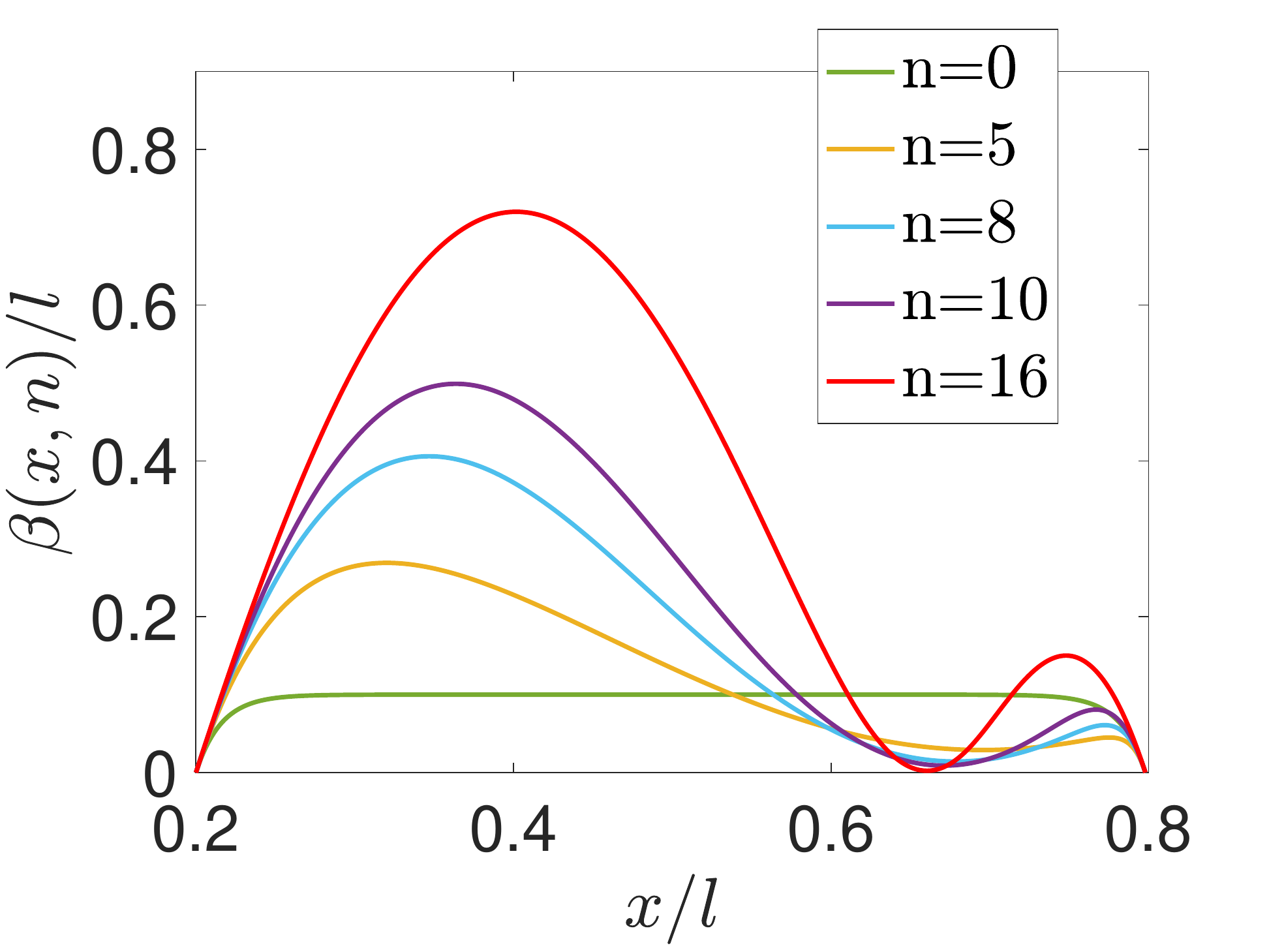}};
        \node[inner sep=0pt] (russell) at (195pt,-85pt)
    {\includegraphics[width=.35\textwidth]{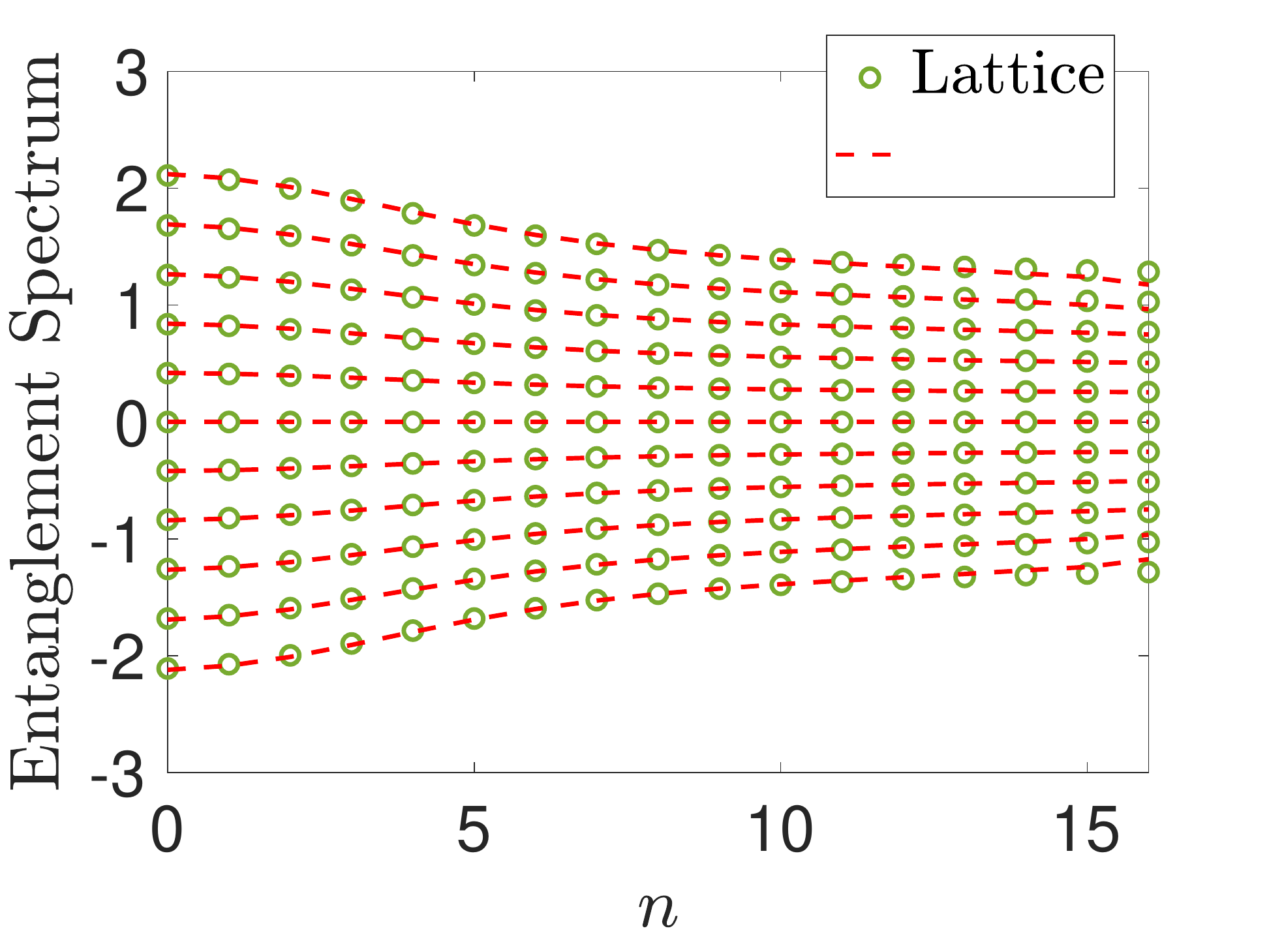}};
    
                                \node at (245pt, -40pt){\scriptsize Eq.\ref{EH_finiteT_floquet}};

    \end{tikzpicture}
\caption{
The effective inverse temperature $\beta(x,n)$ of the (chiral component of) entanglement Hamiltonian (left) and entanglement spectrum (right) evolution in a subsystem that contains the heating region 
with the initial temperature $\beta/l=1/10$.
The local temperature $\beta^{-1}(x,n)$ at the location of energy density peak grows in time.
For the entanglement spectrum, the green circles are lattice results based on the calculation of correlation matrix.
The red dashed lines are obtained from diagonalizing $K_A$ in \eqref{KA_n} and \eqref{EH_finiteT_floquet} (which is obtained in CFT calculations) on the lattice directly.
}
\label{Fig:EH2}
\end{figure}

A sample plot of $\beta(x,n)$ in the subsystem that contains the heating region can be found in Fig.\ref{Fig:EH2}.
One can find that near the location of (chiral) energy density peak, the local temperature $\beta(x,n)^{-1}$ grows in time.
Now let us evaluate this local temperature explicitly.
First, it is noted that the location of (chiral) energy density peak corresponds to the unstable fixed point $x_\circ$ in the operator evolution
(which corresponds to $\gamma_2$ in \eqref{FixedPoint}).
Second, at the unstable fixed point, based on \eqref{Diff_unstable}, one has $g'_n(x_\circ)=e^{-2\lambda_L n}$.
Then based on \eqref{EH_finiteT_floquet_SM}, the local temperature at the energy density peak becomes:
\be
\beta(x_\circ, n)=
2\beta\cdot \frac{\sinh\frac{\pi(x_\bullet-x_\circ+l)}{\beta}\sinh\frac{\pi(x_\circ-x_\bullet)}{\beta}}{ \sinh\frac{\pi l}{\beta}}\cdot e^{2\lambda_l n},
\ee
where $0<x_\circ-x_\bullet<l$.
In other words, the `local temperature' at the energy-density peak grows exponentially fast in time. 
This is because there is a strong entanglement, which grows in time, between this peak and its two nearest two peaks (with the same chirality).

\subsection{Non-heating phase}

\begin{figure}[h]
\begin{center} 
\includegraphics[width=2.2in]{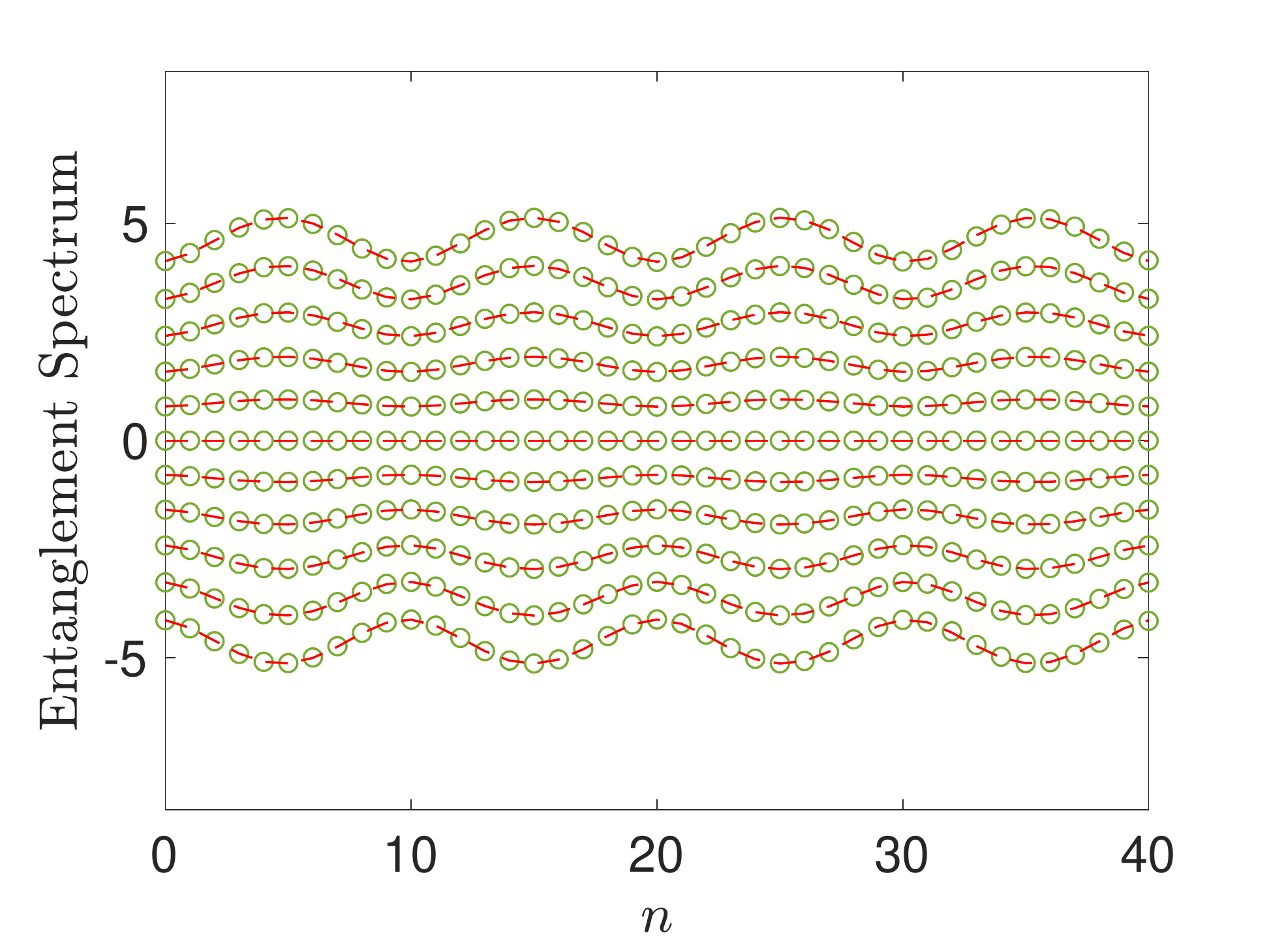}
\includegraphics[width=2.2in]{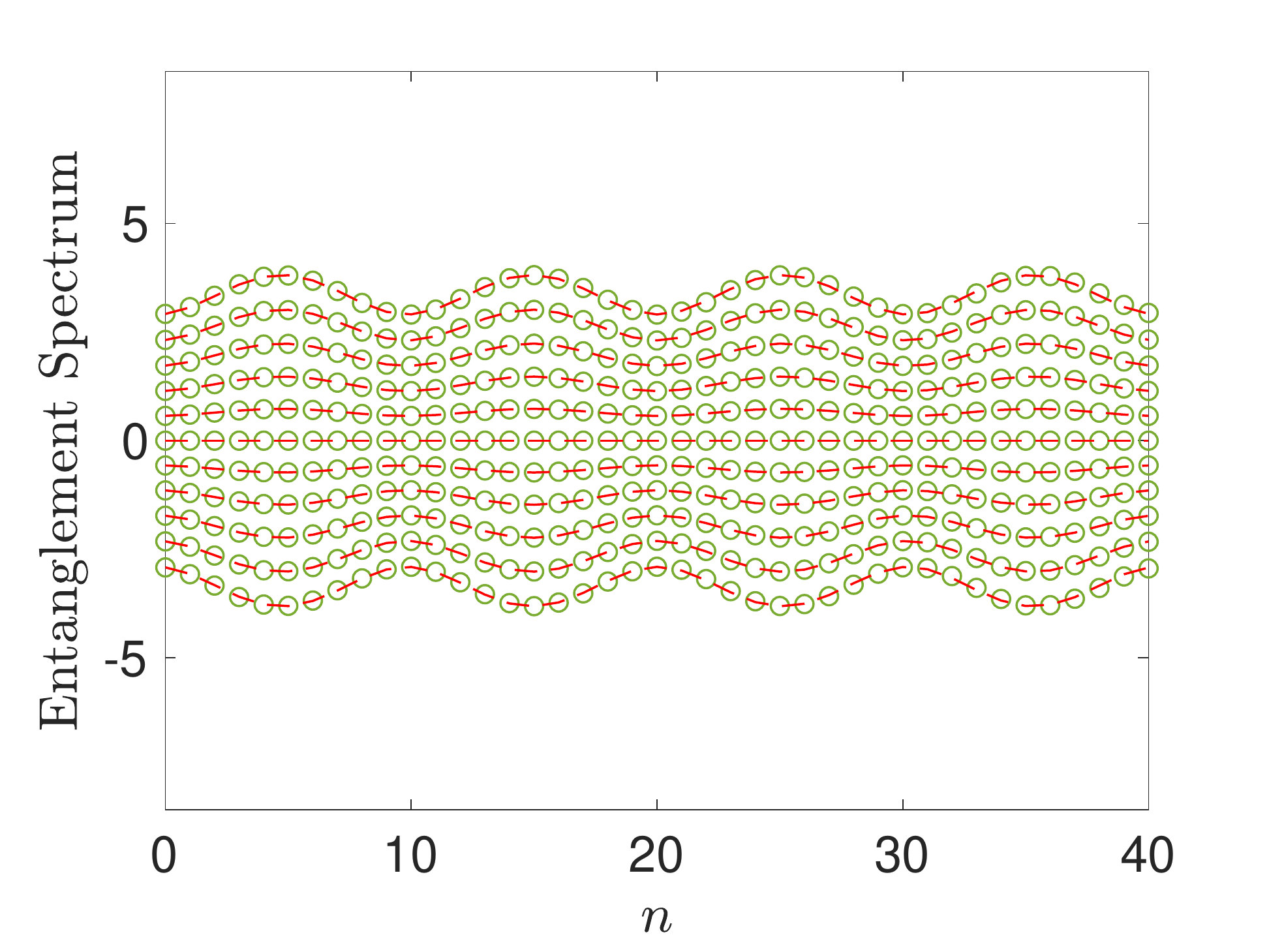}
\end{center}
\caption{
Entanglement spectrum evolution in the non-heating phase of a Floquet CFT starting from a thermal state. Here we choose $\beta/l=1/10$ (left) and $\beta/l=3/50$ (right).
In the lattice model we choose $L=2l=1000$. 
The subsystem of length $l_A=l/4$ is chosen symmetric about $x=l$.
The driving times are $T_0/l=1/10$ and $T_1/l=1/50$ respectively.
The green circles are lattice results based on the calculation of correlation matrix.
The red dashed lines are obtained from diagonalizing $K_A$ in \eqref{KA_n} and \eqref{EH_finiteT_floquet} (which is obtained from CFT calculations) on the lattice directly.
}
\label{ES_Floquet_nonHeating}
\end{figure}

In the non-heating phase, since $g_n(x_2)$, $g_n(x_1)$ and $g_n(x)$ in \eqref{EH_finiteT_floquet_SM}  will oscillate in time, both the entanglement Hamiltonian and the entanglement spectrum will oscillate during the driving.
See, e.g., the entanglement spectrum evolution in Fig.\ref{ES_Floquet_nonHeating} starting from different initial temperatures.
The oscillating period is determined by the driving parameters $(H_i, T_i)$.

Before we leave this section, we note the close relation between our setup and the moving mirror setup in conformal field theory, where the associated time evolution is described by a conformal transformation.
It will be interesting to study whether the CFC found in our setup can be realized in the moving mirror problem \cite{1994Holzhey,2019Martin,2021Takayanagi}.

\section{Conformal cooling via a quantum quench}

As the limit of a time-dependent driving, one can choose only one driving Hamiltonian, which corresponds to a single quantum quench.
Here we show that our method/discussion of conformal cooling in driven CFTs also works for this case.
Similar to the case of periodic driving, the conformal cooling can be observed if and only if there are fixed points in the operator evolution.
For example, for the three types of deformed Hamiltonians as listed in \eqref{HamiltonianType}, one can realize conformal cooling if the quenched Hamiltonian is of parabolic or hyperbolic type.

The operator evolution in \eqref{w_n} is now modified as
\be
\label{w_t}
\small
w(t)=\frac{l}{2\pi }\log \left(\Pi(t)\cdot e^{\frac{2\pi}{l}w}\right)\,,\, \Pi(t)=
\begin{pmatrix}
\alpha(t) &\beta(t)\\
\beta^\ast(t) &\alpha^\ast(t)
\end{pmatrix},
\ee
where $\Pi(t)$ is again a SU$(1,1)$ matrix in the real time evolution.
In the single quench problem, now we have a continuous time rather than the stroboscopic time (which is discrete).

For a subsystem $A=[x_1,x_2]$, the time evolution of entanglement entropy and entanglement Hamiltonian can be described by
\be
\label{SA_quench}
S_A(t)-S_A(t=0)=\frac{c}{12}\cdot
\log\left(\frac{
\left|\sinh\frac{\pi }{\beta}(x_1(t)-x_2(t))\right|^2 }{
\left|\sinh\frac{\pi }{\beta}(x_{1}-x_{2})\right|^2
\frac{\partial x_1(t)}{\partial x_1}
\frac{\partial x_2(t)}{\partial x_2}
}
\right)
+\text{anti-chiral},
\ee
and 
\be
\label{KA_t_SM2}
K_A(t)=\int_{x_1}^{x_2}\beta(x,t)\, T(x)+\text{anti-chiral},
\ee
with
\be
\label{EH_finiteT_quench_SM2}
\begin{split}
\beta(x,t)=2\beta\cdot \frac{\sinh\frac{\pi(x_2(t)-x(t))}{\beta}\sinh\frac{\pi(x(t)-x_1(t)}{\beta}}{\frac{\partial x(t)}{\partial x}\cdot \sinh\frac{\pi (x_2(t)-x_1(t))}{\beta}}. 
\end{split}
\ee
The analysis of conformal cooling process is similar to the case of periodic drivings.
As a concrete example, in the following discussion, we will consider the quenched Hamiltonian of the form in \eqref{H_deform} with $v(x)=2\sin^2\frac{\pi x}{l}$. This choice corresponds to the sine-square-deformation (SSD), as extensively studied in recent literatures
\cite{Nishino2011prb,2011freefermionssd,katsura2012sine,ishibashi2015infinite,ishibashi2016dipolar, Okunishi:2016zat, Wen_2018,Ryu1604,
Tamura:2017vbx, Tada:2017wul,
MacCormack_2019,
tada2019time,caputa2020geometry,liu2020analysis,2021Hotta}.
In this case, we have $\alpha(t)=1+i\frac{\pi t}{l}$ and $\beta(t)=-i\frac{\pi t}{l}$ in \eqref{w_t} in the real time evolution.
$x(t)$ is related to $x(t=0)$ through 
\be
x(t)=\frac{l}{2\pi i}\log\left(
1+\frac{1}{\frac{1}{e^{\frac{2\pi i x}{l}}-1}+\frac{i\pi t}{l}}
\right).
\ee
In the long time driving limit $t/l\gg 1$, one can find that $x(t)$ approaches the fixed point $x_\bullet =n l$ where $n\in \mathbb Z$ as
\be
x(t)\simeq nl-\frac{l^2}{2\pi^2} \cdot \frac{1}{t}.
\ee
In other words, the operator evolution flows to the fixed point $x_\bullet =nl$ polynomially fast, which is different from our Floquet example where the approaching to the fixed point is exponentially fast. The reason is that SSD Hamiltonian belongs to the parabolic type in \eqref{HamiltonianType}. Due to this feature, later we will see the cooling effect is also polynomially fast (rather than exponentially fast).

As a remark, the reason the cooling effect with a SSD Hamiltonian is polynomially fast is because the SSD Hamiltonian is of parabolic type (See \eqref{HamiltonianType}). If one uses a hyperbolic type quenched Hamiltonian, then one can also realize a cooling that is exponentially fast.
See also Ref.\cite{kuzmin2021probing} for a closely related discussion.

\subsection{Entanglement entropy evolution}

\begin{figure}
\begin{center} 
\includegraphics[width=2.2in]{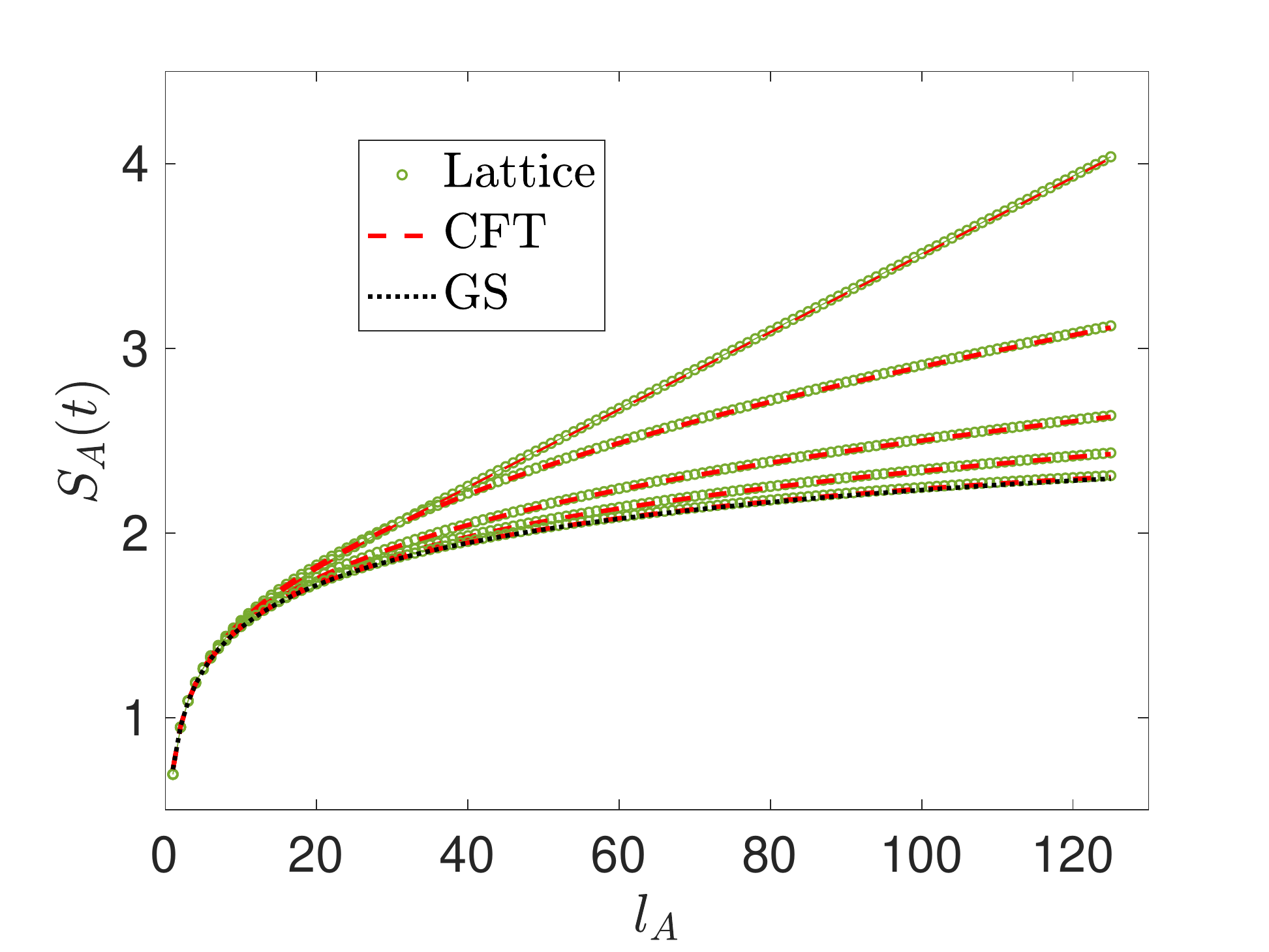}
\includegraphics[width=2.2in]{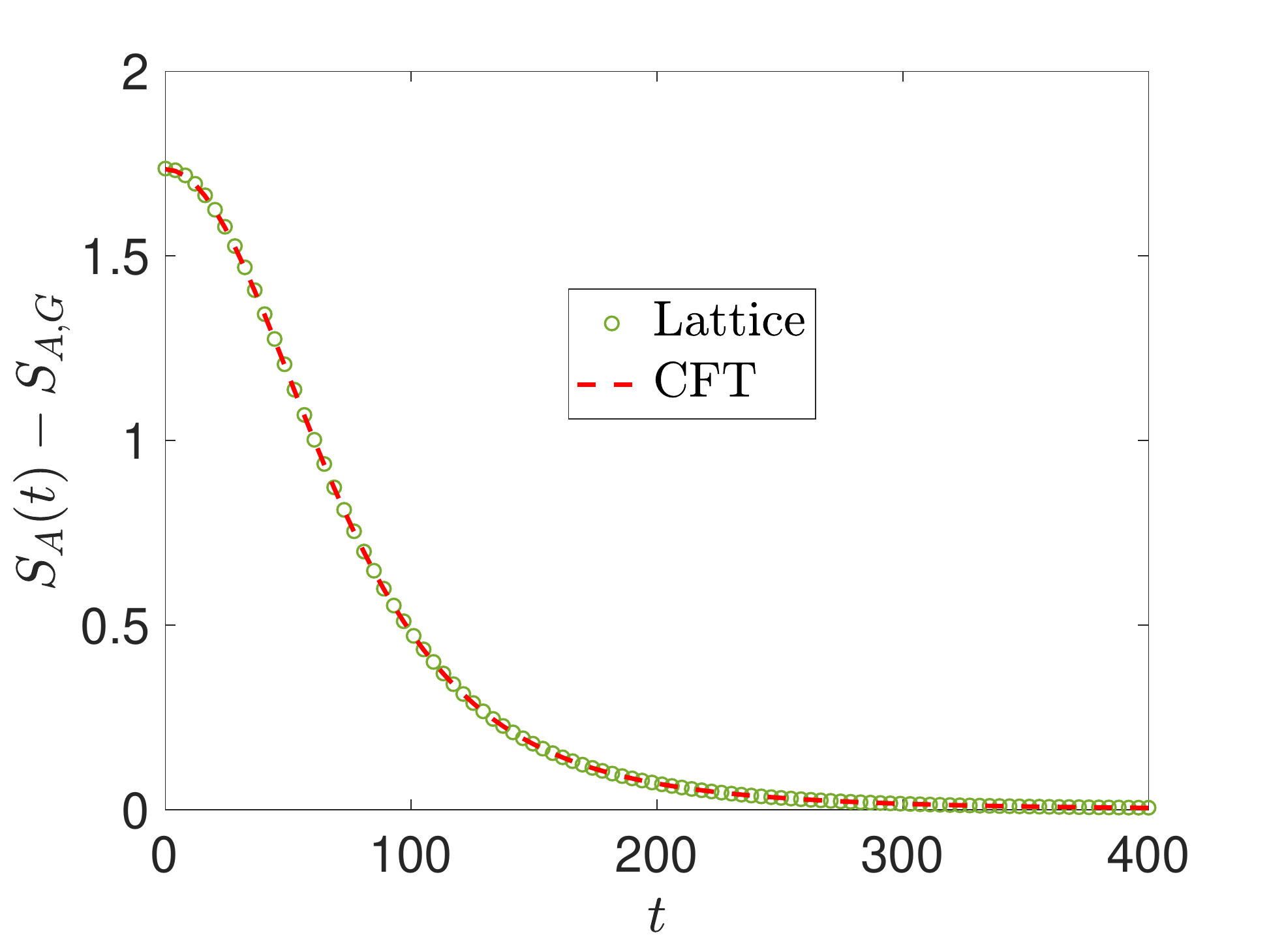}
\includegraphics[width=2.2in]{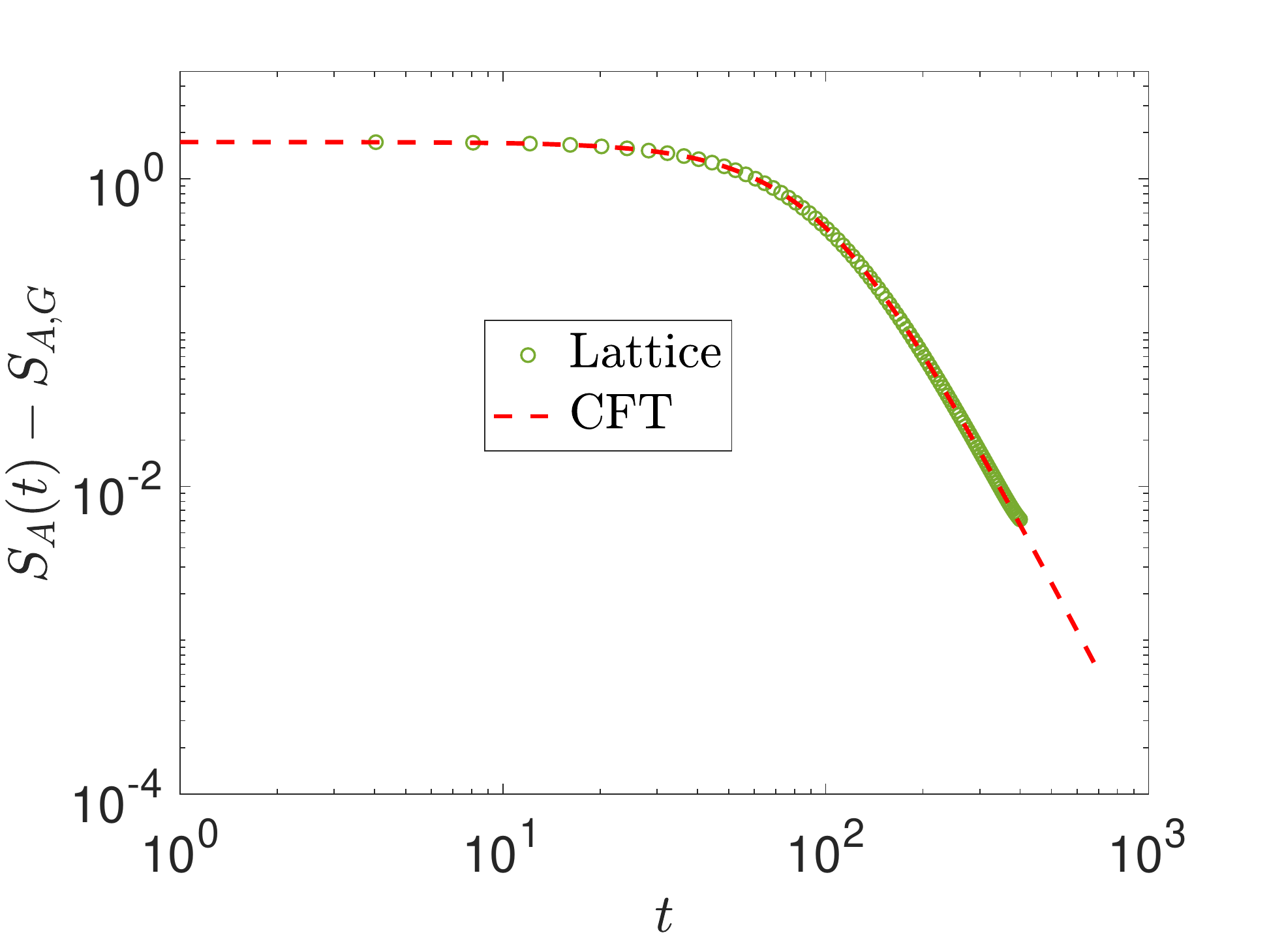}
\end{center}
\caption{Left:
Entanglement entropy evolution as a function of $l_A=|x_2-x_1|$ centered at $x=l/2$ in the cooling region.
From top to bottom, we choose 
$t/l=0$, $1/5$, $3/10$, $2/5$, and $4/5$, where $l=500$. The initial temperature is $\beta/l=1/10$.
Middle: Entanglement entropy evolution as a function of time by fixing $l_A=l/4=125$.
Right: A re-plot of entanglement entropy evolution in log-log scale.
In the free-fermion lattice model, we choose $L=l=500$ with periodic boundary conditions.
The CFT result is plotted according to \eqref{SA_quench}.
}
\label{EE_cooling_Quench}
\end{figure}

For the SSD Hamiltonian, the heating regions are located near $n\cdot l$ where $n\in \mathbb Z$.
Now let us consider the cooling region between two heating regions.
As shown in Fig.\ref{EE_cooling_Quench}, we consider the entanglement entropy evolution for a subsystem of length $l_A$ centered at $x=l/2$.
One can find the entanglement entropy evolves to the ground state value in time.
In particular, since the quenched Hamiltonian is of the parabolic type, $S_A(t)$ approaches the ground state value $S_{A,G}=\frac{c}{3}\log(\frac{l}{\pi}\sin\frac{\pi l_A}{l})$ polynomially fast in time, as shown in Fig.\ref{EE_cooling_Quench} (right plot).

\subsection{Entanglement Hamiltonian/spectrum evolution}

\begin{figure}
\begin{center} 
\includegraphics[width=2.2in]{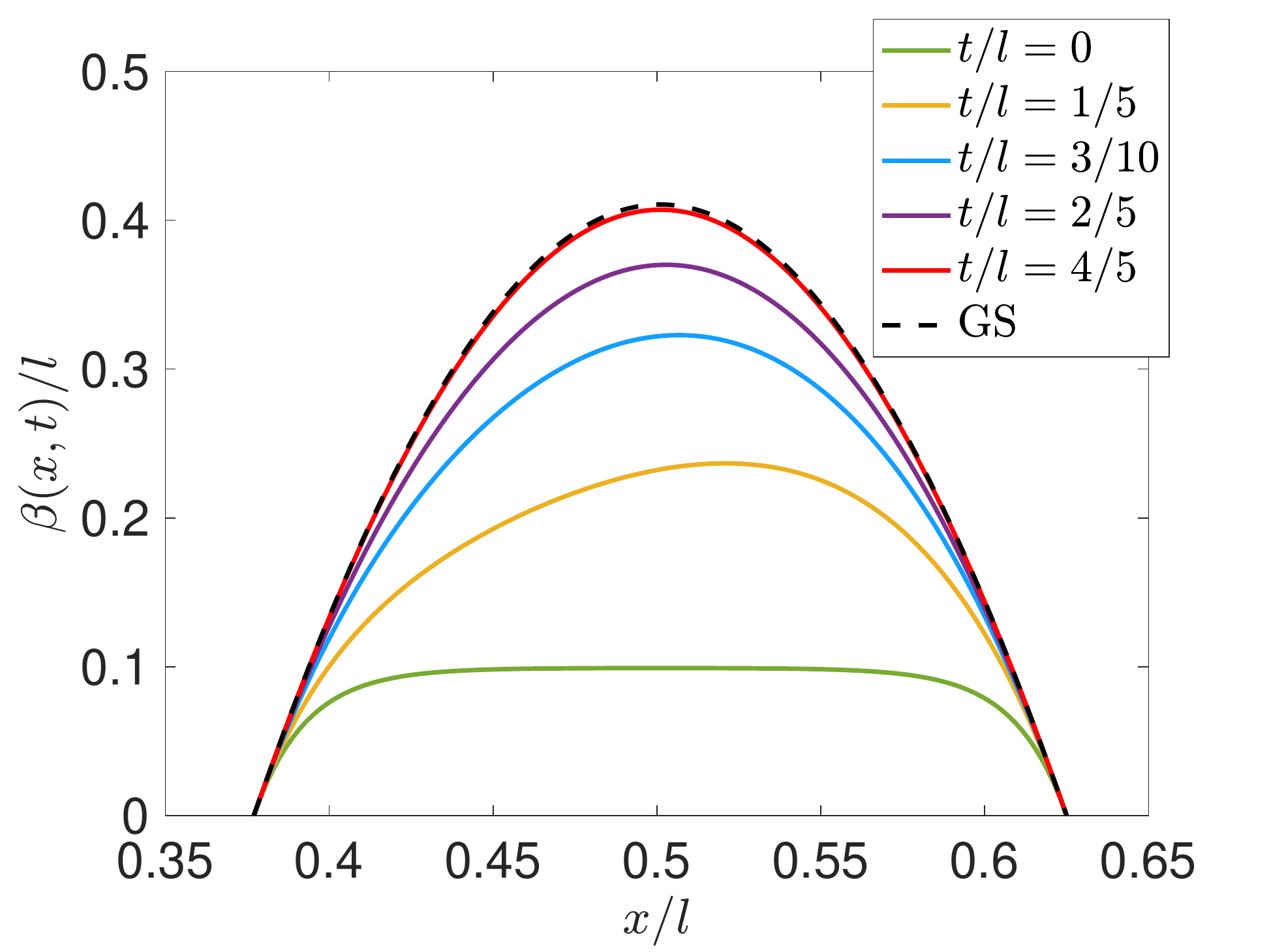}
\includegraphics[width=2.2in]{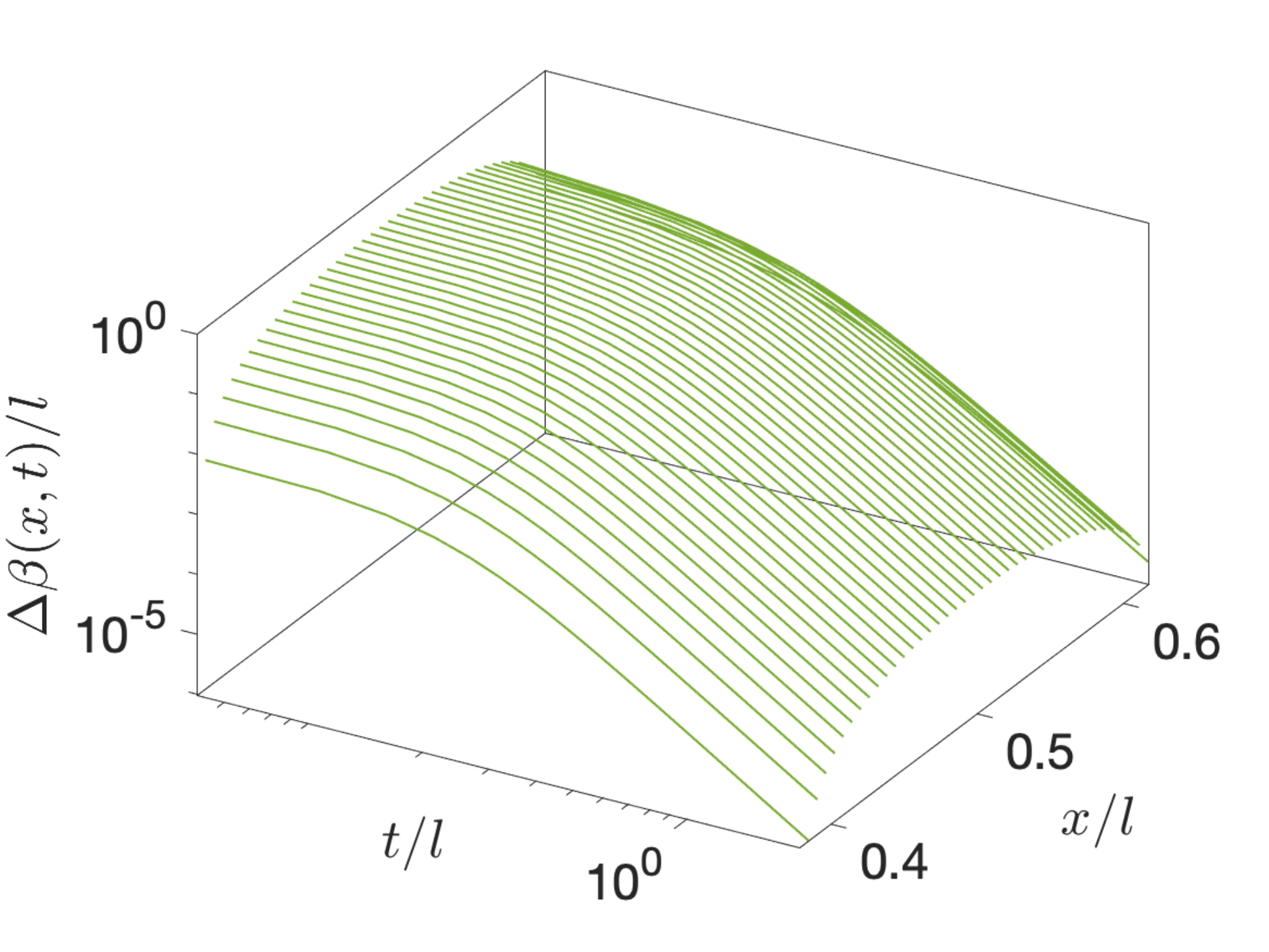}
\includegraphics[width=2.2in]{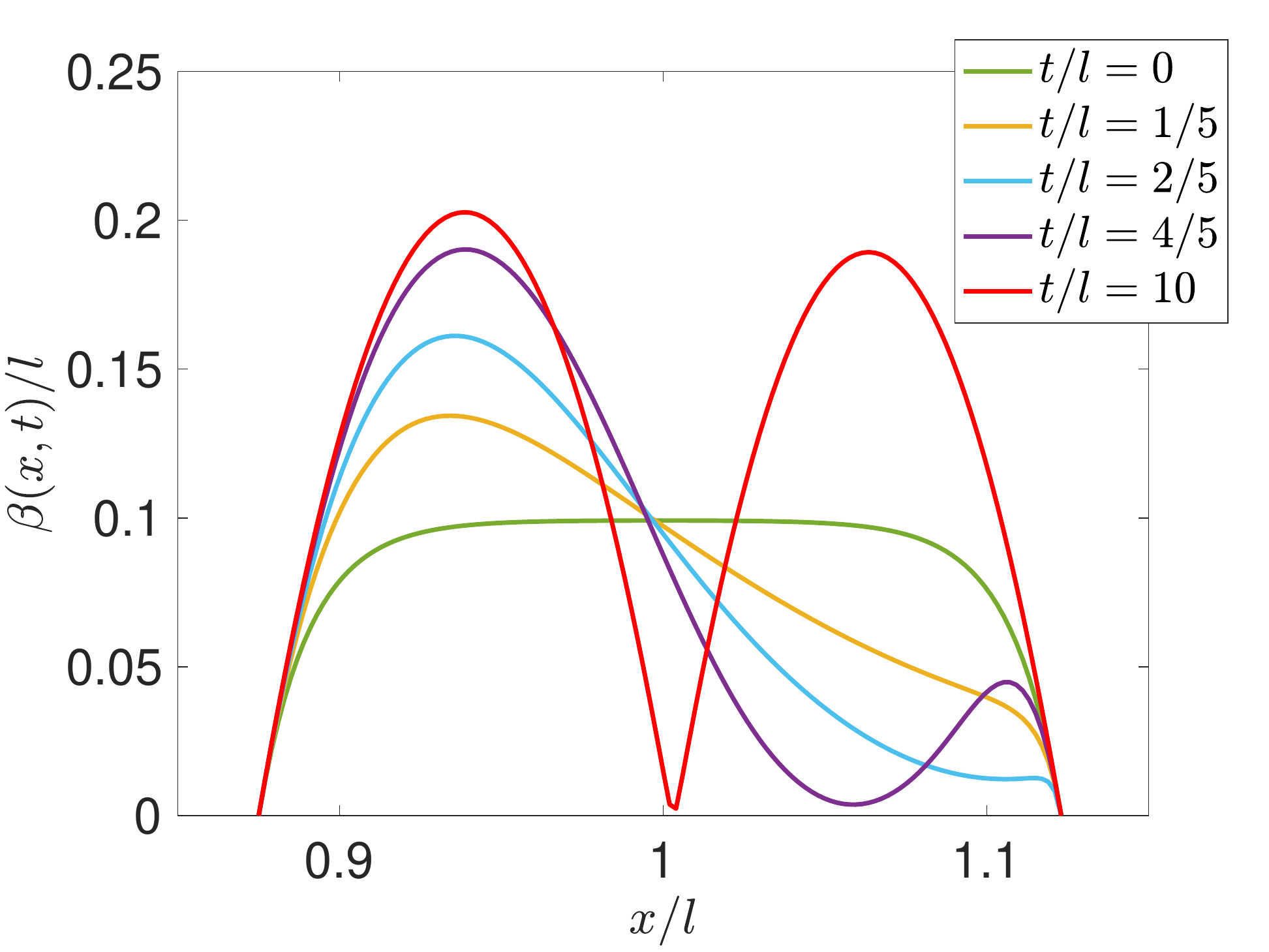}
\end{center}
\caption{
Left: Time evolution of $\beta(x,t)$ in \eqref{EH_finiteT_quench_SM2} in (part of) the cooling region after a quantum quench.
The initial temperature is $\beta/l=1/10$. The dashed line corresponds to the ground state value in \eqref{GS_EH_quench}.
Middle: The difference $\Delta \beta(x,t)=\beta_{GS}(x)-\beta(x,n)$ in the cooling region. 
Right: 
Time evolution of $\beta(x,t)$ in \eqref{EH_finiteT_quench_SM2} in the subsystem that contains the heating region after a quantum quench. The local temperature near the hot spot ($x/l=1$) grows in time.
The initial temperature is $\beta/l=1/10$. 
}
\label{EH_cooling_Quench}
\end{figure}

The entanglement Hamiltonian evolution is described by \eqref{KA_t_SM2} and \eqref{EH_finiteT_quench_SM2}.
In the cooling region, similar to the case of Floquet driving, $x_1(t)$, $x_2(t)$ and $x(t)$ will flow to the same fixed point.
Following the same discussion in Sec.\ref{SM:Cooling_EH}, one can find that $\beta(x,n)$ will evolve to the ground state value in a CFT of length $l$ with periodic boundary conditions, i.e., 
\be
\label{GS_EH_quench}
\beta_{\text{GS}}(x)=2l \cdot\frac{
\sin\frac{\pi(x_2-x)}{l}\cdot \sin\frac{\pi(x-x_1)}{l}
}{
\sin\frac{\pi(x_2-x_1)}{l}
}.
\ee
However, different from the example of Floquet driving, here $\beta(x,n)$ approaches $\beta_{\text{GS}}$ polynomially fast in time, as shown in Fig.\ref{EH_cooling_Quench} (where $\beta_{\text{GS}}-\beta(x,t)$ decreases polynomially fast as a function of time $t$).

For the subsystem that contains a heating region, similar to the Floquet case, one can observe the local temperature $\beta(x,t)$ grows in time at the hot spot, as shown in Fig.\ref{EH_cooling_Quench}.
Again, different from the Floquet case where the local temperature grows exponentially in time, here one can show that the local temperature at the hot spot grows linearly in $t$.

\begin{figure}
\begin{center} 
\includegraphics[width=2.2in]{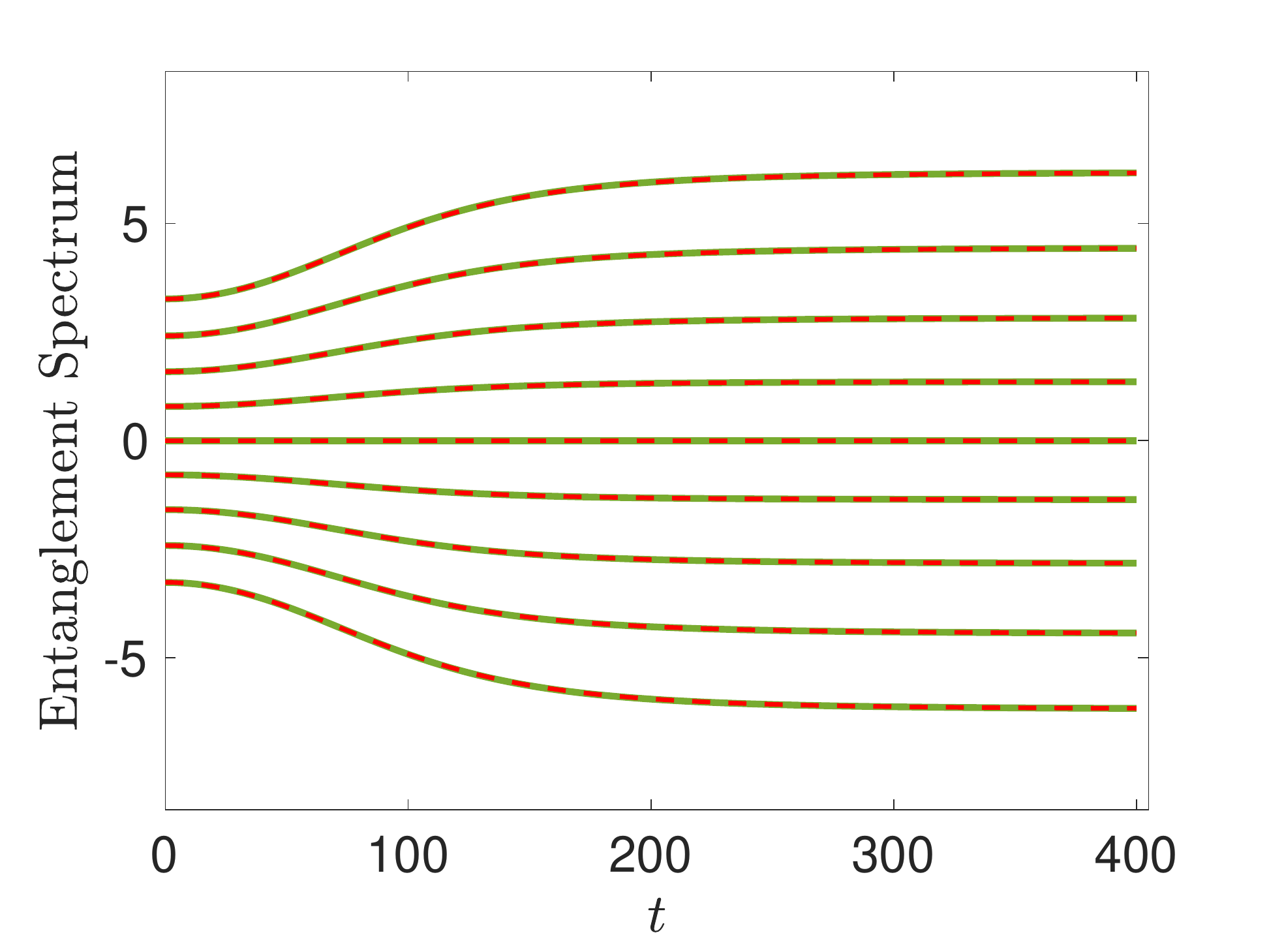}
\includegraphics[width=2.2in]{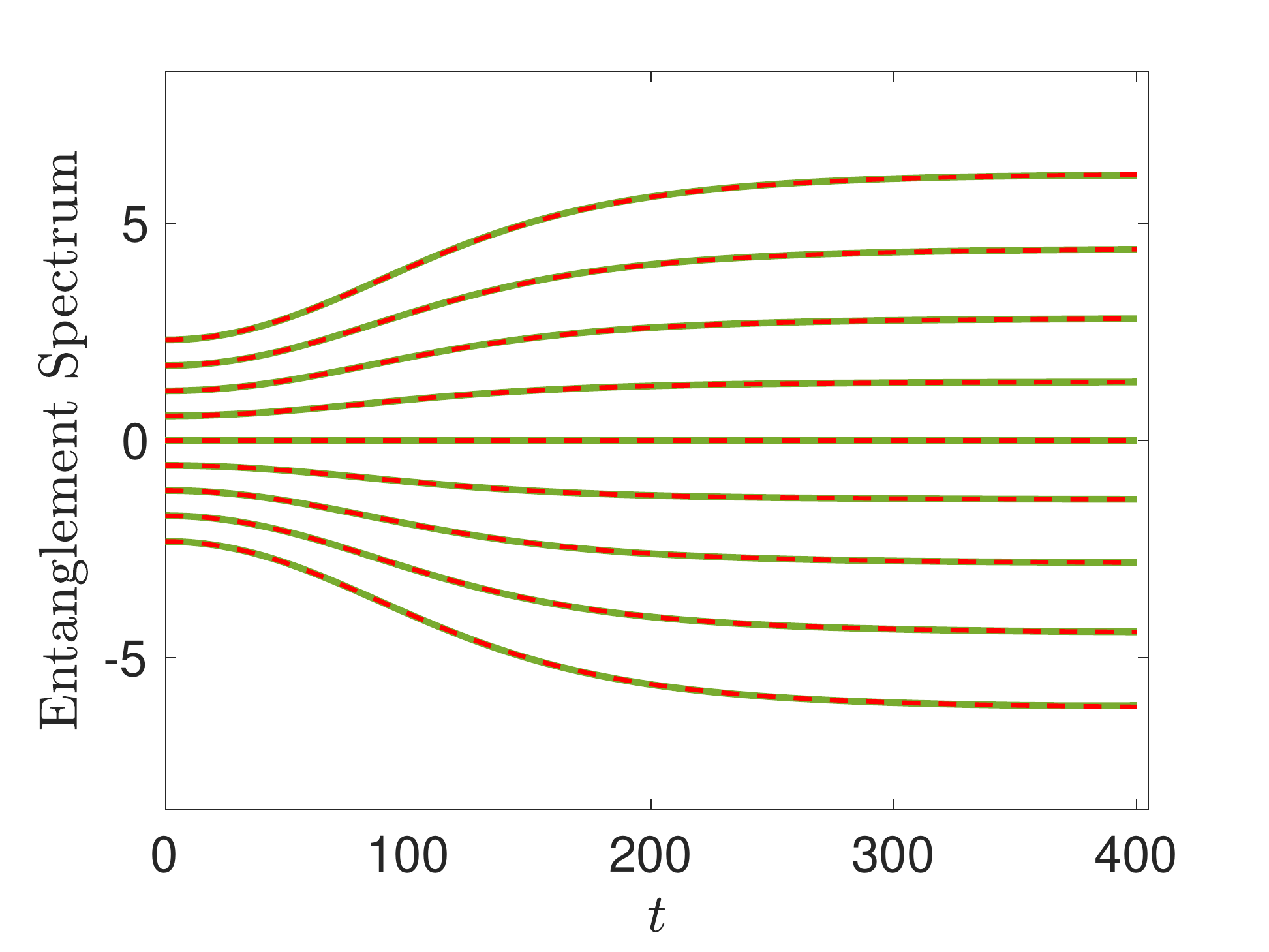}
\\
\includegraphics[width=2.2in]{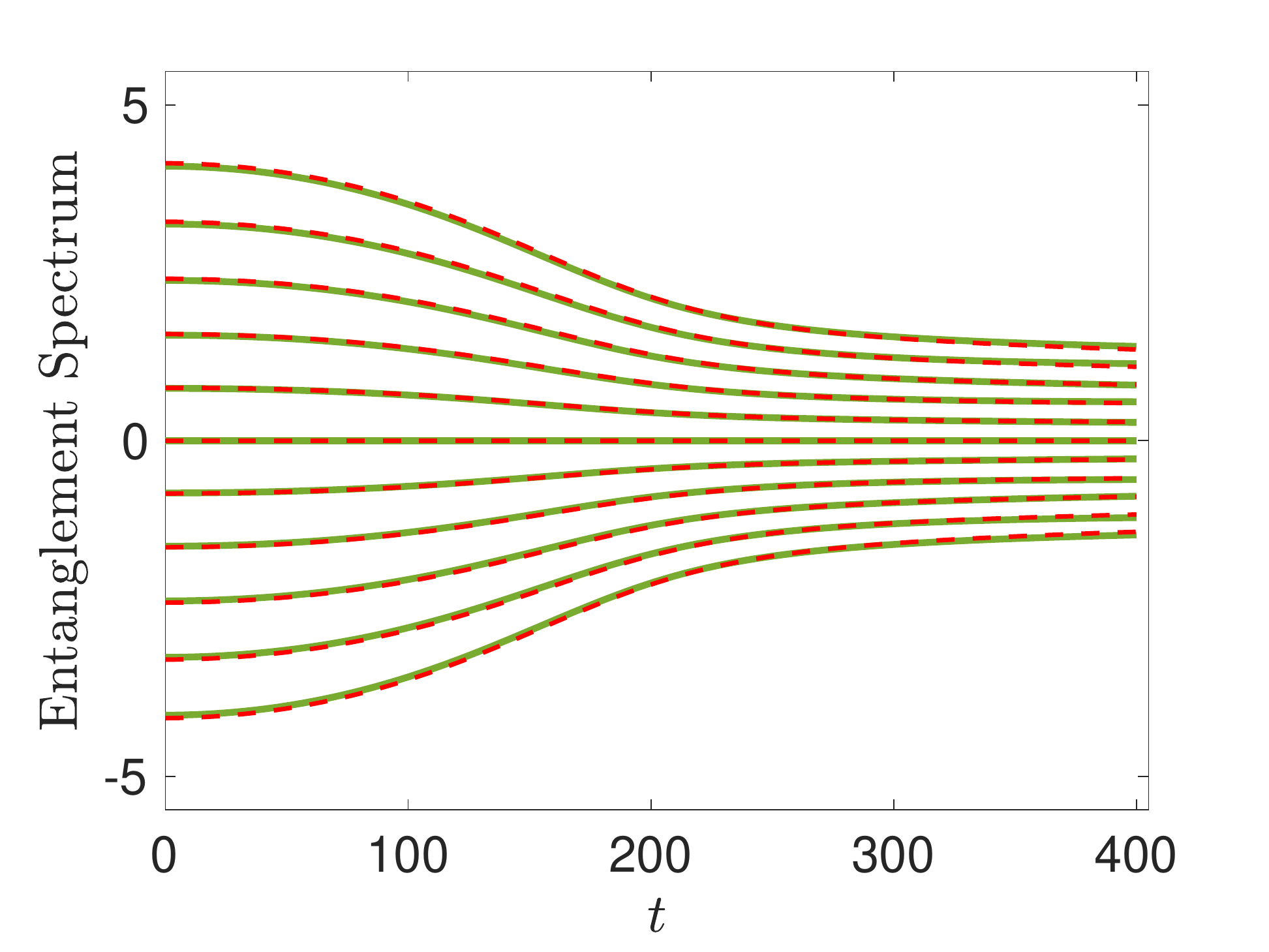}
\includegraphics[width=2.2in]{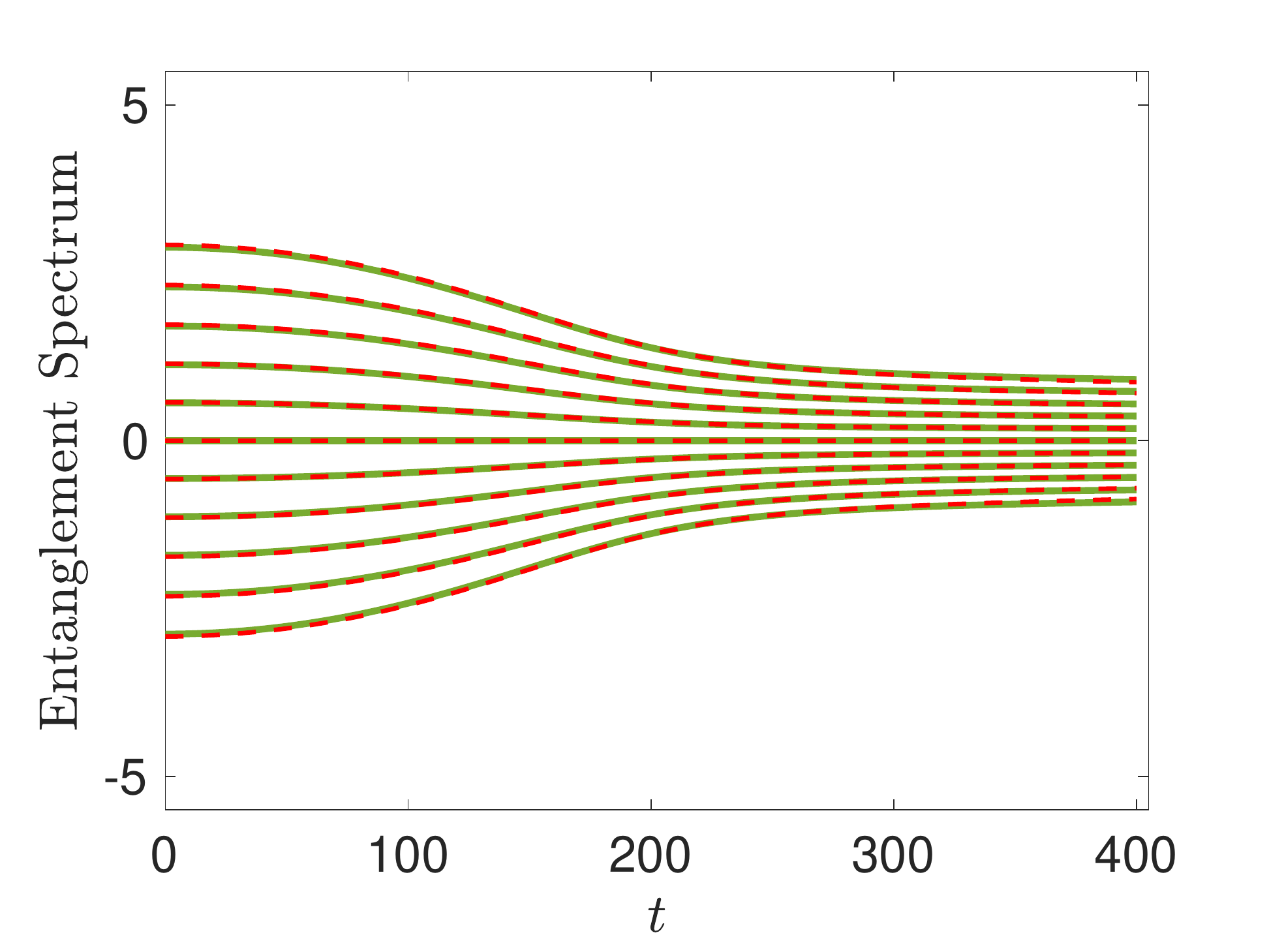}
\end{center}
\caption{
Top:
Entanglement spectrum evolution in (part of) the cooling region. 
We choose $l_A=l/4$ centered at $x=l/2$. The initial temperature is $\beta/l=1/10$ (left) and $\beta/l=3/50$ (right). 
Bottom:
Entanglement spectrum evolution in the subsystem that contains the heating region. 
We choose $l_A=l/4$ centered at $x=l$. The initial temperature is $\beta/l=1/10$ (left) and $\beta/l=3/50$ (right). 
The green circles are lattice results based on the calculation of correlation matrix.
The red dashed lines are obtained from diagonalizing $K_A$ in \eqref{KA_t_SM2} and \eqref{EH_finiteT_quench_SM2} (which is obtained from CFT calculations) on the lattice directly.
}
\label{ES_cooling_Quench}
\end{figure}

The time evolution of entanglement spectrum in both the cooling and heating regions are shown in Fig.\ref{ES_cooling_Quench}, with similar features as the Floquet case. In the cooling region, the spacing of entanglement spectrum grows and saturates because the subsystem is cooled down to the ground state. In the heating region, the spacing of entanglement spectrum decreases in time, because the entanglement entropy in the subsystem grows in time.

\bigskip

The above discussion can be generalized to the quenched Hamiltonian with a general deformation of the form $H=\int v(x) T_{00}(x)\, dx$, 
as long as there are fixed points in the operator evolution. See the appendix in \cite{fan2020General} for more discussions.
It is also an interesting future problem to check other setups of inhomogeneous quantum quenches \cite{2017Dubail,2019Moosavi,2019Moosavi2}.

\section{Free-fermion lattice calculations}

Here we present details on the lattice free-fermion calculations of entanglement entropy, energy density and entanglement spectrum evolution.
All these quantities can be calculated based on the single-particle correlation matrix.

Let us first consider the case without driving.
The single particle correlation matrix is
\be
C_{ij}=
\text{Tr}\left(\rho_{\text{th}}\, c_i^\dag c_j\right)=\sum_n [U^\dag]_{ni} U_{jn} \, f(\epsilon_n)
\ee
where 
\be
f(\epsilon_n)=(1+e^{\beta (\epsilon_n-\mu)})^{-1}
\ee
is the Fermi-Dirac distribution.
For the ground state, the Fermi-Dirac distribution at zero temperature automatically enforces the summation to be over the lowest occupied states. $U$ is the unitary matrix that diagonalizes $H_0$ and $\epsilon_n$ are the single-particle eigen-energies.
More concretely, for $H_0=\sum_{ij}h_{ij}c_i^\dag c_j+h.c.$, it can be diagonalized as
$H_0=\sum_{i=1}^L \epsilon_i \gamma_i^\dag \gamma_i$, with
$c_n=\sum_i U_{ni}\gamma_i$ and $c_n^\dag=\sum_i\gamma_i^\dag U^\ast_{ni}=\sum_i\gamma_i^\dag (U^\dag)_{in}$.

\subsection{A single quench}

In the single quantum quench, we consider the following simple choice of quenched Hamiltonian:
\begin{equation}
\label{H_theta}
    H_1=-\frac{1}{2}\sum_{j=1}^{L-1}v(j)
c_j^{\dag}c_{j+1}+h.c.
\end{equation}
where $v(j)$ characterizes the deformation of Hamiltonian.
The single particle correlation function is
\be
C_{ij}=\text{Tr}\left(\rho(t)\, c_i^\dag c_j\right)=
\text{Tr}\left(U(t) \rho_{\text{th}} U^\dag(t) \, c_i^\dag c_j\right)
=
\text{Tr}\left( \rho_{\text{th}} U^\dag(t) \, c_i^\dag c_j U(t)\right)
=
\text{Tr}\left(\rho_{\text{th}}\, c_i^\dag(t) c_j(t)\right),
\ee
where $c_i^\dag(t)=U^\dag(t) c_i^\dag U(t)$ denotes the operator evolution in the
Heisenberg picture, and similarly for $c_j(t)$.
In the single quantum quench, suppose we diagonalize $H_1$ with a unitary transformation $V$,
then one can obtain
\be
e^{iH_1t} c_i^\dag e^{-iH_1t}=\sum_{m,n}c_n^\dag V_{nm} \cdot e^{i\epsilon_m t}\cdot   [V^\dag]_{mi}
=:\sum_n c_n^\dag [W^\dag]_{ni},
\ee
and 
\be
e^{iH_1t} c_j e^{-iH_1t}=\sum_{m'n'}V_{jm'} e^{-i\epsilon_{m'} t}[V^\dag]_{m'n'} c_{n'}
=:\sum_n W_{jn'} c_{n'},
\ee
where $\epsilon^1_m$ is the single-particle eigen-energy of $H_1$, and we have defined
$
W_{ij}=\sum_k V_{ik}\cdot  e^{-i\epsilon^1_k t}\cdot [V^\dag]_{kj}.
$
Then one can obtain 
\be
\label{rho_t_thermal2}
\begin{split}
C_{ij}=&\text{Tr}\left(\rho(t)\, c_i^\dag c_j\right)=
\text{Tr}\left(\rho_{\text{th}}\, c_i^\dag(t) c_j(t)\right)\\
=&\text{Tr}\left(\rho_{\text{th}}\, \sum_n c_n^\dag [W^\dag]_{ni}\sum_n W_{jn'} c_{n'}
\right)
=
 \sum_n [W^\dag]_{ni}\sum_{n'} W_{jn'} 
\sum_m [U^\dag]_{mn} U_{n'm} f(\epsilon^0_m)\\
=&
\sum_m [W U]_{jm} [U^\dag W^\dag]_{mi}\cdot f(\epsilon_m^0).
\end{split}
\ee
Note that $\epsilon^0_m$ denotes the single particle eigen-energy of $H_0$.

\subsection{Time-dependent driving}

It is straightforward to generalize the quantum quench to the case of time-dependent drivings, such as periodic drivings. For example, let us first consider the 2-step periodic driving with Hamiltonians $H_1$ and $H_2$ with time durations $T_1$ and $T_2$ respectively.
Suppose we diagonalize $H_1$ ($H_2$) with a unitary transformation $V_1$ ($V_2$), one needs to consider 
$U^\dag(T_1,T_2) \,c_i^\dag \, U (T_1,T_2)$ and $U^\dag (T_1,T_2)\, c_j \, U (T_1,T_2)$, where 
$U(T_1,T_2)=e^{-iH_2T_2}e^{-iH_1T_1}$.
Note that
\be
e^{iH_2T_2} c_i^\dag e^{-iH_2T_2}=\sum_{m,n}c_n^\dag V_{2,nm} \cdot e^{i\epsilon_{2,m} T_2}\cdot   [V_2^\dag]_{mi}
=:\sum_n c_n^\dag [W_2^\dag]_{ni},
\ee
and 
\be
e^{iH_2T_2} c_j e^{-iH_2T_2}=\sum_{m'n'}V_{2,jm'} e^{-i\epsilon_{2,m'} T_2}[V^\dag]_{2,m'n'} c_{n'}
=:\sum_{n'} W_{2,jn'} c_{n'},
\ee
where $\epsilon_{2,m}$ is the single-particle eigen-energy of $H_2$, and we have defined
$
W_{2,ij}=\sum_k V_{2,ik}\cdot  e^{-i\epsilon_{2,k} T_2}\cdot [V_2^\dag]_{kj}.
$
Similarly, one has
\be
\begin{split}
&e^{iH_1T_1}e^{iH_2T_2} c_i^\dag e^{-iH_2T_2}e^{-iH_1T_1}
=\sum_n e^{iH_1T_1}c_n^\dag e^{-iH_1T_1} [W_2^\dag]_{ni}=\sum_{n,m}c_m^\dag [W_1^\dag]_{mn}[W_2^\dag]_{ni}
=\sum_m c_m^\dag [W_1^\dag\cdot W_2^\dag]_{mi},
\end{split}
\ee
\be
\begin{split}
&e^{iH_1T_1}e^{iH_2T_2} c_j e^{-iH_2T_2}e^{-iH_1T_1}
=\sum_{n} [W_2]_{jn} e^{iH_1T_1} c_{n} e^{-iH_1T_1}=
\sum_{n,m} [W_2]_{jn} [W_1]_{nm}  c_{m}= \sum_m [W_2\cdot W_1]_{jm} c_m,
\end{split}
\ee
where we have defined
\be
W_{\alpha,ij}=\sum_k [V_{\alpha}]_{ik}\cdot  e^{-i\epsilon_{\alpha,k} T_\alpha}\cdot [V_\alpha^\dag]_{kj},\quad \alpha=1,\,2.
\ee
Then one can obtain 
\be
\label{rho_t_thermal2}
\begin{split}
C_{ij}=&\text{Tr}\left(\rho(t)\, c_i^\dag c_j\right)=
\text{Tr}\left(\rho_{\text{th}}\, c_i^\dag(t) c_j(t)\right)
=
\sum_m [W U]_{jm} [U^\dag W^\dag]_{mi}\cdot f(\epsilon_m^0),
\end{split}
\ee
where $W=W_2\cdot W_1$.
Note that $\epsilon^0_m$ denotes the single particle eigen-energy of $H_0$.
Then in the two-step periodic driving, one simply has 
\be
W=[W_2\cdot W_1]^n,\quad n\in \mathbb Z.
\ee
If we consider a $k$-step periodic driving, where there are $k$ different driving Hamiltonians within one period, then one has 
$
W=[W_k\cdots W_1]^n.
$

\bigskip
Once we obtain the correlation matrix $C_{ij}$, we can obtain the entanglement entropy and entanglement spectrum
\cite{Peschel2002,Eisler2017}. The time evolution of the energy density $\langle T_{00}(x)\rangle$ can also be obtained from $C_{ij}$ considering that $T_{00}(x)$ is of the form $T_{00}(x)=-\frac{1}{2}c_x^\dag c_{x+1}+h.c.$.

\setcounter{equation}{0}
\setcounter{figure}{0}
\setcounter{table}{0}
\setcounter{page}{1}
\makeatletter
\renewcommand{\theequation}{S\arabic{equation}}
\renewcommand{\thefigure}{S\arabic{figure}}
\renewcommand{\bibnumfmt}[1]{[S#1]}
\renewcommand{\citenumfont}[1]{S#1}

\end{document}